\documentclass[11pt,reqno, a4paper]{amsart} 

\usepackage{epsfig}
\usepackage{graphicx}
\usepackage{enumerate}
\usepackage{color} 
\usepackage[toc,page]{appendix}
\usepackage[utf8]{inputenc}

\usepackage[dvipsnames]{xcolor}
\usepackage{bm,bbm}
\usepackage{cancel}
\usepackage{enumitem}

\usepackage{physics}

\usepackage{amsmath,euscript}
\usepackage{amsthm}
\usepackage{amssymb}
\usepackage{graphicx}
\usepackage{mathrsfs}
\usepackage{epsf}
\usepackage{xcolor}
\usepackage{psfrag}
\usepackage{enumerate}
\usepackage{amsfonts,amssymb,amsthm,amsmath}  
\usepackage{hyperref}

\usepackage[numbers]{natbib}
\usepackage{enumitem}  
\usepackage{xspace}
\usepackage[margin=1.40in]{geometry}

\usepackage{changes}
\definechangesauthor[name={SB}, color=magenta]{sb}
\definechangesauthor[name={JF}, color=blue]{jf}
\definechangesauthor[name={ML}, color=red]{ml}
\definechangesauthor[name={MS}, color=green]{ms}

\newtheorem{thm}{Theorem}

\newtheorem{lemma}[thm]{Lemma}
\newtheorem{prop}[thm]{Proposition}
\newtheorem{cor}[thm]{Corollary}

\theoremstyle{remark}
\newtheorem{remark}[thm]{Remark}

\theoremstyle{definition}

\numberwithin{thm}{section}
\numberwithin{equation}{section}

\definecolor{green}{rgb}{0.0, 0.5, 0.5}
\definecolor{yellow}{rgb}{0.5, 0.5, 0}
\definecolor{lgray}{gray}{0.9}
\definecolor{llgray}{gray}{0.95}
\definecolor{lllgray}{gray}{0.975}

\newcommand{\nc}{\newcommand}

	\nc{\la}{\label}
	\nc{\ba}{\begin{array}}
		\nc{\ea}{\end{array}}
	\nc{\bs}{\begin{split}}
		\nc{\es}{\end{split}}

	\newcommand{\R}{\mathbb{R}}
	\newcommand{\C}{\mathbb{C}}

	\newcommand{\cE}{\mathcal{E}}
	
	\newcommand{\cG}{\mathcal{G}}

	\newcommand{\cK}{\mathcal{K}}
	
	\newcommand{\cP}{\mathcal{P}}         
	\newcommand{\cQ}{\mathcal{Q}}
	
	\newcommand{\cS}{\mathcal{S}}

	\newcommand{\cX}{\mathcal{X}}

	\newcommand{\sD}{\mathscr{D}}

	\newcommand{\fH}{\mathfrak{H}}

	\newcommand{\fS}{\mathfrak{S}}

	\newcommand{\lam}{\lambda}

	\newcommand{\inte}{\int_{\mathbb{R}^3}}
	\newcommand{\iinte}{\iint_{\R^3\times\R^3}}

	\newcommand{\ga}{\gamma}

	\def\R{{\mathbb R}}
	\def\C{{\mathbb C}}

	\newcommand{\bsub}{\begin{subequations}}
		\newcommand{\esub}{\end{subequations}$\!$}

	\nc{\ran}{\rangle}
	\nc{\lan}{\langle}

	\newcommand{\supp}{\operatorname{supp}}

	\renewcommand{\Re}{\mathrm{Re}} 

	\nc{\bfone}{{\bf 1}}
	\nc{\1}{{\bf 1}}

	\usepackage{centernot}

	\usepackage[normalem]{ulem}

	\newcommand{\Ex}{\text{Ex}}

	\newcommand{\DETAILS}[1]{}

	\nc{\den}{\text{den}}
	\nc{\ex}{\text{xc}}

	\usepackage{soul}

	\newcommand{\nn}{{\nonumber}}
	
	\allowdisplaybreaks

	\title[HF and HFB theories with Yukawa potential]{Pseudo-relativistic fermionic systems with attractive Yukawa potential
	}

	\author[B. Chen]{Bin Chen}
	\address{(Bin Chen) School of Mathematics and Statistics,  Key Laboratory of Nonlinear Analysis $\&$ Applications (Ministry of Education), Central China Normal University, Wuhan 430079, P. R. China; and Academy of Mathematics and Systems Science, Chinese Academy of Sciences, Beijing 100190, P. R. China
	}
	\email{binchen@amss.ac.cn}

	\author[Y. Guo]{Yujin Guo}
	\address{ (Yujin Guo) School of Mathematics and Statistics,  Key Laboratory of Nonlinear Analysis $\&$ Applications (Ministry of Education), Central China Normal University, Wuhan 430079, P. R. China}
	\email{yguo@ccnu.edu.cn}

	\author[P.T. Nam]{Phan Th\`anh Nam}
	\address{(Phan Th\`anh Nam) Department of Mathematics, LMU Munich, Theresienstrasse 39, 80333 Munich, Germany}
	\email{nam@math.lmu.de}

	\author[D.H. Ou Yang]{Dong Hao Ou Yang}
	\address{(Dong Hao Ou Yang) Department of Mathematics, LMU Munich, Theresienstrasse 39, 80333 Munich, Germany}
	\email{ouyang@math.lmu.de}
	
\begin{document}
	
\begin{abstract} We study the Hartree-Fock and Hartree-Fock-Bogoliubov theories for a large fermionic system with the pseudo-relativistic kinetic energy and an attractive Yukawa interaction potential. We prove that the system is stable if and only if the total mass does not excess a critical value, and investigate the existence and properties of ground states in both sub-critical and critical mass regimes. 
\end{abstract}

\date{\today}

\maketitle

\tableofcontents

\section{Introduction and main results}\label{sec:intro}

In 1931, Chandrasekhar \cite{Chandrasekhar-31} predicted gravitational collapse for neutron stars and white dwarfs: a pseudo-relativistic system with attractive Newtonian interaction has a {\em critical mass} such that the ground state energy of the system is finite. In standard units, this critical mass, also called the {\em Chandrasekhar limit}, is approximately 1.4 times the mass of the Sun. In \cite{LieYau-87}, Lieb and Yau derived the Chandrasekhar theory as a Thomas--Fermi-type variational problem from the many-fermion Schrödinger theory with the pseudo-relativistic kinetic energy operator
\begin{align}
T:=\sqrt{-\Delta+m^{2}}-m,
\end{align}
where $m>0$ is the mass of a single fermion. Within the Hartree--Fock (HF) and Hartree--Fock--Bogoliubov (HFB) theories, the existence and properties of the ground states have been studied by Lenzmann and Lewin \cite{LenLew-10}. They are in particular interesting because the many-body Schrödinger ground state does not exist,  due to the translation-invariant nature of the system, but the nonlinear ground states still exist in the whole subcritical mass regime. The existence problem was also  investigated recently  in the critical mass regime in \cite{CGNO-25}.

In the present paper, we are interested in extending the studies in \cite{LenLew-10,CGNO-25} to a pseudo-relativistic system with a Yukawa-like correction to the Newtonian potential. Typically, the Yukawa potential serves as a natural regularization of the gravitational interaction at long distances, making it a convenient model for studying the stability and structure of self-gravitating quantum systems under the effect of screening; see, e.g., \cite{CST-07,FFCCV-20} for discussions of the relevance of such a correction. To be precise, we will restrict our consideration to the standard attractive Yukawa potential of the form
\[
V(x) = - \frac{e^{-\theta |x|}}{|x|}, \quad x \in \mathbb{R}^3,
\]
where $\theta \ge 0$ is a given parameter. The case $\theta = 0$ corresponds to the standard Newtonian potential for neutron stars and white dwarfs studied in \cite{LenLew-10,CGNO-25}, while the case $\theta > 0$ models a short-range interaction. As we will see, the critical mass is independent of $\theta$, and it is interesting to understand how this affects the existence and properties of ground states in both the subcritical mass and critical mass regimes. We will represent the HFB and HF theories separately.

\subsection{Hatree--Fock--Bogoliubov theory} \label{sec:main-HFB}
The HFB energy functional under the Yukawa potential is defined in terms of the \textit{one-body density operator} $\gamma$ (positive, trace-class) and the \textit{pairing matrix} $\alpha$ (anti-symmetric, Hilbert-Schmidt), acting on $L^{2}(\R^{3};\C^{q})$, by
\begin{align}\label{HFB-energy}
\cE^{\rm HFB}_{m,\kappa,\theta}(\gamma,\alpha)&=\Tr(T\gamma)-\frac{\kappa}{2}\iinte\frac{e^{-\theta|x-y|}}{|x-y|}\Big(\rho_{\gamma}(x)\rho_{\gamma}(y)-|\gamma(x,y)|^2\Big)dxdy\\
&\quad-\frac{\kappa}{2}\iinte\frac{e^{-\theta|x-y|}|\alpha(x,y)|^{2}}{|x-y|}dxdy. \nn
\end{align}
Here $T:=\sqrt{-\Delta+m^{2}}-m$ with rest mass $m\geq 0$, and $q\in\mathbb{N}^+$ describes the internal degrees (e.g., spin) of freedom for particles. The coupling constant $\kappa>0$ is used to model the strength of the interaction, and  $\rho_{\gamma}(x):=|\gamma(x,x)|^{2}$ denotes the density of the particles. In particular,
$$
\Tr \gamma = \int_{
\R^3} \rho_\gamma(x) d x 
$$
is the total number of particles for the system. The three last terms on the right-hand side of \eqref{HFB-energy} are called the \textit{direct, exchange, and  pairing terms}, respectively \cite{BacLieSol-94}. 

In order to ensure that the energy functional in \eqref{HFB-energy} is well-defined, following the formulation in \cite{LenLew-10} we restrict our attention to a class of quasi-free states with finite pseudo-relativistic kinetic energy. To this end, we introduce the (real) Banach space of density matrices
\begin{align}\label{X}
\cX &= \big\{(\gamma,\alpha) \in \fS_{1} \times \fS_{2}:\,  \gamma^{*} = \gamma,\ \alpha^{T} = -\alpha,\ \|(\gamma,\alpha)\|_{\cX} < \infty \big\},
\end{align}
equipped with the norm
\begin{align}\label{X-norm}
	\|(\gamma,\alpha)\|_{\cX} &:= \|(1-\Delta)^{1/4} \gamma (1-\Delta)^{1/4}\|_{\fS_{1}} + \|(1-\Delta)^{1/4} \alpha\|_{\fS_{2}}.
\end{align}
Here $\fS_{1}$ and $\fS_{2}$ denote the spaces of trace-class and Hilbert--Schmidt operators on $L^{2}(\R^{3};\C^{q})$, respectively. Moreover, due to the fermionic property of the particles, we have to assume a supplementary condition which takes  into account Pauli's exclusion principle. This leads to the following
variational space
\begin{align*} 
\cK&:=\Big\{(\gamma,\alpha)\in\cX:\,\ \begin{pmatrix}
0	&	0\\
0	&	0
\end{pmatrix}\leq \begin{pmatrix}
\gamma	&	\alpha\\
\alpha^{*}	&	1-\bar{\gamma}
\end{pmatrix}\leq\begin{pmatrix}
1	&	0\\
0	&	1
\end{pmatrix}, \,\ \Tr(\gamma)<\infty \Big\}.
\end{align*}
Thus, for every given parameters $m\geq0$, $\kappa>0$ and $\theta\ge0$, the corresponding HFB minimization problem with  the total mass $\lambda>0$ satisfies
\begin{align} \label{eq:HFB-variational-problem}
I^{\rm HFB}_{m,\kappa,\theta}(\lambda)&:=\inf\limits_{(\gamma,\alpha)\in\cK_\lambda} \cE^{\rm HFB}_{m,\kappa,\theta}(\gamma,\alpha),
\end{align}
where
\begin{align}\label{kl}
\cK_\lambda:=\big\{(\gamma,\alpha)\in\cK:\, \Tr (\gamma)= \lambda \big\}.
\end{align}
	
We begin with the following elementary result concerning the critical mass of the HFB theory, whose proof is left to Appendix~\ref{app:pf-Chandra-Yukawa}.
	
\begin{prop}[{Chandrasekhar Limit}]\label{prop:Chandrasekhar}
Let $m\geq 0$, $0< \kappa<4/\pi$ and $\theta\geq 0$ be given.  Then there exists a unique critical mass  $\lambda^{\rm HFB}=\lambda^{\rm HFB}(\kappa)>0$  of $I^{\rm HFB}_{m,\kappa,\theta}(\lambda)$, which is independent of $m$ and $\theta$, such that the following statements hold:
\begin{enumerate}
\item [$(i)$] If $0\leq \lambda\leq \lambda^{\rm HFB}$, then $I^{\rm HFB}_{m,\kappa,\theta}(\lambda)>-\infty$.
\item [$(ii)$] If $\lambda>\lambda^{\rm HFB}$, then $I^{\rm HFB}_{m,\kappa,\theta}(\lambda)=-\infty$.
\end{enumerate}
Furthermore, the function $\lambda^{\rm HFB}(\kappa)$ is strictly decreasing, continuous with respect to $\kappa$, and satisfies the asymptotic estimate
\begin{align}\label{prop1.1:A}
\lambda^{\rm HFB}(\kappa)&\sim\left(\frac{\tau_{c}}{\kappa}\right)^{3/2}q^{-1/2}\quad\text{as }\ \kappa\rightarrow0^+,
\end{align}
where the constant $\tau_c$ is defined by
\begin{equation}\label{eq:def-tauc}
\tau_c= \inf \Big\{\frac{2 \int_{\R^3} f^{4/3}(x) dx \Big( \int_{\R^3} f(x) dx \Big)^{2/3} }{\iinte\frac{f(x)f(y)}{|x-y|}dxdy}:\,\  0\le f \in
L^{4/3}(\R^3) \cap L^1(\R^3)  \Big\}.
\end{equation}
\end{prop}

\noindent The computation of $\tau_{c}\cong 2.677$ in Proposition \ref{prop:Chandrasekhar} can be traced back to Chandrasekhar in \cite{Chandrasekhar-31}, and the formula \eqref{eq:def-tauc} can be found in \cite{LieYau-87}. The continuity and (non-strict) monotonicity of $\kappa\mapsto \lambda^{\rm HFB}(\kappa)$ were proved in \cite{LenLew-10}, while the strict monotonicity was recently justified  in \cite{CGNO-25}. By  the proof of Proposition \ref{prop:Chandrasekhar},   it follows from \cite[Remark 3]{LenLew-10} that  $\lambda^{\rm HFB}(\kappa)=0$ holds for any $\kappa>4/\pi$. This further yields that if  $\kappa>4/\pi$, then  $I^{\rm HFB}_{m,\kappa,\theta}(\lambda)=-\infty$ holds for any $m>0,\ \theta\geq0$ and $\lambda>0$.

Our first main result of the HFB theory is concerned with  the following existence of minimizers for $I^{\rm HFB}_{m,\kappa,\theta}(\lambda)$ when $0<\kappa<4/\pi$ and  $0<\lambda<\lambda^{\rm HFB}$.
	

\begin{thm}[Existence in HFB theory]\label{thm:main1}
Let $m>0$, $0<\kappa<4/\pi$ and $0<\lambda<\lambda^{\rm HFB}$, where $\lambda^{\rm HFB}=\lambda^{\rm HFB}(\kappa)$ is the critical mass of $I^{\rm HFB}_{m,\kappa,\theta}(\lambda)$. Then there exists a constant $\theta_{c}=\theta_{c}(m,\kappa,\lambda)>0$ such that  the  problem $I^{\rm HFB}_{m,\kappa,\theta}(\lambda)$  has a minimizer for any $\theta\in [0,\theta_{c}]$.  Moreover, every minimizing sequence $\{(\gamma_{n},\alpha_{n})\}$ of $I^{\rm HFB}_{m,\kappa,\theta}(\lambda)$ is, up to translations, relatively compact in $\cX$, and the following binding inequality holds:
\begin{align*}
I^{\rm HFB}_{m,\kappa,\theta}(\lambda)<I^{\rm HFB}_{m,\kappa,\theta}(\lambda')+I^{\rm HFB}_{m,\kappa,\theta}(\lambda-\lambda'),\ \   \forall\ 0<\lambda'<\lambda.
\end{align*}

\end{thm}

Theorem \ref{thm:main1} extends \cite[Theorem 1]{LenLew-10} from the case $\theta=0$ to the general case $\theta\in[0, \theta_c]$. One cannot however expect that Theorem \ref{thm:main1} holds for sufficiently  large $\theta>0$, because an eigenvalue-counting method can be used to show that the corresponding HF minimization problem has no minimizer, if $\theta>0$ is sufficiently large. While the proof of Theorem~\ref{thm:main1}  follows the same general approach of Lions' concentration-compactness principle \cite{LenLew-10,Lions1}, as illustrated in Section \ref{sec:existence-nonexistence-HFB}, its implementation is rather different from \cite{LenLew-10}, due to the strong asymptotic decay of the Yukawa potential (see Lemma \ref{lem:shell-Yukawa}). For example,  the proof of Theorem~\ref{thm:main1} needs essentially the following  exponential decay of minimizers  for $I^{\rm HFB}_{m,\kappa,\theta}(\lambda)$ when $\theta > 0$.


\begin{thm}[Properties of HFB minimizers]\label{thm:main2}
Let $m>0$, $0<\kappa<4/\pi$, $0<\lambda<\lambda^{\rm HFB}$ and $0<\theta\leq \theta_c$,  where $\theta_c>0$ is as in Theorem \ref{thm:main1}, and suppose $(\gamma,\alpha)$ is a minimizer of $I^{\rm HFB}_{m,\kappa,\theta}(\lambda)$. Then there exists a sufficiently large constant $R_0>0$ such that 
\begin{align}\label{HFB-min-decay}
\sup\limits_{R\geq R_0}\big\|\eta_{R}(e^{3\theta|\cdot|}\gamma e^{3\theta|\cdot|},e^{3\theta|\cdot|}\alpha)\eta_{R}\big\|_{\cX}<\infty,
\end{align}
where $\eta_R\in C^\infty(\R^3;[0,1])$ is a smooth cutoff function satisfying  $\eta_R(x)=1$ for $|x|>2R$ and $\eta_R(x)=0$ for $|x|< R$.
In particular, the density function $\rho_{\gamma}(x)$ satisfies
\begin{align}\label{HFB-min-decay2}
\int_{|x|\geq R}\rho_{\gamma}(x)dx\leq Ce^{-6\theta R}\ \  \text{for sufficiently large}\ R>0,
\end{align}
where $C>0$ is  independent of $R>0$.
\end{thm}


The restrictions $m >0$ and $\theta > 0$ are crucial for deriving the exponential decay (\ref{HFB-min-decay}) and (\ref{HFB-min-decay2}). For example, when $m = 0$, one may expect that the exponential decay (\ref{HFB-min-decay}) fails in view of \cite[Theorem 1.4]{CGNO-25}. Moreover, when $m>0$ and $\theta = 0$, Lenzmann and Lewin \cite{LenLew-10} proved that the right-hand side of \eqref{HFB-min-decay2} is bounded by $O(R^{-k})$ for any $k>0$. However, the decaying analysis in \cite{LenLew-10}  cannot be directly used when $\theta>0$.

We now sketch the main idea of proving \eqref{HFB-min-decay}, from which \eqref{HFB-min-decay2} follows. We will use the mean-field (Euler--Lagrange) equation of $(\gamma,\alpha)$ to derive the pointwise inequalities that propagate at the exponential decay, in the spirit of Dall'Acqua, S{\o}rensen and Stockmeyer \cite{DalSoeSto-08} for Hartree--Fock atomic ground states. The novelty here is that the HFB equation couples $\gamma$ and $\alpha$ in a nontrivial way. Suppose $(\gamma,\alpha)\neq(\gamma, 0)$ is a minimizer of $I^{\rm HFB}_{m,\kappa,\theta}(\lambda)$ for sufficiently small $\theta>0$, and denote $L_{\gamma,\theta}:=\sqrt{-\Delta+m^{2}}-m-\kappa (|\cdot|^{-1}e^{-\theta|\cdot|})*\rho_{\gamma}$.  
By analyzing the mean field equation of $(\gamma,\alpha)$, together with some technical calculations of Proposition \ref{prop:esti-g-a-pos-a-wave} and \eqref{decay14},  we first prove that there exists $\tau>0$ such that for any sufficiently large $R>0$,    
$$\text{Tr}\big(L_{\gamma,\theta}\eta_Re^{\tau|\cdot|}\gamma e^{\tau|\cdot|}\eta_R\big)\le N(R)+C  \big\|\eta_Re^{\tau |x|}\rho_{\gamma}^{1/2}\big\|_{L^2}.$$
On the other hand,  we also prove that there exists a constant $\mu<0$ such that  $\mu_\theta\leq\mu<0$ holds for any sufficiently small $\theta>0$, where  $\mu_\theta$  denotes the corresponding Lagrange multiplier of the minimizer $(\gamma,\alpha)$ for $I^{\rm HFB}_{m,\kappa,\theta}(\lambda)$. By proving that 
for any compact domain $\Omega\subset\R^3$,
\begin{align}\label{A:1.7a}
	\inf\big\{\langle\psi,L_{\gamma,\theta}\psi\rangle/\|\psi\|_{L^2}^{2}:\, \psi\in H^{1/2}(\R^{3}),\psi=0\text{ on }\Omega\big\}\leq0,
\end{align}
we shall further derive that    $$\big\|\eta_Re^{\tau |x|}\rho_{\gamma}^{1/2}\big\|_{L^2}\le C N(R).$$
As a consequence of above results, in Section \ref{sec:pf-main2} we are finally able to deduce that \eqref{HFB-min-decay} holds for some appropriately small $\theta>0$.

\subsection{Hartree--Fock theory} Now for every fixed $m>0, \kappa>0,\ \theta\ge 0,\ 2\leq N\in\mathbb{N}$, we consider the Hartree-Fock (HF) minimization problem  
\begin{equation}\label{problem}
I^{\rm HF}_{m,\kappa,\theta}(N):=\inf\limits_{\gamma \in \mathcal{P}_N} \mathcal{E}^{\rm HF}_{m,\kappa,\theta}(\gamma)
\end{equation}
with the  HF energy functional  
\begin{equation*}\label{functional}
\mathcal{E}^{\rm HF}_{m,\kappa,\theta}(\gamma):=\mathrm{Tr} (T\gamma) -\frac{\kappa}{2}\iinte\frac{e^{-\theta|x-y|}}{|x-y|}\Big(\rho_{\gamma}(x)\rho_{\gamma}(y)-|\gamma(x,y)|^2\Big)dxdy,
\end{equation*}
and  the set of  Slater determinants 
\begin{equation}\label{set}
\cP_{N}:=\Big\{\gamma=\sum\limits_{j=1}^{N}|u_{j}\rangle\langle u_{j}|:\	u_{j}\in H^{1/2}(\R^3, \C),\ \langle
u_{j},u_k\rangle=\delta_{jk},\ j, k=1,\cdots, N\Big\}.
\end{equation}
Here we have ignored the spin of the particles for simplicity.

In our recent work \cite{CGNO-25}, we studied the problem $I^{\rm HF}_{m,\kappa,\theta}(N)$ for the case $\theta=0$ and found that it is fully characterized by  the following Gagliardo--Nirenberg type inequality: for $2\leq N\in\mathbb{N}$,
\begin{equation}\label{eq:GN-HF}
\begin{split}
2\|\gamma\|\mathrm{Tr}\big(\sqrt{-\Delta}\, \gamma\big)
\geq \kappa^{\rm HF}_N \iinte\frac{\rho_{\gamma}(x)\rho_{\gamma}(y)-|\gamma(x, y)|^2}{|x-y|}dxdy,
\ \   \forall\ \gamma\in\mathcal{R}_N,
\end{split}
\end{equation}
where the set  $\mathcal{R}_N$ is defined by 
\begin{equation}\label{rn1}
	\mathcal{R}_N:=\Big\{\gamma:\, 0\leq\gamma=\gamma^*,\ \mathrm{Rank}(\gamma)\leq N, \ \mathrm{Tr}\big(\sqrt{-\Delta}\, \gamma\big)<\infty\Big\}.
\end{equation}
We proved in \cite{CGNO-25} that the optimal constant  $\kappa^{\rm HF}_N\in(0, \infty)$ of \eqref{eq:GN-HF} can be attained among the set  $\mathcal{R}_N$, and  any optimizer of \eqref{eq:GN-HF} is an orthogonal projection operator of rank $N$. More precisely, following \cite[Theorem 1.4]{CGNO-25} and its proof, we have:

\begin{prop}[Gagliardo--Nirenberg inequality in HF theory \cite{CGNO-25}]\label{th2.1} For every $2\le N\in\mathbb{N}$, the best constant $\kappa^{\rm HF}_N$ of (\ref{eq:GN-HF}) can be attained.
Moreover, we have
\begin{enumerate}
\item[$(i)$]  The sequence $\{\kappa^{\rm HF}_N\}_{N\ge 2}$ is strictly decreasing in $N$, and  $\kappa^{\rm HF}_N\sim
\tau_c N^{-2/3}$ as $N\to \infty$, where $\tau_c>0$ is given in \eqref{eq:def-tauc}.
	
\item[$(ii)$]  Any optimizer $\gamma$ of $\kappa^{\rm HF}_N$ satisfying $\|\gamma\|=1$ can be written in the form
$\gamma=\sum_{j=1}^{N}|w_{j}\rangle\langle w_{j}|$, where the orthonormal  functions $w_1, \cdots, w_N$ are the
eigenfunctions of
\begin{equation*}\label{h8}
H_{\gamma}:=\sqrt{-\Delta}-\kappa^{\rm HF}_N\big(\rho_{\gamma}\ast|\cdot|^{-1}\big)(x)
+\kappa^{\rm HF}_N\frac{\gamma(x,y)}{|x-y|}\ \,\ \text{on} \ \, L^2(\R^3,\C),
\end{equation*}
together with negative eigenvalues $\nu_{1}\leq\nu_{2}\leq\cdots \leq \nu_{N}<0$,  and $w_{j}\in C^\infty(\R^3,\C)$ satisfies the decaying property
\begin{equation}\label{decay}
\rho_{\gamma}(x)=\sum_{j=1}^{N}|w_{j}(x)|^{2}\leq C\big(1+|x|\big)^{-8}\ \ \text{in}\ \  \R^3.
\end{equation}
\end{enumerate}
\end{prop}

Our last main result is the following existence and nonexistence of minimizers for the problem $I^{\rm HF}_{m,\kappa,\theta}(N)$ defined in \eqref{problem}.

\begin{thm}[Existence and nonexistence in HF theory]\label{th1}
For any fixed $m>0$ and  $2\leq N\in\mathbb{N}$, let $\kappa_N^{\rm HF}$  and $I^{\rm HF}_{m,\kappa,\theta}(N)$ be defined by  \eqref{eq:GN-HF} and \eqref{problem}, respectively. Then there exists a constant $C(N)>0$, depending only on $N\in \mathbb{N}$, such that the following conclusions hold true:
\begin{enumerate}
\item[$(i)$]   If  $0<\kappa<\kappa_N^{\rm HF}$, then  $I^{\rm HF}_{m,\kappa,\theta}(N)$ has a minimizer for any $0\leq \theta\leq m\kappa C(N)$. Moreover, any minimizer $\gamma$ of $I^{\rm HF}_{m,\kappa,\theta}(N)$ can be written in the form  $\gamma=\sum_{j=1} ^N|u_j\rangle \langle u_j|$, where $\langle u_i, u_j\rangle =\delta_{ij}$ and $(u_1, \cdots, u_N)$ solves the following  system
\begin{equation}\label{equation}
\begin{split}
H_{\gamma}u_j=\mu_j u_j\ \ \text{in}\, \ \R^3, \ \ j=1,\cdots, N.
\end{split}
\end{equation}
Here  $\mu_1 \leq \mu_2\leq\cdots\leq\mu_N< 0$ are  $N$ negative  eigenvalues of the operator
\begin{equation*}\label{operator} H_{\gamma}:=T-\kappa\inte\frac{e^{-\theta|x-y|}\rho_\gamma(y)}{|x-y|}dy +\kappa \, \frac{e^{-\theta|x-y|}}{|x-y|}\gamma(x,y)\ \   on\ \ L^2(\R^3,\C).
\end{equation*}

\item[$(ii)$]  If $\kappa= \kappa^{\rm HF}_N$, then there exists a small constant $\theta^*=\theta^*(m, N)\in(0, m\kappa^{\rm HF}_NC(N)]$ such that $I^{\rm HF}_{m,\kappa,\theta}(N)\in (-\infty,0)$ and $I^{\rm HF}_{m,\kappa,\theta}(N)$ does not admit any minimizer for all $0\leq\theta\leq \theta^{*}$.

\item[$(iii)$]  If $\kappa> \kappa^{\rm HF}_N$, then $I^{\rm HF}_{m,\kappa,\theta}(N)=-\infty$ for any $\theta\geq 0$.
\end{enumerate}
\end{thm}

The existence of Theorem \ref{th1} (i) extends the results of \cite[Theorem 4]{LenLew-10} from the case $\theta=0$ to the general case $\theta\in [0,m\kappa C(N)]$, and the nonexistence of Theorem \ref{th1} (ii) extends \cite[Theorem 1.5]{CGNO-25}.  
Moreover, our method of  proving  Theorem \ref{th1} (ii) is also valid essentially for the bosonic case, which then addresses a question left  after \cite[Theorem 1.2]{Guo17}.

\begin{remark}[Blow-up profile] For every fixed $m>0$, $2\leq N\in\mathbb{N}$ and $\theta \in \big[0, \theta^{**}\big)$ with $\theta^{**}=\theta^{**}(m, N)>0$ sufficiently small, we can also study the limiting behavior of minimizers for   $I^{\rm HF}_{m,\kappa,\theta}(N)$ as $\kappa \nearrow \kappa_N^{\rm HF}$.  To be precise, up to a subsequence as $\kappa_n\nearrow \kappa^{\rm HF}_N$, the HF minimizer $\gamma_{\kappa_n}=\sum_{j=1}^N|u^{\kappa_n}_j\rangle \langle u^{\kappa_n}_j|$ established in Theorem \ref{th1} satisfies 
\begin{equation}\label{1.23a}
\epsilon_{n}^{\frac{3}{2}}u_j^{\kappa_n}(\epsilon_{n}(x+z_n)) \rightarrow w_j(x) 
\end{equation}
strongly in $H^{\frac{1}{2}}(\R^3,\C)\cap L^\infty(\R^3,\C)$ as $n\rightarrow\infty$, where  $\gamma^*:=\sum_{j=1} ^N|w_j\rangle \langle w_j|$  is an optimizer of the interpolation inequality \eqref{eq:GN-HF} and  
\begin{align}\label{1.12b}
\frac{ \kappa_N^{\rm HF}-\kappa_n}{\kappa_N^{\rm HF}\epsilon_{n}^2} =& \frac{m^2}{2} \Tr \frac{\gamma^*}{\sqrt{-\Delta}}\\
&-\frac{\theta^2 \kappa_N^{\rm HF}}{4}\iinte|x-y|\Big(\rho_{\gamma^*}(x)\rho_{\gamma^*}(y)-|\gamma^*(x,y)|^2\Big)dxdy.\nonumber
\end{align}
The limiting behavior of \eqref{1.23a}--\eqref{1.12b} can be compared with the result of the bosonic system addressed in \cite{Guo17}, where the Yukawa potential $e^{-\theta|x|}|x|^{-1}$ is replaced by a function $K(x)$ satisfying $\lim\limits_{|x|\to0}\frac{1-K(x)}{|x|^p}\in(0,\infty)$, as well as our recent result \cite[Theorem 1.6]{CGNO-25} for $\theta=0$, where $\theta>0$ leads to an interesting effect in the coefficient of \eqref{1.12b}. This blow-up result can be proved by following the analysis in \cite[Theorem 1.6]{CGNO-25} and we omit  the details for simplicity. 
\end{remark}

\bigskip
\noindent{\bf Organization of the paper}.  In Section \ref{sec:HFB-min}, we recall some fundamental properties of the HFB theory. The decaying property of Theorem \ref{thm:main2} is first  proved in Section \ref{sec:pf-main2}.  In Section \ref{sec:existence-nonexistence-HFB}, we then prove  Theorem \ref{thm:main1} on the existence of  HFB minimizers, using the concentration-compactness method.  The existence and nonexistence of HF minimizers in Theorem \ref{th1}  are further proved in  Section \ref{sec:existence-nonesistence-HF}. 
We also defer the proofs of some technical results to the appendices.

\bigskip
\noindent
{\bf Acknowledgments.} B. Chen  is partially supported by NSF of China (Grant 12501151).  Y. J. Guo is partially supported by NSF of China (Grants 12225106 and 12371113) and National Key R $\&$ D Program of China (Grant 2023YFA1010001).
P. T. Nam and D. H. Ou Yang are partially supported by the European Research Council (ERC Consolidator Grant RAMBAS, Project Nr. 10104424) and the Deutsche Forschungsgemeinschaft (TRR 352, Project Nr. 470903074).

\section{Preliminary results on HFB theory}\label{sec:HFB-min}

In this section, we collect some elementary properties of the HFB  problem $I^{\rm HFB}_{m,\kappa,\theta}(\lambda)$ defined in \eqref{eq:HFB-variational-problem} and the associated Lagrange multipliers, where $m\geq0 $, $ 0<\kappa<4/\pi$, $0<\lambda\leq\lambda^{\rm HFB}(\kappa)$, and  the constant $\lambda^{\rm HFB}(\kappa)$ is given by  Proposition \ref{prop:Chandrasekhar}. These properties are essentially similar to those of the case $\theta=0$ studied in \cite{LenLew-10}, and they are helpful  for the proofs of  Theorems \ref{thm:main2} and \ref{thm:main1}  in the coming sections.

We begin by generalizing \cite[Lemmas 4.1--4.2]{LenLew-10} from the  Newtonian potential to the Yukawa potential.

\begin{lemma}\label{lem:prop-Imktheta-lamb}
Suppose $m\geq0$, $0<\kappa<4/\pi$, $\theta\geq 0$ and $0\leq\lambda<\lambda^{\rm HFB}(\kappa)$.  Then the following statements hold:
\begin{enumerate}
\item[$(i)$] $-m\lambda\leq I^{\rm HFB}_{m,\kappa,\theta}(\lambda)\leq 0$, and 
\begin{align}\label{2.6}
I^{\rm HFB}_{m,\kappa,\theta}(\lambda)&\leq I^{\rm HFB}_{m,\kappa,\theta}(\lambda')+I^{\rm HFB}_{m,\kappa,\theta}(\lambda-\lambda'),\ \   \forall\ 0<\lambda'<\lambda.
\end{align}
\item[$(ii)$] The function $\lambda\mapsto I^{\rm HFB}_{m,\kappa,\theta}(\lambda)$ is continuous and non-increasing.
\end{enumerate}
\end{lemma}

\noindent \textbf{Proof.}
(i). Since the subadditivity property follows from the same argument in \cite{LenLew-10}, we omit its proof for simplicity. It can be checked   that  $I^{\rm HFB}_{0,\kappa,\theta}(\lambda)=0$ and
$$
I^{\rm HFB}_{0,\kappa,\theta}(\lambda)-m\lambda\leq I^{\rm HFB}_{m,\kappa,\theta}(\lambda)\leq I^{\rm HFB}_{0,\kappa,\theta}(\lambda).
$$
Thus, we have $-m\lambda\leq I^{\rm HFB}_{m,\kappa,\theta}(\lambda)\leq0$.

(ii). The continuity of $\lambda\mapsto I^{\rm HFB}_{m,\kappa,\theta}(\lambda)$ can be established by employing trial states.
The (non-strict) monotonicity of $I^{\rm HFB}_{m,\kappa,\theta}(\lambda)$ follows from \eqref{2.6} and the non-positivity of $I^{\rm HFB}_{m,\kappa,\theta}(\lambda)$. The detailed proof is left to the interested reader.
\qed

\begin{lemma}\label{lem:coercive}
Suppose $m\geq 0$, $0<\kappa<4/\pi$ and $\theta\geq0$.  Then for any $0\leq\lambda<\lambda^{\rm HFB}(\kappa)$, the energy $\cE^{\rm HFB}_{m,\kappa,\theta}$ is coercive on $\cK_{\lambda}$.  In particular, every minimizing sequence $\{(\gamma_{n},\alpha_{n})\}$ of $I^{\rm HFB}_{m,\kappa,\theta}(\lambda)$ on $\cK_{\lambda}$ is bounded in the  $\cX$-norm of (\ref{X-norm}).
\end{lemma}

\noindent \textbf{Proof.}
For any $(\gamma,\alpha)\in\cK_{\lambda}$ and $0<\varepsilon<1$, we have
\begin{align*}
	\cE^{\rm HFB}_{m,\kappa,\theta}(\gamma,\alpha)
	&\geq \varepsilon\Tr(T\gamma)+(1-\varepsilon)I^{\rm HFB}_{m,\kappa/(1-\varepsilon),\theta}(\lambda),
\end{align*}
where $I^{\rm HFB}_{m,\kappa/(1-\varepsilon),\theta}(\lambda)$ is the infimum of $\cE^{\rm HFB}_{m,\kappa,\theta}(\gamma,\alpha)$ on $\cK_{\lambda}$ with the coupling constant $\kappa$ replaced by $\kappa/(1-\varepsilon)$.  For any $0\leq \lambda<\lambda^{\rm HFB}(\kappa)$, since the function $\kappa\mapsto\lambda^{\rm HFB}(\kappa)$ is continuous, there exists a sufficiently small $\varepsilon>0$ such that $0\leq\lambda\leq \lambda^{\rm HFB}(\kappa/(1-\varepsilon))$. This then implies from Proposition \ref{prop:Chandrasekhar} that for any $0\leq \lambda<\lambda^{\rm HFB}(\kappa)$,
\begin{align*}
	I^{\rm HFB}_{m,\kappa/(1-\varepsilon),\theta}(\lambda)>-\infty.
\end{align*}
Thus, if $(\gamma_{n},\alpha_{n})\in\cK_{\lambda}$ satisfies $\|(\gamma_{n},\alpha_{n})\|_{\cX}\rightarrow\infty$ as $n\rightarrow\infty$, using  the fact that $\alpha_n^{*}\alpha_n\leq \gamma_n$,
then we conclude that  $\Tr(T\gamma_{n})\rightarrow\infty$ as $n\rightarrow\infty$, which implies that $\cE^{\rm HFB}_{m,\kappa,\theta}(\gamma_{n},\alpha_{n})\rightarrow\infty$ as $n\rightarrow\infty$.  We hence obtain that the energy $\cE^{\rm HFB}_{m,\kappa,\theta}$ is coercive on $\cK_{\lambda}$. Together with the boundedness of $I^{\rm HFB}_{m,\kappa,\theta}(\lambda)$ in Lemma \ref{lem:prop-Imktheta-lamb} (i), we therefore complete the proof of Lemma \ref{lem:coercive}.
\qed

\subsection{Auxiliary variational problem $G_{\theta}(\lambda)$}\label{sec:auxiliary-functional}

In this subsection, we study the following auxiliary variational problem:
\begin{align}\label{min-Gtheta}
G_{\theta}(\lambda)&:=\inf_{(\gamma,\alpha)\in\cK_{\lambda}}\cG_{\theta}(\gamma,\alpha),\ \   \theta\geq0,\ \lambda>0,
\end{align}
where
\begin{align}\label{G-func}
\cG_{\theta}(\gamma,\alpha)&:=\Tr(K\gamma)-\frac{\kappa}{2}\iinte\frac{e^{-\theta|x-y|}|\alpha(x,y)|^2}{|x-y|}dxdy,\ \   0\leq\kappa\leq4/\pi ,
\end{align}
and $K:=\sqrt{-\Delta+m^{2}}$.

We begin with some preliminaries. Recall that the Schatten space $\fS_{p}$ is reflexive for $1<p<\infty$, whose dual space is $\fS_{p'}$ with $p^{-1}+p'^{-1}=1$. Moreover, the space $\fS_{1}$ of trace class operators on $L^{2}(\R^{3};\C^{q})$ is the dual space of $\fS_{\infty}$, the space of compact operators.  This induces a \textit{weak-$*$ topology} on $\cX$ as follows: We call $(\gamma_{n},\alpha_{n})\rightharpoonup (\gamma,\alpha)$ weakly-$*$ in $\cX$, if
\begin{align*}
&\Tr\big((1-\Delta)^{1/4}\gamma_{n}(1-\Delta)^{1/4}K_{1}\big)\rightarrow\Tr\big((1-\Delta)^{1/4}\gamma(1-\Delta)^{1/4}K_{1}\big)\ \ \text{as}\  n\rightarrow\infty,\\
&\Tr\big((1-\Delta)^{1/4}\alpha_{n}K_{2}\big)\rightarrow\Tr\big((1-\Delta)^{1/4}\alpha K_{2}\big)\ \ \text{as}\  n\rightarrow\infty,
\end{align*}
hold for all $K_{1}\in\fS_{\infty}$ and $K_{2}\in\fS_{2}$.  Note that $\cK_{\lambda}$
is \textit{not} weakly-$*$ closed, i.e., there exists a sequence $\{(\gamma_{n},\alpha_{n})\}\subset\cK_{\lambda}$  that converges weakly-$*$ to some $(\gamma,\alpha)$ satisfying $\Tr(\gamma)<\lambda$.

For $(\gamma,\alpha)\in\cK$, the density function $\rho_{\gamma}$ satisfies
\begin{align}\label{kin-density}
\big\langle\sqrt{\rho_{\gamma}},\, \sqrt{-\Delta}\sqrt{\rho_{\gamma}}\big\rangle\leq \Tr(\smash{\sqrt{-\Delta}\gamma}).
\end{align}
This implies from  Sobolev's embedding theorem (cf. \cite{embeding}) that $\sqrt{\rho_{\gamma}}\in L^{p}(\R^{3})$ holds for  any $2\leq p\leq 3$, and thus the map $(\gamma,\alpha)\mapsto\sqrt{\rho_{\gamma}}$ is continuous from $\cK$ to $L^{p}(\R^{3})$ for all $2\leq p<3$.  Moreover,  we denote for $\theta\geq 0$,
\begin{align}
\label{Dtheta}D_{\theta}(\rho_{\gamma},\rho_{\gamma}):=\iinte\frac{e^{-\theta|x-y|}\rho_{\gamma}(x)\rho_{\gamma}(y)}{|x-y|}dxdy,
\end{align}
\begin{align}
\label{Extheta}\Ex_\theta(a,b):=\iinte\frac{e^{-\theta|x-y|}\overline{a(x,y)}b(x,y)}{|x-y|}dxdy,\ \ \Ex_\theta(a):=\Ex_\theta(a,a).
\end{align}
For simplicity, we drop the subscripts of $D_{\theta}$ and $\Ex_\theta$ for the case $\theta=0$. Since the constraint of $\cX$ yields that $\alpha\alpha^{*}+\gamma^{2}\leq \gamma$,
appying the  Hardy-Littlewood-Sobolev and Hardy-Kato inequalities, we obtain that
\begin{align}
\label{inter-bdd1}&\Ex_\theta(\gamma)\leq D_{\theta}(\rho_{\gamma},\rho_{\gamma})\lesssim\|\sqrt{\rho_{\gamma}}\|_{L^{12/5}}^{4},
\end{align}
\begin{align}
\label{inter-bdd3}&\Ex_\theta(\alpha)\leq \frac{\pi}{2}\Tr(\smash{\sqrt{-\Delta}\alpha\alpha^{*}})\leq \frac{\pi}{2}\Tr(\smash{\sqrt{-\Delta}\gamma}).
\end{align}
The relation \eqref{inter-bdd3} then gives from (\ref{G-func}) that where $0\leq \kappa\leq 4/\pi$,
\begin{align*}
\cG_{\theta}(\gamma,\alpha)\geq \cG_{0}(\gamma,\alpha)\geq 0,\ \  \forall\ (\gamma,\alpha)\in\cK.
\end{align*}
Employing above preliminaries, we are able to obtain the following properties of $G_\theta(\lambda)$ defined in (\ref{min-Gtheta}).

\begin{prop}\label{prop:G-w*-lc}
Suppose $m>0$, $0<\kappa<4/\pi$ and $\theta\geq 0$.  Then the energy functional $\cG_{\theta}$ is weakly-$*$ lower semi-continuous on $\cK$, i.e., if $(\gamma_{n},\alpha_{n})\in\cK$ has a weakly-$*$ limit $(\gamma,\alpha)\in\cX$, then we have
$$
\liminf_{n\rightarrow\infty}\cG_{\theta}(\gamma_{n},\alpha_{n})\geq \cG_{\theta}(\gamma,\alpha).
$$
Moreover, the equality holds if and only if $(\gamma_{n},\alpha_{n})\rightarrow(\gamma,\alpha)$ strongly in $\cX$ as $n\to\infty$.
\end{prop}

\begin{cor}\label{cor:conservation-mass}
Suppose  $m>0$, $0<\kappa<\pi/4$, $\theta\geq 0$, and $0<\lambda<\lambda^{\rm HFB}(\kappa)$.  Consider a minimizing sequence $\{(\gamma_{n},\alpha_{n})\}$ of $I^{\rm HFB}_{m,\kappa,\theta}(\lambda)$ satisfying  $(\gamma_{n},\alpha_{n})\rightharpoonup(\gamma,\alpha)$ weakly-$*$ in $\cX$ as $n\to\infty$.  Then $(\gamma_{n},\alpha_{n})\rightarrow(\gamma,\alpha)$ strongly in $\cX$ as $n\to\infty$, if and only if $\Tr(\gamma)=\lambda$.
\end{cor}

Proposition \ref{prop:G-w*-lc} and Corollary  \ref{cor:conservation-mass} can be proved by the same arguments of \cite{LenLew-10}, and we omit the details. In Appendix \ref{app:pf-Chandra-Yukawa} we shall establish the following proposition.

\begin{prop}\label{prop:min-Gtheta}
Suppose  $m>0$, $0<\kappa<4/\pi$ and $0\leq \lambda\leq\lambda^{\rm HFB}$, and define $K_{\theta}=K-\frac{\kappa e^{-\theta|x|}}{2|x|}$ in the sense of quadratic forms on $L^2(\R^3)$ with form domain $H^{1/2}(\R^{3})$. Then there exists a constant  $\widetilde{\theta}=\widetilde{\theta}(m,\kappa, \lambda)>0$ such that for any  $\theta\in [0,\widetilde{\theta}]$,
\begin{align}\label{Gtheta-eigen}
G_{\theta}(\lambda)&=\beta_{\theta}\lambda,
\end{align}
where $\beta_{\theta}\in\big(m(1-\kappa\pi/4), m\big)$ is given by
\begin{align}\label{2.10}
\beta_{\theta}:&=\begin{cases}
\inf_{\psi\in L_{\rm odd}^{2}(\R^{3}),\|\psi\|_{L^{2}}=1}\langle\psi,K_{\theta}\psi\rangle,&\ \text{if }q=1;\\
\inf_{\psi\in L^{2}(\R^{3}),\|\psi\|_{L^{2}}=1}\langle\psi,K_{\theta}\psi\rangle,&\text{ if }q\geq 2;
\end{cases}
\end{align}
and $\beta_{\theta}$ is continuous and non-decreasing with respect to $\theta$.  In addition, if $0<\lambda<\lambda^{\rm HFB}$, then
\begin{align}\label{min-Gtheta-relation}
I^{\rm HFB}_{m,\kappa,\theta}(\lambda)<G_{\theta}(\lambda)-m\lambda.
\end{align}
\end{prop}
Following the argument in \cite[Lemma 4.1(iii)]{LenLew-10}, one has an immediate corollary.

\begin{cor}\label{cor:non-weak*-continuity}
Suppose $m>0$, $0<\kappa<4/\pi$ and $0<\lambda<\lambda^{\rm HFB}$.  If $\theta\in [0,\widetilde{\theta}]$, where $\widetilde{\theta}>0$ is given by Proposition \ref{prop:min-Gtheta}, then the HFB energy functional $\cE^{\rm HFB}_{m,\kappa,\theta}$ is \textit{not} weakly-$*$ lower semi-continuous on $\cK_{\lambda}$.
\end{cor}

\subsection{Estimates of chemical potential $\mu_{\theta}$}\label{esmu}

In this subsection, we derive some estimates of the Lagrange multiplier corresponding to a minimizer of $I^{\rm HFB}_{m,\kappa,\theta}(\lambda)$, where $m>0 $, $ 0<\kappa<4/\pi$, and  $0<\lambda\leq\lambda^{\rm HFB}(\kappa)$.
Let $(\gamma,\alpha)\in\cK_{\lambda}$ be a minimizer of $I^{\rm HFB}_{m,\kappa,\theta}(\lambda)$, and suppose $\Gamma$  is the associated one-particle density matrix (1-pdm) given by
\begin{align}\label{ga}
\Gamma:=\Gamma_{\gamma,\alpha}:
&=\begin{pmatrix}
\gamma	&	\alpha\\
\alpha^*	& 1-\bar{\gamma}
\end{pmatrix}.
\end{align}
We introduce the corresponding \textit{HFB mean field operator}
\begin{align*}
F_{\Gamma,\theta}:&=
\begin{pmatrix}
H_{\gamma,\theta}	&	-\kappa X_{\alpha,\theta}\\
-\kappa X_{\alpha^{*},\theta}	&	-\overline{H_{\gamma,\theta}}
\end{pmatrix}
\quad\text{acting on }L^{2}(\R^{3};\C^{q})\oplus L^{2}(\R^{3};\C^{q}),
\end{align*}
where  $H_{\gamma,\theta}$ is the HF mean field operator given by
\begin{align}\label{hga}
H_{\gamma,\theta}:=T-\kappa (W_{\theta}*\rho_{\gamma})+\kappa X_{\gamma,\theta},
\end{align}
$W_{\theta}(x)= e^{-\theta|x|}/|x|$ is the Yukawa potential, and $X_{A,\theta}$ denotes the operator with an integral kernel $W_{\theta}(x-y)A(x,y)$.

Since $\cE^{\rm HFB}_{m,\kappa,\theta}\big((1-t)(\gamma, \alpha)+t(\gamma',\alpha')\big)\geq \cE^{\rm HFB}_{m,\kappa,\theta} (\gamma, \alpha)$ holds for any $t\in(0,1)$ and $(\gamma',\alpha')\in\cK_{\lambda}$,
we conclude that $(\gamma,\alpha)$ is a solution of the following \textit{linearized variational problem}:
\begin{align}\label{Ktheta}
K_{\theta}(\lambda):&=\frac{1}{2}\inf_{\substack{(\gamma',\alpha')\in\cK_{\lambda}}}\Tr_{P_{-}^{0}}\big[F_{\Gamma,\theta}(\Gamma'-P_{-}^{0})\big],
\end{align}
where $\Tr_{P_{-}^{0}}$ is the $P_{-}^{0}$-trace defined in \cite{HaiLewSer-05}, and
\begin{align*}
P_{-}^{0}:=\begin{pmatrix}
0	&	0\\
0	&	1
\end{pmatrix},\quad \  \Gamma':=\Gamma_{\gamma',\alpha'}=\begin{pmatrix}
\gamma'	&	\alpha'\\
\alpha'^{*}	&	1-\overline{\gamma'}
\end{pmatrix}.
\end{align*}
Using the convexity of the function $\lambda\mapsto K_{\theta}(\lambda)$, 
we can deduce  that  $K_{\theta}(\lambda)$ satisfies
\begin{align}\label{convex-Ktheta}
K_{\theta}(\lambda')&\geq K_{\theta}(\lambda)+\mu_{\theta}(\lambda'-\lambda)\ \   \text{ for all }\lambda'>0,
\end{align}
where $\mu_{\theta}\in [K_{\theta,-}'(\lambda),K_{\theta,+}'(\lambda)]$,  and $K_{\theta,-}'(\lambda)$ and $K_{\theta,+}'(\lambda)$ denote the left and right derivatives of $K_{\theta}(\lambda)$, respectively. We note that these derivatives exist, since $K_{\theta}(\lambda)$ is convex in $\lambda$.

Since $K_{\theta}(\lambda)$ is non-increasing, we get from (\ref{convex-Ktheta}) that $\mu_{\theta}\leq 0$ holds for any $\theta\geq 0$.  Moreover, note that
\begin{align*}
\Tr(\smash{\gamma'})&=\frac{1}{2}\Tr_{P_{-}^{0}}\big[\mathcal{N}(\Gamma'-P_{-}^{0})\big],
\end{align*}
where
$\mathcal{N}:=\begin{pmatrix}
1	&	0\\
0	& 	-1
\end{pmatrix}.$
Together with \eqref{convex-Ktheta}, we can remove the constraint $\Tr(\smash{\gamma'})=\lambda$ in \eqref{Ktheta}, so that  $(\gamma,\alpha)$ is also a minimizer of the following minimization problem  
\begin{align*}
\frac{1}{2}\inf_{(\gamma',\alpha')\in\cK}\Tr_{P_{-}^{0}}\big[(F_{\Gamma,\theta}-\mu_{\theta}\mathcal{N})(\Gamma'-P_{-}^{0})\big].
\end{align*}
The same argument of \cite{HaiLewSer-05} hence gives that if $(\gamma, \alpha)$ is a minimizer of $I^{\rm HFB}_{m,\kappa,\theta}(\lambda)$, then $\Gamma=\Gamma_{\gamma,\alpha}$  is a solution of the following self-consistent equation
\begin{align}\label{2.14}
\Gamma&=\chi_{(-\infty,0)}(F_{\Gamma,\theta}-\mu_{\theta} \mathcal{N})+D,
\end{align}
where $\chi_{(-\infty,0)}$ denotes the characteristic function on $(-\infty,0)$, and $D$ satisfying $ran(D)\subseteq\ker(F_{\Gamma,\theta}-\mu_{\theta} \mathcal{N})$ is a finite-rank operator of the same form as $\Gamma$.  This further implies that
\begin{align}\label{1pdm-min}
\Gamma (F_{\Gamma,\theta}-\mu_{\theta}\mathcal{N})&\leq 0\ \   \text{and}\ \    [F_{\Gamma,\theta}-\mu_{\theta}\mathcal{N},\Gamma]=0.
\end{align}

We now estimate the chemical potential $\mu_{\theta}$ given by  \eqref{convex-Ktheta}.

\begin{prop}\label{prop:chemical-pot}
Let $(\gamma,\alpha)$ be a minimizer of $I^{\rm HFB}_{m,\kappa,\theta}(\lambda)$, and suppose $\mu_{\theta}\leq 0$ is the associated chemical potential 
given by  \eqref{convex-Ktheta}, where $m>0 $, $ 0<\kappa<4/\pi$, $0<\lambda\leq\lambda^{\rm HFB}(\kappa)$, $0\leq\theta\leq\widetilde{\theta}$, and $\widetilde{\theta}>0$ is as in Proposition \ref{prop:min-Gtheta}. Then we have $\mu_{\theta}<0$. Moreover, if $\alpha\neq 0$, then $\mu_{\theta}<\beta_{\theta}-m$, where $\beta_{\theta}$ is as in Proposition \ref{prop:min-Gtheta}. 
\end{prop}

\noindent \textbf{Proof.}
By the argument of \cite{LenLew-10}, which handles the case $\theta=0$, it suffices  to consider the case $\theta>0$.


We first consider the case where  $\alpha=0$. By \eqref{1pdm-min}, we have
\begin{align*}
(H_{\gamma,\theta}-\mu_{\theta})\gamma&\leq 0\ \   \text{and}\ \    [H_{\gamma,\theta},\gamma]=0.
\end{align*}
This implies that there exists  an orthonormal system $\{u_{j}\}$ in $L^{2}(\R^{3};\C^{q})$ such that
\begin{align*}
\gamma&=\sum_{j\geq1}\lambda_{j}|u_{j}\rangle \langle u_{j}|\ \   \text{and}\ \    H_{\gamma,\theta}u_{j}=\nu_{j}u_{j},
\end{align*}
where $\lambda_{j}\in (0,1]$ and $\nu_{j}\leq \mu_{\theta} \leq 0$ hold for each $j$.  By a standard argument as in the HF model, we have $M:=\text{Rank}(\gamma)<\infty$, whose range is spanned by the eigenfunctions  corresponding to the lowest $M$ eigenvalues of $H_{\gamma,\theta}$.  If $\nu_{j}=0$ holds for some $j$, then we define the trial state
\begin{align}\label{trial-HF-min}
\gamma_{j}^{(\delta)}:&=\gamma-\delta|u_{j}\rangle \langle u_{j}|,\ \   \delta\in(0, \lambda_j),
\end{align}
so that $(\gamma_{j}^{(\delta)}, 0)\in\cK_{\lambda-\delta}$.  By a straightforward computation, we have
\begin{equation}\label{HF-en-min}
\begin{split}
I^{\rm HFB}_{m,\kappa,\theta}(\lambda-\delta)&\leq\cE^{\rm HFB}_{m,\kappa,\theta}(\gamma_{j}^{(\delta)},0)=\cE^{\rm HFB}_{m,\kappa,\theta}(\gamma,0)-\delta\langle u_{j},H_{\gamma,\theta}u_{j}\rangle\\
&=\cE^{\rm HFB}_{m,\kappa,\theta}(\gamma,0)-\delta\nu_{j}
=I^{\rm HFB}_{m,\kappa,\theta}(\lambda).
\end{split}\end{equation}
Together with Lemma \ref{lem:prop-Imktheta-lamb}, we thus obtain that
\begin{align*}
I^{\rm HFB}_{m,\kappa,\theta}(\lambda-\delta)=I^{\rm HFB}_{m,\kappa,\theta}(\lambda),\ \   I^{\rm HFB}_{m,\kappa,\theta}(\delta)\leq 0,
\end{align*}
and
$$
I^{\rm HFB}_{m,\kappa,\theta}(\lambda)\leq I^{\rm HFB}_{m,\kappa,\theta}(\lambda-\delta)+I^{\rm HFB}_{m,\kappa,\theta}(\delta),
$$
which show that $I^{\rm HFB}_{m,\kappa,\theta}(\delta)=0$.  However, it then follows from Proposition \ref{prop:min-Gtheta} that  $I^{\rm HFB}_{m,\kappa,\theta}(\delta)<G_{\theta}(\delta)-m\delta<0$  for any  $\theta\in [0,\widetilde{\theta}]$, which leads to a contradiction. Hence, $\nu_j<0$ holds for every $j$. Together with \eqref{2.14}, we therefore obtain  that  $\mu_{\theta}<0$ holds true.
\DETAILS{
\begin{align}\label{ga}
	\Gamma
	&=\begin{pmatrix}
		\gamma	&	\alpha\\
		\alpha^*	& 1-\bar{\gamma}
	\end{pmatrix}.
\end{align}
\begin{align}\label{ga}
H\Gamma
&=\begin{pmatrix}
	h	&	k\\
	k^{*}	&	-\overline{h}
\end{pmatrix}
\begin{pmatrix}
		\gamma	&	\alpha\\
		\alpha^*	& 1-\bar{\gamma}
	\end{pmatrix}
=\begin{pmatrix}
	h\gamma+k\alpha^*	&	h\alpha+k(1-\bar{\gamma})\\
	k^*\gamma-\bar{h}\alpha^*	& k^*\alpha-\bar{h}(1-\bar{\gamma})
\end{pmatrix}.
\end{align}
\begin{align}\label{ga}
	\Gamma H
	&=
	\begin{pmatrix}
		\gamma	&	\alpha\\
		\alpha^*	& 1-\bar{\gamma}
	\end{pmatrix}
\begin{pmatrix}
	h	&	k\\
	k^{*}	&	-\overline{h}
\end{pmatrix}
	=\begin{pmatrix}
		\gamma h+\alpha k^*	&	\gamma k-\alpha\bar{h}\\
		\alpha^*h+(1-\bar{\gamma})k^*& \alpha^*k-(1-\bar{\gamma})\bar{h}
	\end{pmatrix}.
\end{align}
\begin{align}\label{ga}
	H\Gamma H
	&=
	\begin{pmatrix}
		h	&	k\\
		k^{*}	&	-\overline{h}
	\end{pmatrix}
\begin{pmatrix}
		\gamma h+\alpha k^*	&	\gamma k-\alpha\bar{h}\\
		\alpha^*h+(1-\bar{\gamma})k^*& \alpha^*k-(1-\bar{\gamma})\bar{h}
	\end{pmatrix}\\
&=\begin{pmatrix}
h\gamma h+h\alpha k^*+k\alpha^*h+k(1-\bar{\gamma})k^*	&	h\gamma k-h\alpha\bar{h}+k\alpha^*k-k(1-\bar{\gamma})\bar{h}\\
k^*\gamma h+k^*\alpha k^*-h\alpha^*h-h(1-\bar{\gamma})k^*& k^*\gamma k-k^*\alpha\bar{h}-\bar{h}\alpha^*k+\bar{h}(1-\bar{\gamma})\bar{h}
\end{pmatrix}.
\end{align}
\begin{align}\label{ga}
	\Gamma H^2
	&=\begin{pmatrix}
		\gamma h+\alpha k^*	&	\gamma k-\alpha\bar{h}\\
		\alpha^*h+(1-\bar{\gamma})k^*& \alpha^*k-(1-\bar{\gamma})\bar{h}
	\end{pmatrix}
\begin{pmatrix}
	h	&	k\\
	k^{*}	&	-\overline{h}
\end{pmatrix}\\
&=\begin{pmatrix}
	\gamma h^2+\alpha k^*h+	\gamma kk^*-\alpha\bar{h}k^*&	\gamma hk+\alpha k^*k-\gamma k\bar{h}+\alpha\bar{h}^2\\
	\alpha^*h^2+(1-\bar{\gamma})k^*h+\alpha^*kk^*-(1-\bar{\gamma})\bar{h}k^*
	&\alpha^*hk+(1-\bar{\gamma})k^*k -\alpha^*k\bar{h}+(1-\bar{\gamma})\bar{h}^2
\end{pmatrix}.
\end{align}
\begin{align}\label{ga}
	H^2\Gamma
	&=\begin{pmatrix}
		h	&	k\\
		k^{*}	&	-\overline{h}
	\end{pmatrix}
\begin{pmatrix}
		h\gamma+k\alpha^*	&	h\alpha+k(1-\bar{\gamma})\\
		k^*\gamma-\bar{h}\alpha^*	& k^*\alpha-\bar{h}(1-\bar{\gamma})
	\end{pmatrix}\\
&=\begin{pmatrix}
	h^2\gamma+hk\alpha^* +k k^*\gamma-k\bar{h}\alpha^*	&	h^2\alpha+hk(1-\bar{\gamma})+kk^*\alpha-k\bar{h}(1-\bar{\gamma})\\
	k^*h\gamma+k^*k\alpha^*-\bar{h}k^*\gamma+\bar{h}^2\alpha^*	
	& k^*h\alpha+k^*k(1-\bar{\gamma})-\bar{h}k^*\alpha+\bar{h}^2(1-\bar{\gamma})
\end{pmatrix}.
\end{align}}

For the case $\alpha\neq 0$, we consider the Bogoliubov transformation $e^{i\varepsilon H}$ with
\begin{equation*}\label{Bogoliubov}
\begin{split}
H&=\begin{pmatrix}
h	&	k\\
k^{*}	&	-\overline{h}
\end{pmatrix},\ \ \text{ where}\,\ h=h^{*}\in\fS_{1},\ k=-k^{T}\in\fS_{2}.
\end{split}
\end{equation*}
Set $\Gamma(\varepsilon):=e^{i\varepsilon H}\Gamma e^{-i\varepsilon H}=\begin{pmatrix}
\gamma(\varepsilon)	&	\alpha(\varepsilon)\\
\alpha(\varepsilon)^{*}	&	1-\overline{\gamma(\varepsilon)}
\end{pmatrix}$, where $\Gamma $ is as in \eqref{ga}. We can calculate that
\begin{align*}
	\gamma(\varepsilon)=&\gamma+\varepsilon i(h\gamma+k\alpha^*-\gamma h-\alpha k^*)\\
	&+\varepsilon^2\big[h\gamma h+h\alpha k^*+k\alpha^*h+k(1-\bar{\gamma})k^*-\frac{1}{2}(\gamma h^2+\alpha k^*h+	\gamma kk^*-\alpha\bar{h}k^*)\\
	&\ \ \ \ \ \ \ \ -\frac{1}{2}(h^2\gamma+hk\alpha^* +k k^*\gamma-k\bar{h}\alpha^*	)\big]+O(\varepsilon^3)\ \ \  \text{as}\ \ \varepsilon\to0,
\end{align*}
\begin{align*}
	\alpha(\varepsilon)=&\alpha+\varepsilon i\big(h\alpha+k(1-\bar{\gamma})-\gamma k+\alpha\bar{h}\big)\\
	&+\varepsilon^2\big[h\gamma k-h\alpha\bar{h}+k\alpha^*k-k(1-\bar{\gamma})\bar{h}-\frac{1}{2}\big(\gamma hk+\alpha k^*k-\gamma k\bar{h}+\alpha\bar{h}^2\big)\\
	&\ \ \ \ \ \ \ \ -\frac{1}{2}\big(h^2\alpha+hk(1-\bar{\gamma})+kk^*\alpha-k\bar{h}(1-\bar{\gamma})\big)\big]+O(\varepsilon^3)\ \ \ \text{as}\ \ \varepsilon\to0,
\end{align*}
and  
\begin{align*}\label{Bogoliubov-exp}
	\Gamma(\varepsilon)&=\Gamma+\varepsilon\Gamma_{1}+\varepsilon^{2}\Gamma_{2}+O(\varepsilon^{3})\ \ \ \text{as}\ \ \varepsilon\to0,
\end{align*}
where
\begin{align*}
\Gamma_{1}&=i[H,\Gamma],\quad\Gamma_{2}=H\Gamma H-\frac{1}{2}\Gamma H^{2}-\frac{1}{2}H^{2}\Gamma.
\end{align*}
Taking $h=0$ and $k=-ia$ with $a=-a^{T}\in\fS_{2}$, it then follows that
\begin{equation*}
\begin{split}
\Gamma_{1}&=\begin{pmatrix}
0	&	a\\
a^{*}	&	0
\end{pmatrix}+\begin{pmatrix}
\gamma_{1}	&	\alpha_{1}\\
\alpha_{1}^{*}	&	-\overline{\gamma_{1}}
\end{pmatrix},\quad
\Gamma_{2}=\begin{pmatrix}
aa^{*}	&	0\\
0&	-a^{*}a
\end{pmatrix}+\begin{pmatrix}
\gamma_{2}	&	\alpha_{2}\\
\alpha_{2}^{*}	&	-\overline{\gamma_{2}}
\end{pmatrix},	
\end{split}\end{equation*}
where
\begin{equation}\label{g1-a1}
\gamma_{1}=a\alpha^{*}+\alpha a^{*},\quad\alpha_{1}=-\gamma a-a\overline{\gamma},
\end{equation}
and
\begin{equation}\label{g2-a2}
\gamma_{2}=-a\overline{\gamma}a^{*}-\frac{1}{2}aa^{*}\gamma-\frac{1}{2}\gamma aa^{*},\quad\alpha_{2}=-a\alpha^{*}a-\frac{1}{2}aa^{*}\alpha-\frac{1}{2}\alpha a^{*}a.
\end{equation}
By substituting these into $\cE^{\rm HFB}_{m,\kappa,\theta}$, together with the fact that $[F_{\Gamma,\theta}-\mu_{\theta} \mathcal{N},\Gamma]=0$, we have
\begin{align}\label{E-epsi-exp}
\cE^{\rm HFB}_{m,\kappa,\theta}(\gamma(\varepsilon),\alpha(\varepsilon))&=\cE^{\rm HFB}_{m,\kappa,\theta}(\gamma,\alpha)+\varepsilon\mu_{\theta}\Tr(\gamma_{1})+\varepsilon^{2}\cQ_{\theta}(a,a)+O(\varepsilon^{3})\ \   \text{as}\ \ \varepsilon\to0,
\end{align}
where
\begin{align}
\cQ_{\theta}(a,a)&=\Tr\big(T(a^{*}a+\gamma_{2})\big)-\kappa D_{\theta}(\rho_{\gamma},\rho_{a^{*}a+\gamma_{2}})\nonumber\\
&\quad\quad+\kappa\Re \Ex_\theta(\gamma,a^{*}a+\gamma_{2})-\kappa\Re \Ex_\theta(\alpha,\alpha_{2})\\
&\quad\quad-\frac{\kappa}{2}D_{\theta}(\rho_{\gamma_{1}},\rho_{\gamma_{1}})+\frac{\kappa}{2}\Ex_\theta(\gamma_{1},\gamma_{1})-\frac{\kappa}{2}\Ex_\theta(a+\alpha_{1},a+\alpha_{1}),\nonumber
\end{align}
and  $\Ex_\theta(a,b)$ is as in  \eqref{Extheta}.

Denote  $\fH$ to be either $L_{\rm odd}^{2}(\R^{3})$ or $L^{2}(\R^{3})$. Before ending the proof of Proposition \ref{prop:chemical-pot}, we next introduce the following lemma.

\begin{lemma}\label{lem:Q(a,a)<0}
There exists a matrix $a=-a^{T}\in\fS_{2}$ satisfying $\Tr(Ka^{*}a)<\infty$ such that $\Tr(\gamma_{1})<0$ and $\cQ_{\theta}(a,a)<(\beta_{\theta}-m)\Tr(\gamma_{2})$, where $\gamma_1$ and $\gamma_2$ are given by  \eqref{g1-a1} and \eqref{g2-a2}, respectively, and $\beta_{\theta}$ given by  \eqref{2.10} is the minimum eigenvalue of the operator $K_{\theta}:=K-\frac{\kappa e^{-\theta|x|}}{2|x|}$ on $\fH$.
\end{lemma}

Let  $(\gamma,\alpha)\neq (\gamma,0)$ be a minimizer of $I^{\rm HFB}_{m,\kappa,\theta}(\lambda)$ given by Proposition \ref{prop:chemical-pot}. To prove Lemma \ref{lem:Q(a,a)<0}, we recall (cf. \cite {LenLew-10}) the following observation: Suppose  there exists a matrix $a=-a^{T}\in\fS_{2}$ such that 
$$\Tr(Ka^{*}a)<\infty\ \ \ \text{and}\ \ \  \cQ_{\theta}(a,a)<(\beta_{\theta}-m)\Tr(\gamma_{2})<0.$$  If $\Tr(\gamma_{1})>0$, then we can replace $a$ with $-a$ for $\gamma_{1}$ to obtain the opposite inequality.  If $\Tr(\gamma_{1})=0$, then we can replace $a$ with $a'=a-\epsilon\alpha$ and $\epsilon>0$, so that
\begin{align*}
\Tr(\gamma_{1}')&=-2\epsilon\Tr(\alpha\alpha^{*})<0,
\end{align*}
and
\begin{align*}
\cQ(a',a')-(\beta_{\theta}-m)\Tr(\gamma_{2}')&=\cQ(a,a)-(\beta_{\theta}-m)\Tr(\gamma_{2})+O(\epsilon)\\
&<0\ \   \text{for sufficiently small}\ \, \epsilon>0.
\end{align*}
In view of the above observation, it therefore suffices to construct such a matrix $a\in\fS_{2}$.

Let  $\tilde{\theta}>0$  and $\theta^*>0$ be given by Proposition \ref{prop:min-Gtheta} and Lemma \ref{lem:Gtheta-eigen}, respectively. Recall from the proof of  Proposition  \ref{prop:min-Gtheta} that $\tilde{\theta}\leq\theta^*$.
This implies   from Lemma \ref{lem:Gtheta-eigen} that for every $\theta\in [0,\tilde{\theta}]$, the operator $K_{\theta}:=K-\frac{\kappa e^{-\theta|x|}}{2|x|}$ has a real-valued eigenfunction $f_{\theta}\in\fH$ with an eigenvalue $\beta_{\theta}$, where $\fH=L_{\rm odd}^{2}(\R^{3})$ when $q=1$ and  $\fH=L^{2}(\R^{3})$ when $q\geq 2$. Define for each $L>0$,
\begin{align}\label{c.1}
a_{L,U}(x,y):&=\chi_{L}\big(U(x-3L\vec{v})\big)f_{\theta}\big(U(x-y)\big)\chi_{L}\big(U(y-3L\vec{v})\big),
\end{align}
where $U\in SO(3)$  denotes a rotation  in $\R^3$ centered at the origin, $\vec{v}$ is a fixed unit vector in $\R^{3}$, and $\chi_{L}(x)=L^{-3/4}\chi(x)\in C_{0}^{\infty}(\R^{3})$ is a radial function with support in $B_1(0)$.  We then have the following estimates of $a_{L,U}$ for sufficiently large $L>0$.

\begin{lemma}\label{lem:aUL-esti}
There exists a constant $C>0$, independent of $U$, $\vec{v}$ and $L$, such that for sufficiently large $L>0$,
\begin{align}\label{aUL-esti}
\|a_{L,U}\|_{\fS_{2}}+\|K^{1/2}a_{L,U}\|\leq CL^{-3/2},
\end{align}
\begin{align}\label{KXtheta-esti}
\Tr\big((K-\beta_{\theta})a_{L,U}a_{L,U}^{*}\big)-\frac{\kappa}{2}\Ex_\theta(a_{L,U},a_{L,U})=O(L^{-2}).
\end{align}
\end{lemma}

The proof of Lemma \ref{lem:aUL-esti} is very similar to that of \cite{LenLew-10} and we omit the details. Now we are ready to conclude: 

\vspace{0.15cm}
\noindent \textbf{Proof of Lemma \ref{lem:Q(a,a)<0}.} Let $\gamma_{1},\alpha_{1}, \gamma_{2}$ and $\alpha_{2}$ be defined by \eqref{g1-a1} and \eqref{g2-a2} in terms of the matrix $a:=a_{L,U}$ constructed by (\ref{c.1}). Since $K^{1/2}\alpha\in \fS_{2}$ and $K^{1/2}\gamma\in\fS_{1}$,  where $(\gamma,\alpha)\neq (\gamma,0)$  be a minimizer of $I^{\rm HFB}_{m,\kappa,\theta}(\lambda)$, we obtain from Lemma \ref{lem:aUL-esti} that there exists a constant $C>0$ such that  for any sufficiently large $L>0$,
\begin{align*}
\big\|K^{1/2}\gamma_{1}\big\|_{\fS_{2}}+\big\|K^{1/2}\alpha_{1}\big\|_{\fS_{1}}\leq CL^{-3/2},\quad \big\|K^{1/2}\gamma_{2}\big\|_{\fS_{1}}+\big\|K^{1/2}\alpha_{2}\big\|_{\fS_{2}}\leq CL^{-3}.
\end{align*}
Together with the fact that
\begin{align*}
|\Ex_\theta(f,g)|\leq |\Ex_{0}(f,g)|\leq C\big\|K^{1/2}f\big\|_{\fS_{2}}\big\|K^{1/2}g\big\|_{\fS_{2}},
\end{align*}
one can calculate from \eqref{KXtheta-esti} that
\begin{equation}\label{Qtheta-Dtheta-esti}
\begin{split}
&\cQ_{\theta}(a_{L,U},a_{L,U})-(\beta_{\theta}-m)\Tr(\gamma_{2})\\
\leq& \Tr\big((K-\beta_{\theta})a_{L,U}a_{L,U}^{*}\big)-\frac{\kappa}{2}\Ex_\theta(a_{L,U},a_{L,U})\\
&-\frac{\kappa}{2}D_{\theta}(\rho_{\gamma},\rho_{a_{L,U}a_{L,U}^{*}})+O(L^{-3/2})\\
\leq& -\frac{\kappa}{2}D_{\theta}(\rho_{\gamma},\rho_{a_{L,U}a_{L,U}^{*}})+O(L^{-3/2})\ \   \text{as}\ \ L\to\infty.
\end{split}
\end{equation}

Set $g_{\theta}(x)=\frac{1-e^{-\theta|x|}}{\theta|x|}$, so that $\|g_{\theta}\|_{L^{\infty}}=\sup_{x\in\R^{3}}g_{\theta}(x)=1$ holds for any $\theta>0$. We then get from \eqref{aUL-esti} that
\begin{align*}
-D_{\theta}(\rho_{\gamma},\rho_{a_{L,U}a_{L,U}^{*}})&=-D(\rho_{\gamma},\rho_{a_{L,U}a_{L,U}^{*}})+\theta\int_{\R^{3}}g_{\theta}(x-y)\rho_{\gamma}(x)\rho_{a_{L,U}a_{L,U}^{*}}(y)dxdy\nonumber\\
&\leq -D(\rho_{\gamma},\rho_{a_{L,U}a_{L,U}^{*}})+\theta\|g_{\theta}\|_{L^{\infty}}\|\rho_{\gamma}\|_{L^{1}}\|\rho_{a_{L,U}a_{L,U}^{*}}\|_{L^{1}}\nonumber\\
&\leq-D(\rho_{\gamma},\rho_{a_{L,U}a_{L,U}^{*}})+O(L^{-3})\ \   \text{as}\ \ L\to\infty.
\end{align*}
This further implies from \eqref{Qtheta-Dtheta-esti} that
\begin{align}\label{c.5}
\cQ_{\theta}(a_{L,U},a_{L,U})&-(\beta_{\theta}-m)\Tr(\gamma_{2})\leq -\frac{\kappa}{2}D(\rho_{\gamma},\rho_{a_{L,U}a_{L,U}^{*}})+O(L^{-3/2})
\end{align}
as $L\to\infty$. Moreover, averaging over $U$ and applying Newton's theorem, we deduce that there exists a constant $\delta>0$ such that  for sufficiently large $L>0$,
\begin{align}\label{c.6}
\int_{SO(3)}D(\rho_{\gamma},\rho_{a_{L,U}a_{L,U}^{*}})dU&\geq \frac{\Tr(a_{1,L}a_{1,L}^{*})\int_{|x|\leq L}\rho_{\gamma}dx}{4L}\geq \frac{\delta}{L},
\end{align}
where $dU$ denotes the normalized Haar measure on $SO(3)$.  Therefore, we conclude from \eqref{c.5} and \eqref{c.6} that
\begin{align*}
\cQ_{\theta}(a_{L,U},a_{L,U})&-(\beta_{\theta}-m)\Tr(\gamma_{2})<0\ \    \text{for sufficiently large}\ L>0.
\end{align*}
This completes the proof of Lemma \ref{lem:Q(a,a)<0}. \qed

Now we come back to Proposition \ref{prop:chemical-pot}. Since $\Tr(\gamma_{1})<0$, we derive that  for sufficiently small $\varepsilon>0$,
\begin{align}\label{2.29}
\Tr(\gamma(\varepsilon))<\Tr(\gamma)=\lambda.
\end{align}
Recall from  Proposition \ref{prop:min-Gtheta} that  
\begin{align}\label{2.31}
I^{\rm HFB}_{m,\kappa,\theta}(\lambda')<(\beta_\theta-m)\lambda'<0\ \   \text{for any}\,\ 0<\lambda'<\lambda^{\rm HFB}(\kappa),\ \theta\in [0,\widetilde{\theta}].
\end{align}
It then follows from \eqref{2.29}, \eqref{2.31} and Lemma \ref{lem:prop-Imktheta-lamb} (i) that
\begin{align}\label{2.30}
&\ \     \cE^{\rm HFB}_{m,\kappa,\theta}\big(\gamma(\varepsilon),\alpha(\varepsilon)\big)\nonumber\\
&\geq I^{\rm HFB}_{m,\kappa,\theta}\big(\Tr(\gamma(\varepsilon))\big)
\geq I^{\rm HFB}_{m,\kappa,\theta}(\lambda)-I^{\rm HFB}_{m,\kappa,\theta}\big(\lambda-\Tr(\gamma(\varepsilon))\big)\nonumber\nonumber\\
&\geq I^{\rm HFB}_{m,\kappa,\theta}(\lambda)-(\beta_{\theta}-m)\big(\lambda-\Tr(\gamma(\varepsilon))\big)\\
&\geq I^{\rm HFB}_{m,\kappa,\theta}(\lambda)+(\beta_{\theta}-m)\big[\varepsilon\Tr(\gamma_{1})+\varepsilon^{2}\Tr(\gamma_{2})+O(\varepsilon^{3})\big]\ \   \text{as}\ \ \varepsilon\to0^+.\nonumber
\end{align}
We thus deduce from \eqref{E-epsi-exp}, \eqref{2.30} and   Lemma \ref{lem:Q(a,a)<0}  that for sufficiently small $\varepsilon>0$,
\begin{align*}
(\beta_{\theta}-m-\mu_{\theta})\Tr(\gamma_{1})
&\leq \varepsilon\big[\cQ_{\theta}(a,a)-(\beta_{\theta}-m)\Tr(\gamma_{2})\big]+O(\varepsilon^{3})<0,
\end{align*}
which  implies that $\mu_{\theta}<\beta_{\theta}-m<0$, and the proof of   Proposition \ref{prop:chemical-pot} is therefore complete.  \qed

\section{Decaying estimate of HFB minimizers: Proof of Theorem \ref{thm:main2}}\label{sec:pf-main2}
In this section,  we prove Theorem \ref{thm:main2} on  the decaying estimate of minimizers (if they exist) for the problem $I^{\rm HFB}_{m,\kappa,\theta}(\lambda)$, where $m>0,\ 0<\kappa<4/\pi,\ 0<\lambda<\lambda^{\rm HFB}$ and $0<\theta\leq\theta_c$. Here $\lambda^{\rm HFB}=\lambda^{\rm HFB}(\kappa)$
given by proposition \ref{prop:Chandrasekhar} denotes the critical total mass,  and $\theta_c=\theta_c(m, \kappa, \lambda)$ 
is a positive constant to be determined later.

Throughout this section, we suppose  $(\gamma,\alpha)\in\mathcal{K}_\lambda$ is  a minimizer of $I^{\rm HFB}_{m,\kappa,\theta}(\lambda)$, and define
\begin{align}\label{g-a-Phi}
\gamma_{\Phi}&:=e^{\Phi}\gamma e^{\Phi},\ \ \alpha_{\Phi}:=e^{\Phi}\alpha,
\end{align}
where $\Phi(x)=\tau|x|$ for $\tau\geq 0$.  Define $\eta_{R}(x):=\eta(x/R)$ for $R>0$, where $\eta\in C^\infty(\R^3;[0,1])$ is a smooth cutoff function such that  $\eta(x)=1$ for $|x|>2$ and
$\eta(x)=0$ for $|x|< 1$.  We shall show that for sufficiently small $\theta>0$,  $(\eta_{R}\gamma_{\Phi}\eta_{R},\eta_{R}\alpha_{\Phi})\in\cX$ holds for sufficiently large $R>0$, where $\tau=3\theta$.
For simplicity,  throughout this section we always denote by $\delta(R)$ any generic positive function on $(0,\infty)$ that vanishes at infinity.

\subsection{Estimates from the mean-field equations}\label{sec:esti-g-a-pos-a-wave}


Let $H_{\gamma,\theta},\  X_{\alpha,\theta}$ and $\mu_\theta<0$ be as in Subsection \ref{esmu}.
It then follows from \eqref{1pdm-min} that
\begin{align}
\label{g-a-pos}&(H_{\gamma,\theta}-\mu_{\theta})\gamma-\kappa X_{\alpha,\theta}\alpha^{*}=\gamma (H_{\gamma,\theta}-\mu_{\theta})-\kappa \alpha X_{\alpha^{*},\theta}\leq 0,\\
\label{a-wave}&\Big((H_{\gamma,\theta})_{x}+(H_{\gamma,\theta})_{y}-\frac{\kappa e^{-\theta|x-y|}}{|x-y|}-2\mu_{\theta}\Big)\alpha=-\kappa(\gamma_{x}+\gamma_{y})X_{\alpha,\theta},
\end{align}
where  $(H_{\gamma,\theta})_{x}$ means that the operator $H_{\gamma,\theta}$  on $L^2(\R^3;\C^q)$ acts on the $x$-variable in $\alpha(x, y)$.
Applying \eqref{g-a-pos} and \eqref{a-wave},  in this subsection we mainly  establish the following proposition, which provides some key estimates of the minimizers for $I^{\rm HFB}_{m,\kappa,\theta}(\lambda)$.

\begin{prop}\label{prop:esti-g-a-pos-a-wave}
Let $(\gamma,\alpha)$ be a minimizer of $I^{\rm HFB}_{m,\kappa,\theta}(\lambda)$ satisfying $\alpha\neq 0$, and  suppose $0<\tau\leq \min\{1,m/6,\theta\}$ for $\theta\in (0,\widetilde{\theta}]$, where $m>0,\ 1<\kappa<4/\pi,\  0<\lambda<\lambda^{\rm HFB}$, and $\widetilde{\theta}=\widetilde{\theta}(m, \kappa,\lambda)>0$ is given in Proposition \ref{prop:min-Gtheta}.  Then there exists a constant $C_1>0$, which is independent of $R>0$, such that for  sufficiently large $R>0$, 
\begin{align}\label{esti-g-a-pos}
\Tr\Big(\big(L_{\gamma,\theta}-\mu_{\theta}\big)\eta_{R}\gamma_{\Phi}\eta_{R}-\big(C_1\tau^2m^{-1}+\delta(R)\big)\gamma_{\Phi}\Big)\leq \kappa \Ex_\theta(\eta_{R/4}\alpha_{\Phi}\eta_{R/4}),
\end{align}
and
\begin{align}\label{esti-a-wave}
\Ex_\theta(\eta_{R}\alpha_{\Phi}\eta_{R})
\leq \delta(R)\Tr\big((L_{\gamma,\theta}-\mu_{\theta})\eta_{R}\gamma_{\Phi}\eta_{R}\big) +C_1\tau^{2}m^{-1}\Tr(\eta_{R}\gamma_{\Phi}\eta_{R})+M_{R}\lambda,
\end{align}
where $L_{\gamma,\theta}:=T-\kappa V_\theta:=T-\kappa (W_{\theta}*\rho_{\gamma})$ is related to \eqref{hga},  $\delta(R)>0$ satisfies $\delta(R)\to0$ as $R\to\infty$, and  the constant $M_{R}>0$ depends on $R>0$.
\end{prop}

In order to prove Proposition \ref{prop:esti-g-a-pos-a-wave},  we start with the following elementary lemma.

\begin{lemma}\label{lem:IMS}
Suppose $\chi\in C^\infty(\R^3)$, and set $K=\sqrt{-\Delta+m^{2}}$ with $m>0$.  Then we have
\begin{align*}\label{local-kinetic}
\frac{1}{2}(K\chi^{2}+\chi^{2}K)&=\chi K\chi-\frac{1}{\pi}\int_{0}^{\infty}\frac{1}{s+K^{2}}|\nabla\chi|^{2}\frac{1}{s+K^{2}}\sqrt{s}ds+L_{\chi},
\end{align*}
where $L_{\chi}$ is a non-negative  operator given by
\begin{equation}\label{3.6b}
L_{\chi}:=\frac{1}{\pi}\int_{0}^{\infty}A_{\chi}^{*}(s)A_{\chi}(s)\sqrt{s}ds,\ \ A_{\chi}(s):=\frac{1}{(s+K^{2})^{1/2}}[\chi,K^{2}]\frac{1}{s+K^{2}}.
\end{equation}
Moreover,  if $\nabla\chi\in L^{\infty}(\R^{3})$, then there exists a constant $C>0$, which is independent of $\chi$, such that
\begin{equation*}
\|L_{\chi}\|\leq C\|\nabla\chi\|_{L^{\infty}}^{2},\ \ \Big\|\int_{0}^{\infty}\frac{1}{s+K^{2}}|\nabla\chi|^{2}\frac{1}{s+K^{2}}\sqrt{s}ds\Big\|\leq C\|\nabla\chi\|_{L^{\infty}}^{2}.
\end{equation*}
\end{lemma}

Since the proof of Lemma \ref{lem:IMS} follows from \cite[Lemma A.2]{LenLew-10}, we omit the detailed proof for simplicity. The proof of Proposition \ref{prop:esti-g-a-pos-a-wave} also needs the following lemma.

\begin{lemma}\label{lem:resolv-s-bdd}
For $m>0$, let $\Psi$ be any Lipschitz function on $\R^{3}$ having a Lipschitz bound $\zeta\leq m/2$.  Then for each $s\geq 0$, the operator $e^{\Psi}(s+K^{2})^{-1}e^{-\Psi}$ is bounded on $L^{2}(\R^{3})$ and satisfies
\begin{align}\label{s+K-f-bdd}
\|e^{\Psi}(s+K^{2})^{-1}e^{-\Psi}\|&\leq 4(m^{2}+s)^{-1},
\end{align}
where $K=\sqrt{-\Delta+m^{2}}$ with $m>0$.
\end{lemma}

\noindent \textbf{Proof.}
Denote $Q(s):=e^{\Psi}(s+K^{2})^{-1}e^{-\Psi}$.  Note  from  \cite[Theorem 6.23]{LieLos-01} that the integral kernel of $(s+K^{2})^{-1}$ is
\begin{align}\label{3.5}	(s+K^{2})^{-1}(x,y)&=\frac{1}{4\pi|x-y|}e^{-\sqrt{m^{2}+s}\, |x-y|}.
\end{align}
Since $s\geq 0$ and $\zeta\leq m/2$, we have
\begin{align}\label{3.6a}
\zeta-\sqrt{m^{2}+s}&\leq -\frac{\sqrt{m^{2}+s}}{2}.
\end{align}
By  the Lipschitz property of $\Psi$, we  deduce from \eqref{3.5} and \eqref{3.6a} that for any $\psi\in L^{2}(\R^{3})$,
\begin{align*}\label{f-point-bdd}
\big|(Q(s)\psi)(x)\big|
&=\frac{1}{4\pi}\Big|\int_{\R^{3}}\frac{e^{\Phi(x)-\Phi(y)}e^{-\sqrt{m^{2}+s}|x-y|}\psi(y)}{|x-y|}dy\Big|\nonumber\\
&\leq \frac{1}{4\pi}\int_{\R^{3}}\frac{e^{\zeta|x-y|-\sqrt{m^{2}+s}\, |x-y|}|\psi(y)|}{|x-y|}dy\nonumber\\
&\leq \frac{1}{4\pi}\int_{\R^{3}}\frac{e^{-\frac{1}{2}\sqrt{m^{2}+s}\, |x-y|}|\psi(y)|}{|x-y|}dy\nonumber\\
&=\big(((m^{2}+s)/4-\Delta)^{-1}|\psi|\big)(x)\ \   \text{in}\ \ \R^3.
\end{align*}
This then  implies that
\begin{equation*}
\begin{split}
\|Q(s)\psi\|_{L^{2}}^{2}&\leq \int_{\R^{3}}\Big|\big((m^{2}+s)/4-\Delta\big)^{-1}|\psi|)(x)\Big|^{2}dx\\
&=\big\|((m^{2}+s)/4-\Delta)^{-1}|\psi|\big\|_{L^{2}}^{2}\leq \frac{16}{(m^{2}+s)^{2}}\|\psi\|_{L^{2}}^{2},
\end{split}\end{equation*}
and thus  \eqref{s+K-f-bdd} holds true.   This proves Lemma \ref{lem:resolv-s-bdd}. \qed

\medskip

\noindent \textbf{Proof of Proposition \ref{prop:esti-g-a-pos-a-wave}.}
We first prove that  \eqref{esti-g-a-pos} holds true. For this purpose, we follow (\ref{hga}) to denote $A:=(H_{\gamma,\theta}-\mu_{\theta})\gamma-\kappa X_{\alpha,\theta}\alpha^{*}$. It then follows from (\ref{hga}) and \eqref{g-a-pos} that
$$
A=A^{*},\ \      \Tr\big(e^{\Phi}\eta_{R}(A+A^{*})\eta_{R}e^{\Phi}\big)\leq 0,
$$
and thus
\begin{equation}\label{decay1.1}
\begin{split}
&\Tr\Big(\frac{\eta_{R}^{2}e^{2\Phi}T+Te^{2\Phi}\eta_{R}^{2}}{2}\gamma\Big)-\mu_{\theta}\int_{\R^{3}}\eta_{R}^{2}\rho_{\gamma_{\Phi}}dx\\
\leq &\kappa \Ex_\theta(\eta_{R}\alpha_{\Phi})+\kappa\int_{\R^{3}}(W_{\theta}*\rho_{\gamma})(x)\eta_{R}^{2}\rho_{\gamma_{\Phi}}dx, \ \  \text{where}\ W_\theta(x)=\frac{e^{-\theta|x|}}{|x|}>0.
\end{split}\end{equation}
By Lemma \ref{lem:IMS}, one can verify that
\begin{align}\label{kinetic-local}
\Tr\Big(\frac{\eta_{R}^{2}e^{2\Phi}T+Te^{2\Phi}\eta_{R}^{2}}{2}\gamma\Big)&\geq\Tr(T\eta_{R}\gamma_{\Phi}\eta_{R})-\Tr\big(S_{\eta,\Phi}(R)\gamma_{\Phi}\big),
\end{align}
where
\begin{align}\label{sr}
S_{\eta,\Phi}(R)&:=\frac{1}{\pi}\int_{0}^{\infty}e^{-\Phi}\frac{1}{s+K^{2}}|\nabla(\eta_{R}e^{\Phi})|^{2}\frac{1}{s+K^{2}}e^{-\Phi}\sqrt{s}ds.
\end{align}
Note that 
\begin{equation*}\label{esti-Lzetaf}
|\nabla(\eta_{R}e^{\Phi})|^{2}\leq 2e^{2\Phi}\Big( \tau^{2}\eta_{R}^{2}+|\nabla\eta_{R}|^{2}\Big),\,\ \tau\geq 0,
\end{equation*}
since $\Phi(x)=\tau|x|$ holds for $\tau\geq 0$.
This then  implies from Lemma \ref{lem:resolv-s-bdd} that
\begin{equation}\label{sr1}
\begin{split}
\|S_{\eta,\Phi}(R)\|&\leq \frac{2\big(\tau^{2}\|\eta\|_{L^{\infty}}^{2}+\|\nabla\eta\|_{L^{\infty}}^{2}R^{-2}\big)}{\pi}\int_{0}^{\infty}\|Q_{\Phi}(s)\|^{2}\sqrt{s}ds\\
&\leq\frac{32(\tau^{2}+\|\nabla\eta\|_{L^{\infty}}^{2}R^{-2})}{\pi}\int_{0}^{\infty}\frac{\sqrt{s}}{(m^{2}+s)^{2}}ds\\
&=16\tau^{2}m^{-1}+\delta(R),
\end{split}\end{equation}
where $Q_{\Phi}(s):=e^{\Phi}(s+K^{2})^{-1}e^{-\Phi}$.
Together with \eqref{decay1.1} and \eqref{kinetic-local}, we hence deduce that
\begin{align}\label{decay3} \Tr\Big(\big(L_{\gamma,\theta}-\mu_{\theta}\big)\eta_{R}\gamma_{\Phi}\eta_{R}-\big(16\tau^2m^{-1}+\delta(R)\big)\gamma_{\Phi}\Big)\leq \kappa\Ex_\theta(\eta_{R}\alpha_{\Phi}).
\end{align}
Moreover, by the facts that $\alpha_\Phi\alpha_\Phi^{*}\leq \gamma_\Phi$ and $\eta_{R}\leq \eta_{R/4}$ for all $R>0$, we have
\begin{equation*}\label{decay2}
\begin{split}
\Ex_\theta(\eta_{R}\alpha_{\Phi})&=\Ex_\theta(\eta_{R}\alpha_{\Phi}\eta_{R/4})+ \Ex_\theta(\eta_{R}\alpha_{\Phi}\chi_{R/4})\nonumber\\
&\leq \Ex_\theta(\eta_{R/4}\alpha_{\Phi}\eta_{R/4})+\frac{2e^{-\theta R/2}}{R}\Tr(\eta_{R}\alpha_{\Phi}\chi_{R/4}^{2}\alpha_{\Phi}^{*}\eta_{R})\\
&\leq \Ex_\theta(\eta_{R/4}\alpha_{\Phi}\eta_{R/4})+ \delta(R)\Tr(\eta_{R}\gamma_\Phi\eta_{R}),\nonumber
\end{split}
\end{equation*}
where $\chi_R(x)=\sqrt{1-\eta_R^2(x)}$ in $\R^3$. Together with \eqref{decay3}, we therefore obtain  that \eqref{esti-g-a-pos} holds true.

\smallskip

We  next prove that \eqref{esti-a-wave} holds true.  Multiplying $e^{\Phi(x)}\eta_{R}(x)\eta_{R}(y)$ on  both hand sides of  equation \eqref{a-wave} and  projecting onto the wave function $\alpha_{R,\Phi}(x,y):=e^{\Phi(x)}\eta_{R}(x)\alpha(x,y)\eta_{R}(y)$ in $L^{2}(\R^{3}\times\R^{3})$,  we obtain that
\begin{align}\label{decay10}
I+II=&\Big\langle\eta_{R}(y)\alpha,\ \frac{e^{2\Phi(x)}\eta_{R}^{2}(x)T_{x}+T_{x}e^{2\Phi(x)}\eta_{R}^{2}(x)}{2}\eta_{R}(y)\alpha\Big\rangle\nonumber\\
&+\Big\langle\eta_{R}(x)\alpha_{\Phi},\ \frac{\eta_{R}^{2}(y)T_{y}+T_{y}\eta_{R}^{2}(y)}{2}\eta_{R}(x)\alpha_{\Phi}\Big\rangle\\
&-\kappa \Ex_\theta(\alpha_{R,\Phi})-2\mu_{\theta}\int_{\R^{3}\times\R^{3}}|\alpha_{R,\Phi}(x,y)|^{2}dxdy\nonumber\\
&-2\kappa\int_{\R^{3}\times\R^{3}}V_\theta(x)\eta^2_{R}(x)\eta^2_{R}(y)e^{2\Phi(x)}|\alpha(x,y)|^2dxdy,\nonumber
\end{align}
where
\begin{align*}
I&:=-\kappa\Re\big\langle\alpha_{R,\Phi},\ \eta_{R}(x)\eta_{R}(y)e^{\Phi(x)}\big((X_{\gamma,\theta})_{x}+(X_{\gamma,\theta})_{y}\big)\alpha\big\rangle,\\
II&:=-\kappa\Re\big\langle\alpha_{R,\Phi},\ \eta_{R}(x)\eta_{R}(y)e^{\Phi(x)}(\gamma_{x}+\gamma_{y})X_{\alpha,\theta}\big\rangle.
\end{align*}
Applying Lemmas \ref{lem:IMS} and \ref{lem:resolv-s-bdd}, together with the relation $\alpha_{\Phi}\alpha_{\Phi}^{*}\leq \gamma_{\Phi}$,  one can calculate from \eqref{sr1} that 
\begin{align}\label{decay8}
&\ \  \Big\langle\eta_{R}(y)\alpha,\ \frac{e^{2\Phi(x)}\eta_{R}^{2}(x)T_{x}+T_{x}\eta_{R}^{2}(x)e^{2\Phi(x)}}{2}\eta_{R}(y)\alpha\Big\rangle\nonumber\\
&\geq \langle\alpha_{R,\Phi},\, T_{x}\alpha_{R,\Phi}\rangle
-\big\langle\eta_{R}(y)\alpha_{\Phi},\, S_{\eta,\Phi,x}(R)\eta_{R}(y)\alpha_{\Phi}\big\rangle\nonumber\\
&\geq\langle\alpha_{R,\Phi},\, T_{x}\alpha_{R,\Phi}\rangle-\big(16\tau^{2}m^{-1}+\delta(R)\big)\big\langle\eta_{R}(y)\alpha_{\Phi},\, \eta_{R}(y)\alpha_{\Phi}\big\rangle\nonumber\\
&\geq\big\langle\alpha_{R,\Phi},\,  T_{x}\alpha_{R,\Phi}\big\rangle-\big(16\tau^{2}m^{-1}+\delta(R)\big)\langle\alpha_{R,\Phi},\,  \alpha_{R,\Phi}\rangle\\
&\ \     -\big(16\tau^{2}m^{-1}+\delta(R)\big)\big\langle\chi_{R}(x)\eta_{R}(y)\alpha_{\Phi},\, \eta_{R}(x)\eta_{R}(y)\alpha_{\Phi}\big\rangle\nonumber\\
&\geq \langle\alpha_{R,\Phi},\ T_{x}\alpha_{R,\Phi}\rangle-\big(16\tau^{2}m^{-1}+\delta(R)\big)\Tr(\eta_{R}\gamma_{\Phi}\eta_{R})\nonumber\\
&\ \     -\big(16\tau^{2}m^{-1}+\delta(R)\big)e^{4\tau R}\lambda,\nonumber
\end{align}
where   $S_{\eta,\Phi}(R)$ is as in \eqref{sr}, and $S_{\eta,\Phi,x}(R)$ denotes the operator $S_{\eta,\Phi}(R)$ acting on the $x$-variable. Applying Lemma \ref{lem:IMS}, the same argument as above gives that
\begin{align}\label{decay9}
&\ \     \Big\langle\eta_{R}(x)\alpha_{\Phi},\, \frac{\eta_{R}^{2}(y)T_{y}+T_{y}\eta_{R}^{2}(y)}{2}\eta_{R}(x)\alpha_{\Phi}\Big\rangle\nonumber\\
&\geq \langle\alpha_{R,\Phi},T_{y}\alpha_{R,\Phi}\rangle
-\frac{1}{\pi}\int_{0}^{\infty}\Big\langle\eta_{R}(x)\alpha_{\Phi},\, \frac{1}{s+K_{y}^{2}}|\nabla\eta_{R}(y)|^{2}\frac{1}{s+K_{y}^{2}}\eta_{R}(x)\alpha_{\Phi}\Big\rangle\sqrt{s}ds\nonumber\\
&\geq\langle\alpha_{R,\Phi},T_{y}\alpha_{R,\Phi}\rangle-\delta(R)\langle\eta_{R}(x)\alpha_{\Phi},\eta_{R}(x)\alpha_{\Phi}\rangle\\
&\geq\langle\alpha_{R,\Phi},\, T_{y}\alpha_{R,\Phi}\rangle-\delta(R)\int_{\R^3}\big(\eta_R^2(x)+\chi_R^2(x)\big)\rho_{\gamma_{\Phi}}dx\nonumber\\
&\geq\langle\alpha_{R,\Phi},\, T_{y}\alpha_{R,\Phi}\rangle-\delta(R)\text{Tr}(\eta_R\gamma_\Phi\eta_R)-\delta(R)e^{4\tau R}\lambda.\nonumber
\end{align}
Therefore, since $\alpha_{\Phi}\alpha_{\Phi}^{*}\leq \gamma_{\Phi}$ and  $\lim\limits_{|x|\to\infty}(W_{\theta}*\rho_{\gamma})(x)=0$, we  conclude from  \eqref{decay10}--\eqref{decay9} that
\begin{equation}\label{decay10.1}
\begin{split}
&\big\langle\alpha_{R,\Phi},(T_{x}+T_{y})\alpha_{R,\Phi}\big\rangle-\kappa\Ex_\theta(\alpha_{R,\Phi})-2\mu_{\theta}\int_{\R^{3}\times\R^{3}}|\alpha_{R,\Phi}(x,y)|^{2}dxdy\\
\leq&|I|+|II|+\big(16\tau^{2}m^{-1}+\delta(R)\big)\Tr(\eta_{R}\gamma_{\Phi}\eta_{R})+(16\tau^{2}m^{-1}+\delta(R))e^{4\tau R}\lambda.
\end{split}\end{equation}

To estimate $I$ of (\ref{decay10.1}), we first prove that
\begin{align}\label{3.24f}
\|K^{-1/2}e^{\Phi}X_{\eta_{R}\gamma,\theta}e^{-\Phi}\|_{\fS_{2}}\to0\ \   \text{as}\ \ R\to\infty,
\end{align}
where $X_{\eta_{R}\gamma, \theta}$ denotes  the operator defined by the integral kernel
$$
X_{\eta_{R}\gamma,\theta}(x,y)=\frac{e^{-\theta|x-y|}\eta_{R}(x)\gamma(x,y)}{|x-y|}.
$$ 
Note that  the integral kernel of the operator
$K^{-1/2}e^{\Phi}X_{\eta_{R}\gamma,\theta}e^{-2\Phi}X_{\eta_{R}\gamma,\theta}^{*}e^{\Phi}K^{-1/2}$
is as follows:
\begin{align*}
&\big(K^{-1/2}e^{\Phi}X_{\eta_{R}\gamma,\theta}e^{-2\Phi}X_{\eta_{R}\gamma,\theta}^{*}e^{\Phi}K^{-1/2}\big)(x,x')\nonumber\\
=&\int_{(\R^{3})^3}K^{-1/2}(x,y)\Psi(y,z)\Psi(z,w)\frac{(\eta_{R}\gamma)(y,z)(\gamma \eta_{R})(z,w)}{|y-z||z-w|} K^{-1/2}(w,x')dydzdw,
\end{align*}
where $\Psi(x,y):=e^{-\theta|x-y|+\Phi(x)-\Phi(y)}\leq1$ in $\R^3\times\R^3$ follows from the fact  $0<\tau\leq \theta$.
Using the non-negativity of the integral kernel $K^{-1/2}(x,y)$  (see \cite{FL,green}),  we  then conclude that
\begin{align*}
&\ \     \big|\big(K^{-1/2}e^{\Phi}X_{\eta_{R}\gamma,\theta}e^{-2\Phi}X_{\eta_{R}\gamma,\theta}^{*}e^{\Phi}K^{-1/2}\big)(x,x')\big|\nonumber\\[1.5mm]
&\leq \int_{(\R^{3})^3}K^{-1/2}(x,y)\frac{(\eta_{R}\gamma)(y,z)(\gamma \eta_{R})(z,w)}{|y-z||z-w|}K^{-1/2}(w,x')dydzdw\nonumber\\
&= \int_{(\R^{3})^3}K^{-1/2}(x,y)X_{\eta_{R}\gamma,0}(y,z)X_{\eta_{R}\gamma,0}^{*}(z,w)K^{-1/2}(w,x')dydzdw\nonumber\\
&=(K^{-1/2}X_{\eta_{R}\gamma,0}X_{\eta_{R}\gamma,0}^{*}K^{-1/2})(x,x').
\end{align*}
This  implies that
\begin{align}\label{3.16}
\|K^{-1/2}e^{\Phi}X_{\eta_{R}\gamma,\theta}e^{-\Phi}\|_{\fS_{2}}^{2}&=\int_{\R^{3}}\big|\big(K^{-1/2}e^{\Phi}X_{\eta_{R}\gamma,\theta}e^{-2\Phi}X_{\eta_{R}\gamma,\theta}^{*}e^{\Phi}K^{-1/2}\big)\big|(x,x)dx\nonumber\\
&\leq \int_{\R^{3}}(K^{-1/2}X_{\eta_{R}\gamma,0}X_{\eta_{R}\gamma,0}^{*}K^{-1/2})(x,x)dx\\
&=\|K^{-1/2}X_{\eta_{R}\gamma,0}\|_{\fS_{2}}^{2}.\nonumber
\end{align}

We thus deduce from \eqref{hkt} and \eqref{3.16} that
\begin{align}\label{3.26f}
\|K^{-1/2}e^{\Phi}X_{\eta_{R}\gamma,\theta}e^{-\Phi}\|_{\fS_{2}}
\leq \|K^{-1/2}X_{\eta_{R}\gamma,0}\|_{\fS_{2}}
\leq C\|K^{1/2}\eta_{R}\gamma\|_{\fS_{2}},
\end{align}
where $C>0$ is independent of $R>0$.
Moreover,  note from \cite[(A.6)]{LenLew-10} that
\begin{align*}
[K^{1/2},\eta_{R}]=\frac{1}{\pi}\int_0^\infty\frac{1}{s+K}[K,\eta_{R}]\frac{1}{s+K}\sqrt{s}ds.
\end{align*}
This gives that
\begin{align}\label{3.22}
\big\|[K^{1/2},\eta_{R}]\big\|&\leq\frac{\big\|[K,\eta_{R}]\big\|}{\pi}\int_{0}^{\infty}\frac{\sqrt{s}}{(s+m)^{2}}ds\leq \frac{C_1}{2\sqrt{m}R},
\end{align}
where we have  used the fact that
\begin{align*}
\|[K,\eta_{R}]\|\leq C_1R^{-1}\ \ \ \text{for some}\ C_1>0.
\end{align*}
Since $\gamma^{2}\leq\gamma$ and
\begin{equation*}\label{commutator}
\begin{split}
\eta_{R}K\eta_{R}=& K^{1/2}\eta_{R}^{2}K^{1/2}+K^{1/2}\eta_{R}[K^{1/2},\, \eta_{R}]+[\eta_{R},\, K^{1/2}]\eta_{R}K^{1/2}\nonumber\\
&+[\eta_{R}, \, K^{1/2}][K^{1/2},\, \eta_{R}],
\end{split}\end{equation*}
we obtain that
\begin{align}\label{kinetic-decay}
&\ \     \|K^{1/2}\eta_{R}\gamma\|_{\fS_{2}}^{2}
=\Tr(\eta_{R}K\eta_{R}\gamma^{2})\nonumber\\
&\leq\|\eta_{R}K^{1/2}\gamma\|_{\fS_{2}}^{2}+2\Re\big\langle\eta_{R}K^{1/2}\gamma,\, [K^{1/2},\eta_{R}]\gamma\big\rangle+\big\|[K^{1/2},\, \eta_{R}]\gamma\big\|_{\fS_{2}}^2\\
&\leq\|\eta_{R}K^{1/2}\gamma\|_{\fS_{2}}^{2}+2\|\gamma\|_{\fS_{2}}\|[K^{1/2},\eta_{R}]\|\|\eta_{R}K^{1/2}\gamma\|_{\fS_{2}}+\|[K^{1/2},\eta_{R}]\|^{2}\big(\Tr(\gamma)\big)^2\nonumber\\
&\leq\int_{\R^{3}}\eta_{R}^{2}(x)\rho_{K^{1/2}\gamma K^{1/2}}(x)dx+\delta(R),\nonumber
\end{align}
where $\delta(R)>0$ satisfies $\delta(R)\to0$ as $R\to\infty$. Because $\rho_{K^{1/2}\gamma K^{1/2}}\in L^{1}(\R^{3})$, we obtain from \eqref{kinetic-decay} that
\begin{align}\label{3.30a}
\|K^{1/2}\eta_{R}\gamma\|_{\fS_{2}}\rightarrow 0\ \   \text{as}\ \  R\rightarrow\infty.
\end{align}
Together with \eqref{3.26f},  we  therefore derive that  \eqref{3.24f} holds true.

We now calculate from \eqref{3.24f}  that   for  sufficiently large $R>0$,
\begin{align}\label{decay5}
&\ \      \big|\big\langle\alpha_{R,\Phi},\ \eta_{R}(x)\eta_{R}(y)e^{\Phi(x)}(X_{\gamma,\theta})_{x}\alpha\big\rangle\big|\nonumber\\
&=\big|\Tr\big(\alpha_{\Phi}\eta_{R}\alpha_{R,\Phi}^{*}K^{1/2}K^{-1/2}e^{\Phi}X_{\eta_{R}\gamma,\theta}e^{-\Phi}\big)\big|\nonumber\\
&\leq \|\alpha_{\Phi}\eta_{R}\|_{\fS_{2}}\|K^{1/2}\alpha_{R,\Phi}\|_{\fS_{2}}\|K^{-1/2}e^{\Phi}X_{\eta_{R}\gamma,\theta}e^{-\Phi}\|_{\fS_{2}}\nonumber\\
&\leq \delta(R)\|K^{1/2}\alpha_{R,\Phi}\|_{\fS_{2}}\big(\|\eta_{R}\alpha_{\Phi}\eta_{R}\|_{\fS_{2}}+\|\chi_{R}\alpha_{\Phi}\eta_{R}\|_{\fS_{2}}\big)\\
&\leq\delta(R)\sqrt{\text{Tr}(K\eta_{R}\gamma_{\Phi}\eta_{R})}\Big(\sqrt{\text{Tr}(K\eta_{R}\gamma_{\Phi}\eta_{R})}+e^{2\tau R}\|\chi_{R}\alpha\|_{\fS_{2}}\Big)\nonumber\\
&\leq \delta(R)\Big(\text{Tr}(K\eta_{R}\gamma_{\Phi}\eta_{R})+e^{4\tau R}\|\chi_{R}\alpha\|^2_{\fS_{2}}\Big)\nonumber\\
&\leq\delta(R)\Big(\Tr (T\eta_{R}\gamma_{\Phi}\eta_{R})+\inte \eta_R^2\rho_{\gamma_\phi}dx+e^{4\tau R}\lambda\Big)\nonumber\\
&\leq \delta(R)\Big(\Tr\big((L_{\gamma,\theta}-\mu_{\theta})\eta_{R}\gamma_{\Phi}\eta_{R}\big)+e^{4\tau R}\lambda\Big),\nonumber
\end{align}
where the last inequality follows from  the facts that $\mu_{\theta}<0$ and $\lim\limits_{|x|\to\infty}(W_{\theta}*\rho_{\gamma})(x)=0$.
Similarly,  it follows that for  sufficiently large $R>0$,
\begin{align}\label{esti-Xgty}
&\ \     \big|\big\langle\alpha_{R,\Phi},\, \eta_{R}(x)\eta_{R}(y)e^{\Phi(x)}(X_{\gamma,\theta})_{y}\alpha\big\rangle\big|\nonumber\\
&=\big|\Tr(K^{1/2}\alpha_{R,\Phi}^{*}(\eta_{R}\alpha_{\Phi})X_{\overline{\gamma\eta_{R}},\theta}K^{-1/2})\big|\\
&\leq \|K^{1/2}\alpha_{R,\Phi}\|_{\fS_{2}}\|K^{-1/2}X_{\eta_{R}\overline{\gamma},\theta}\|_{\fS_{2}}\|\eta_{R}\alpha_{\Phi}\|_{\fS_{2}}\nonumber\\
&\leq \delta(R)\Big(\Tr\big((L_{\gamma,\theta}-\mu_{\theta})\eta_{R}\gamma_{\Phi}\eta_{R}\big)+e^{4\tau R}\lambda\Big),\nonumber
\end{align}
and thus 
\begin{align}\label{i}
|I|\leq \delta(R)\Big(\Tr\big((L_{\gamma,\theta}-\mu_{\theta})\eta_{R}\gamma_{\Phi}\eta_{R}\big)+e^{4\tau R}\lambda\Big),
\end{align}
where $\delta(R)>0$ satisfies $\delta(R)\to0$ as $R\to\infty$.

We now estimate $II$ of \eqref{decay10.1}.  The same arguments of  \eqref{3.26f} and \eqref{3.30a} yield that
\begin{align*}
&\ \ \,\    \big|\big\langle\alpha_{R,\Phi},\, \eta_{R}(x)\eta_{R}(y)e^{\Phi(x)}\gamma_{x}X_{\alpha,\theta}\big\rangle\big|\nonumber\\
&=\big|\Tr(e^{-\Phi}X_{\alpha\eta_{R},\theta}K^{-1/2}K^{1/2}\alpha_{R,\Phi}^{*}\eta_{R}\gamma_{\Phi})\big|\nonumber\\
&\leq \|K^{-1/2}X_{\eta_{R}\alpha^{*},\theta}\|_{\fS_{2}}\|K^{1/2}\alpha_{R,\Phi}^{*}\|_{\fS_{2}}\|\eta_{R}\gamma_{\Phi}\|_{\fS_{2}}\nonumber\\
&\leq C\|K^{1/2}\eta_{R}\alpha^{*}\|_{\fS_{2}}\|K^{1/2}\alpha_{R,\Phi}^{*}\|_{\fS_{2}}\|\eta_{R}\gamma_{\Phi}\|_{\fS_{2}}\\
&\leq \delta(R)\|K^{1/2}\alpha_{R,\Phi}^{*}\|_{\fS_{2}}\|\eta_{R}\gamma_{\Phi}\|_{\fS_{2}}.
\end{align*}
Hence, by increasing $\delta(R)$ if necessary, we conclude that
\begin{align*}
\big|\langle\alpha_{R,\Phi},\ \eta_{R}(x)\eta_{R}(y)e^{\Phi(x)}\gamma_{x}X_{\alpha,\theta}\rangle\big|\leq \delta(R)\Tr\big((L_{\gamma,\theta}-\mu_{\theta})\eta_{R}\gamma_{\Phi}\eta_{R}\big).
\end{align*}
Similarly, we have
\begin{align*}
&    \big|\langle\alpha_{R,\Phi},\eta_{R}(x)\eta_{R}(y)e^{\Phi(x)}\gamma_{y}X_{\alpha,\theta}\rangle\big|\\
= &\big|\Tr(K^{-1/2}X_{\eta_{R}\alpha_{\Phi},\theta}(\overline{\gamma\eta_{R}})\alpha_{R,\Phi}^{*}K^{1/2})\big|\nonumber\\
\leq & \|K^{-1/2}X_{\eta_{R}\alpha_{\Phi},\theta}\|_{\fS_{2}}\|K^{1/2}\alpha_{R,\Phi}\|_{\fS_{2}}\|\gamma\eta_{R}\|_{\fS_{2}}\nonumber\\
\leq &\delta(R)\Tr\big((L_{\gamma,\theta}-\mu_{\theta})\eta_{R}\gamma_{\Phi}\eta_{R}\big),
\end{align*}
and hence for  sufficiently large $R>0$,
\begin{align}\label{ii}
|II|\leq\delta(R)\Tr\big((L_{\gamma,\theta}-\mu_{\theta})\eta_{R}\gamma_{\Phi}\eta_{R}\big).
\end{align}

As a consequence of  \eqref{decay10.1}, \eqref{i} and \eqref{ii},  one gets that for  sufficiently large $R>0$,
\begin{equation}\label{decay11}
\begin{split}
&\big\langle\alpha_{R,\Phi},\,  (T_{x}+T_{y})\alpha_{R,\Phi}\big\rangle-\kappa \Ex_\theta(\alpha_{R,\Phi})-2\mu_{\theta}\int_{\R^{3}\times\R^{3}}|\alpha_{R,\Phi}(x,y)|^{2}dxdy\\
\leq& 17\tau^{2}m^{-1}\Tr(\eta_{R}\gamma_{\Phi}\eta_{R})+\delta(R)\Tr\big((L_{\gamma,\theta}-\mu_{\theta})\eta_{R}\gamma_{\Phi}\eta_{R}\big)
+17\tau^{2}m^{-1}e^{4\tau R}\lambda.
\end{split}\end{equation}
Let  $\beta_{\theta}>0$ be given by  Proposition \ref{prop:min-Gtheta}. The definition of $\beta_{\theta}$ then gives that  $\beta_{\theta}=\beta_{\theta}(\kappa)$ is continuous in  $\kappa\in(0, 4/\pi)$. Since $\mu_{\theta}<\beta_{\theta}-m$, there exists a sufficiently small  $\varepsilon>0$ such that $\mu_{\theta}<\beta_{\theta}(\kappa+\varepsilon)-m$, which  gives that
\begin{align}\label{decay11.1}
\text{LHS of \eqref{decay11} }&=\Big\langle\alpha_{R,\Phi},\, \Big(T_{x}+T_{y}-\frac{(\kappa+\varepsilon-\varepsilon) e^{-\theta|x-y|}}{|x-y|}-2\mu_{\theta}\Big)\alpha_{R,\Phi}\Big\rangle\nonumber\\
&\geq\Big\langle\alpha_{R,\Phi},\, \Big(2\big(\beta_{\theta}(\kappa+\varepsilon)-m-\mu_{\theta}\big)+\frac{\varepsilon e^{-\theta|x-y|}}{|x-y|}\Big)\alpha_{R,\Phi}\Big\rangle\\
&\geq \varepsilon \Ex_\theta(\alpha_{R,\Phi})=\varepsilon\Ex_\theta(\eta_{R}\alpha_{\Phi}\eta_{R}).\nonumber
\end{align}
We thus  deduce from \eqref{decay11} and \eqref{decay11.1} that  \eqref{esti-a-wave} holds  for sufficiently large $R>0$, and  the proof of Proposition \ref{prop:esti-g-a-pos-a-wave} is therefore complete.\qed


\subsection{Proof of Theorem \ref{thm:main2}}\label{sec:pf-main2-aneq0} 
In this subsection, we complete the proof of Theorem \ref{thm:main2}. We begin with the following analytical properties of the operator
\begin{align}\label{3.31a}
L_{\gamma,\theta}:=T-\kappa V_\theta:=T-\kappa \int_{\R^3}\frac{e^{-\theta|x-y|}\rho_{\gamma}(y)}{|x-y|}dy\ \ \mbox{on}\ \   L^{2}(\R^{3}).
\end{align}

\begin{lemma}\label{lem:spe-mean-field}
Suppose $0<\kappa<4/\pi$ and $\theta\geq 0$. Then  $V_{\theta}>0$ is  bounded and is a $\sqrt{-\Delta}$-relatively compact multiplication operator on $L^{2}(\R^{3})$. 
\end{lemma}

\noindent \textbf{Proof. }
Note from \eqref{hkt} that for any $\theta\geq0$, 
\begin{align*}\label{W-Linf-bdd}
\|V_{\theta}\|_{L^{\infty}}&=\sup_{x\in\R^{3}}\int_{\R^{3}}\frac{e^{-\theta|x-y|}\rho_{\gamma}(y)}{|x-y|}dy\leq \frac{\pi}{2}\langle\sqrt{\rho_{\gamma}},\sqrt{-\Delta}\sqrt{\rho_{\gamma}}\rangle\leq\frac{\pi}{2}\text{Tr}\sqrt{-\Delta}\gamma<\infty.
\end{align*}
This yields that  for any $\theta\geq 0$, $V_{\theta}\in L^{\infty}(\R^{3})$  is a bounded and multiplicative operator on $L^{2}(\R^{3})$.
Since $\sqrt{\rho_{\gamma}}\in H^{1/2}(\R^{3})$,  Sobolev's embedding theorem yields that $\rho_{\gamma}\in L^{p}(\R^{3})$ holds for any $p\in [1,3/2]$.  By the  Hardy-Littlewood-Sobolev inequality, we then derive that  for any $p\in (1,3/2)$ and any $f\in L^{q}(\R^{3})$ satisfying $1/p+1/q=5/3$,
\begin{align*}
\Big|\int_{\R^{3}}f(x)V_{\theta}(x)dx\Big|&=\Big|\int_{\R^{3}\times\R^{3}}e^{-\theta|x-y|}\frac{f(x)\rho_{\gamma}(y)}
{|x-y|}dxdy\Big|\leq \|f\|_{L^{q}}\|\rho_{\gamma}\|_{L^{p}}.
\end{align*}
Applying a duality argument, this then implies that
$V_{\theta}(x)\in L^{r}(\R^{3})$ holds for $r=\frac{3p}{3-2p}$.  Since $p\in (1,3/2)$ is arbitrary, we deduce that
\begin{equation}\label{d.1d}
V_{\theta}(x)\in L^{r}(\R^{3}),\ \    \forall\ r\in (3,\infty).
\end{equation}

Let $\{\varphi_{n}\}\subset L^{2}(\R^{3})$  satisfy $\varphi_{n}\rightharpoonup 0$ weakly in  $L^{2}(\R^{3})$ as $n\to\infty$.
By Fourier transform argument, one can deduce that the sequence $\{\psi_{n}:=(\sqrt{-\Delta}+1)^{-1}\varphi_{n}\}$ is bounded uniformly in $H^{1}(\R^{3})$. This shows that $\{\psi_{n}\}$ is bounded uniformly in $L^{3}(\R^{3})$, and
\begin{equation}\label{local-converg}
\psi_{n}\rightarrow 0\quad\text{strongly in }L_{loc}^{3}(\R^{3})\ \text{ as}\ n\to\infty.
\end{equation}
Note that
\begin{equation*}\label{WR-psi-con}
\begin{split}
\int_{\R^3}|V_{\theta}\psi_{n}|^2dx
=&\int_{|x|\leq R}|V_{\theta}\psi_{n}|^2dx+\int_{|x|\geq R}|V_{\theta}\psi_{n}|^2dx\\
\leq& \|V_{\theta}\|_{L^6}^2\|\psi_{n}\|_{L^3(B_R)}^2
+\|V_{\theta}\|_{L^6(B^c_R)}^2\|\psi_{n}\|_{L^3}^2.
\end{split}
\end{equation*}
We thus deduce from  \eqref{d.1d} and \eqref{local-converg} that $V_{\theta}(\sqrt{-\Delta}+1)^{-1}\varphi_{n}\rightarrow 0$ strongly in $L^{2}(\R^{3})$ as $n\to\infty$, which yields that $V_{\theta}$ is $\sqrt{-\Delta}$-relatively compact on $L^2(\R^3)$. \qed

\begin{lemma}\label{lem:K-ess}
Let $\Omega$ be a compact subset in $\R^{3}$, and define
\begin{align}\label{E1}
E_{1}(L_{\gamma,\theta}\big|_{\R^{3}\setminus \Omega})&:=\inf\big\{\langle\psi,L_{\gamma,\theta}\psi\rangle/\|\psi\|_{L^2}^{2}:\, \psi\in H^{1/2}(\R^{3}),\psi=0\text{ on }\Omega\big\},
\end{align}
where $0<\kappa<4/\pi$ and $\theta\geq 0$. Then we have
\begin{align}\label{Einf}
E_{\infty}&:=\sup_{\Omega\subseteq\R^{3}\text{ is compact}}E_{1}(L_{\gamma,\theta}|_{\R^{3}\setminus \Omega})=0.
\end{align}
\end{lemma}

\noindent \textbf{Proof.}
We first prove that  for every $R>0$,
\begin{align}\label{E1-less0}
E_{1}(L_{\gamma,\theta}|_{B_{R}^{c}})&:=\inf\big\{\langle\psi,L_{\gamma,\theta}\psi\rangle/\|\psi\|_{L^2}^{2}:\, \psi\in H^{1/2}(\R^{3}),\psi=0\text{ on }B_R\big\}\leq 0,
\end{align}
where  $L_{\gamma,\theta}$ is given by \eqref{3.31a}. 
In order to prove (\ref{E1-less0}), let $\ell[\psi]:=\langle\psi,L_{\gamma,\theta}\psi\rangle$ be the quadratic form of $L_{\gamma,\theta}$ with form domain $H^{1/2}(\R^{3})$.   Note from  Lemma \ref{lem:spe-mean-field} that $\sigma_{\rm ess}(L_{\gamma,\theta})=\sigma_{\rm ess}(T)=[0,\infty)$. We then deduce from  \cite[Lemma B.3]{RoosSeir} that  there exists a sequence $\{\psi_{n}\}\subset H^{1/2}(\R^{3})$ such that $\|\psi_{n}\|_{L^2}=1$,
$$\lim\limits_{n\to\infty}\ell[\psi_{n}]=0\ \   \text{and}\ \   \psi_{n}\rightharpoonup 0\ \   \text{weakly in}\ H^{1/2}(\R^{3})\ \text{ as}\,\ n\to\infty.
$$
Denote $\eta_R(x):=\sqrt{1-\chi_R^{2}(x)}$, where $\chi_R\in C_0^\infty(\R^3;[0,1])$ is a smooth cutoff function such that $\chi_R(x)=0$ for $|x|> R$ and $\chi_R(x)=1$ for $|x|\leq 3R/4$. By the definition of $E_{1}(L_{\gamma,\theta}|_{B_{R}^{c}})$,
this implies that for any $n\geq 1$,
\begin{align}\label{d4}
E_{1}(L_{\gamma,\theta}|_{B_{R}^{c}})&\leq \ell\big[\eta_{R}\psi_{n}/\|\eta_{R}\psi_{n}\|_{L^2}\big].
\end{align}
Since
\begin{equation}\label{d6}
\psi_{n}\rightarrow 0\quad\text{strongly in }L_{loc}^{2}(\R^{3})\ \text{ as}\ n\to\infty,
\end{equation}
we have
\begin{align*}
\lim\limits_{n\to\infty}\|\eta_{R}\psi_{n}\|_{L^2}^{2}&=\lim\limits_{n\to\infty}\big(\|\psi_{n}\|_{L^2}^{2}-\|\chi_{R}\psi_{n}\|_{L^2}^{2}\big)=\lim\limits_{n\to\infty}\|\psi_{n}\|_{L^2}^{2}=1,
\end{align*}
which further yields from \eqref{d4} that  for sufficiently large $n>0$,
\begin{align}\label{d5}
E_{1}(L_{\gamma,\theta}|_{B_{R}^{c}})\leq 2\ell[\eta_{R}\psi_{n}].
\end{align}

Moreover,  we obtain from Lemma \ref{lem:IMS}  that
\begin{align}\label{xi-form-decomp}
\ell[\psi_{n}]\geq&\ell[\chi_{R}\psi_{n}]+\ell[\eta_{R}\psi_{n}]\\
&-\frac{1}{\pi}\int_{0}^{\infty}\big\langle\psi_{n},\frac{1}{s+K^{2}}\big(|\nabla\chi_{R}|^{2}+|\nabla\eta_{R}|^{2}\big)\frac{1}{s+K^{2}}\psi_{n}\big\rangle\sqrt{s}ds.\nonumber
\end{align}
It can be checked from \eqref{d6} that
\begin{align}\label{d8}
\lim\limits_{n\rightarrow\infty}\frac{1}{\pi}\int_{0}^{\infty}\big\langle\psi_{n},\frac{1}{s+K^{2}}\big(|\nabla\chi_{R}|^{2}+|\nabla\eta_{R}|^{2}\big)\frac{1}{s+K^{2}}\psi_{n}\big\rangle\sqrt{s}ds=0.
\end{align}
Set $E:=\inf\sigma(L_{\gamma,\theta})$.  Then we have
\begin{align*}
	\ell[\chi_{R}\psi_{n}]&\geq E\|\chi_{R}\psi_{n}\|_{L^2}^{2},
\end{align*}
and hence
\begin{align}\label{chiR-xi-jinf}
\liminf_{n\rightarrow\infty}\ell[\chi_{R}\psi_{n}]&\geq \liminf_{n\rightarrow\infty} E\|\chi_{R}\psi_{n}\|_{L^2}^{2}=0.
\end{align}
As a consequence of  \eqref{xi-form-decomp}--\eqref{chiR-xi-jinf}, we derive that
\begin{align*}
\limsup_{n\rightarrow\infty}\ell[\eta_{R}\psi_{n}]&\leq \limsup_{n\rightarrow\infty}\ell[\psi_{n}]=0,
\end{align*}
which therefore implies that \eqref{E1-less0} holds true in view of \eqref{d5}.

By \eqref{E1-less0} and the fact that $\sigma_{\rm ess}(L_{\gamma,\theta})=[0,\infty)$, we derive that
$$E_{\infty}:=\sup_{\Omega\subseteq\R^{3}\text{ is compact}}E_{1}(L_{\gamma,\theta}|_{\R^{3}\setminus \Omega})\le 0.$$
On the other hand, take the sequences $\{R_{n}\}\subset\R$ and $\{\phi_{n}\}\subset H^{1/2}(\R^{3})$ such that $\lim\limits_{n\to\infty}R_{n}=\infty$,  $\|\phi_{n}\|_{L^2}\equiv 1$,  the  support of $\phi_{n}$ is contained in $\{|x|\geq R_{n}\}$ for each $n\in \mathbb{N}$, and
\begin{align*}
\ell[\phi_{n}]-E_{1}(L_{\gamma,\theta}|_{B_{R_{n}}^{c}})\rightarrow 0\ \   \text{as}\ \ n\to\infty.
\end{align*}
Since  $\phi_{n}\rightharpoonup0$ weakly in $L^{2}(\R^{3})$ as $n\to\infty$, we deduce from \cite[Lemma B.3]{RoosSeir}  that
\begin{align*}
0=\inf\sigma_{\rm ess}(L_{\gamma,\theta})\leq \liminf_{n\rightarrow\infty}\ell[\phi_{n}]=\liminf_{n\rightarrow\infty}E_{1}(L_{\gamma,\theta}|_{B_{R_{n}}^{c}})\leq E_{\infty},
\end{align*}
and hence $E_{\infty}=0$. This therefore completes the proof of Lemma \ref{lem:K-ess}. \qed

Following  above lemmas, we are next able to establish the following proposition.

\begin{prop}\label{pro3.6}
Suppose $m>0$, $0<\kappa<4/\pi$ and $0<\lambda<\lambda^{\rm HFB}$.  Then there exists a constant $0<\theta_{c}=\theta_{c}(m, \kappa, \lambda)\le \min\{1, m/6, \widetilde{\theta}\}$ such that  if $\theta\in (0,\theta_{c}]$ and $I^{\rm HFB}_{m,\kappa,\theta}(\lambda)$ has
a minimizer $(\gamma,\alpha)\neq(\gamma, 0)$, then there exists a  sufficiently large constant $R_0>0$ such that
\begin{align}\label{3.36a}
\sup\limits_{R\geq R_0}\big\|\eta_{R}(e^{3\theta|\cdot|}\gamma e^{3\theta|\cdot|},\ e^{3\theta|\cdot|}\alpha)\eta_{R}\big\|_{\cX}<\infty,
\end{align}
where  $\widetilde{\theta}=\widetilde{\theta}(m, \kappa,\lambda)>0$ is given by Proposition \ref{prop:min-Gtheta}.
\end{prop}


\noindent \textbf{Proof.}
Suppose $(\gamma,\alpha)\neq(\gamma, 0)$ is a minimizer of $I^{\rm HFB}_{m,\kappa,\theta}(\lambda)$, where $0< \theta\leq \min\{1,m/6,\widetilde{\theta}\}$. We shall carry out the proof by four steps.

$Step\ 1$.  We claim that  there exists $\tau_{c_1}>0$  such that for any fixed sufficiently large $R>0$, 
\begin{equation}\label{st1}
(1-\Delta)^{1/4}\eta_{R}\gamma_{\Phi}\eta_{R}(1-\Delta)^{1/4}\in\fS_{1},
\end{equation}
where $\Phi(x)=e^{\tau_{c_1}|x|}$.

To address the above claim (\ref{st1}), take $0<\tau\leq \min\{1,m/6,{\theta}\}$. Note from \eqref{esti-g-a-pos}  and  \eqref{esti-a-wave} that for sufficiently large $R>0$, 
\begin{align}\label{decay13}
&\Tr\Big(\big[L_{\gamma,\theta}-\mu_{\theta}\big]\eta_{R}\gamma_{\Phi}\eta_{R}-\big[C_1\tau^2m^{-1}+\delta(R)\big]
\gamma_{\Phi}\Big)\\
\leq& \delta(R)\kappa\Tr\big([L_{\gamma,\theta}-\mu_{\theta}]\eta_{R/4}\gamma_{\Phi}\eta_{R/4}\big)+ \kappa C_1\tau^{2}m^{-1}\Tr(\eta_{R/4}\gamma_{\Phi}\eta_{R/4})+M_{R/4}\lambda.\nonumber
\end{align}
Since $\chi_R^2+\eta^2_{R}=1$ and  $\eta_{R}\eta_{R/4}=\eta_{R}$,
we conclude from Lemma \ref{lem:IMS} that
\begin{align*}
\eta_{R/4}T\eta_{R/4}&=\eta_{R/4}\frac{(\chi_{R}^{2}+\eta_{R}^{2})T+T(\chi_{R}^{2}+\eta_{R}^{2})}{2}\eta_{R/4}\nonumber\\
&\leq\eta_{R/4}\chi_{R}T\chi_{R}\eta_{R/4}+\eta_{R}T\eta_{R}+\eta_{R/4}L_{\chi_{R}}\eta_{R/4}+\eta_{R/4}L_{\eta_{R}}\eta_{R/4},
\end{align*}
and
$$\|L_{\chi_{R}}\|+\|L_{\eta_{R}}\|\leq C'_1(\|\nabla\chi_{R}\|_{L^{\infty}}^{2}+\|\nabla\eta_{R}\|_{L^{\infty}}^{2})\leq C''_1R^{-2},
$$
where  the operators $L_{\chi_{R}}$ and $L_{\eta_{R}}$ are as in \eqref{3.6b}, and the constant $C''_1>0$ is independent of $R>0$. This implies that
\begin{align}\label{decay15}
&\ \     \Tr\big(\eta_{R/4}(L_{\gamma,\theta}-\mu_{\theta})\eta_{R/4}\gamma_{\Phi}\big)\nonumber\\
&\leq \Tr\big(\eta_{R/4}\chi_{R}(L_{\gamma,\theta}-\mu_{\theta})\chi_{R}\eta_{R/4}\gamma_{\Phi}\big)+\Tr\big(\eta_{R}(L_{\gamma,\theta}-\mu_{\theta})\eta_{R}\gamma_{\Phi}\big)\nonumber\\
&\ \     +C''_1R^{-2}\Tr(\eta_{R/4}\gamma_{\Phi}\eta_{R/4})\\
&\leq \Tr\big(\eta_{R}(L_{\gamma,\theta}-\mu_{\theta})\eta_{R}\gamma_{\Phi}\big)+ \Tr(L_{\gamma,\theta}g_{R}\gamma g_{R})+(-\mu_{\theta}+C''_1R^{-2})e^{4\tau R}\lambda\nonumber\\
&\ \     +C''_1R^{-2}\Tr(\eta_{R}\gamma_{\Phi}\eta_{R}),\nonumber
\end{align}
where $g_{R}:=\eta_{R/4}\chi_{R}e^{\Phi}$.
Since $g_{R}\in C^\infty_0(\R^3)$ has bounded derivatives and compact support, we obtain that  $\Tr(L_{\gamma,\theta}g_{R}\gamma g_{R})<\infty$ holds for  any fixed $R>0$.
Moreover, note that
\begin{align}\label{1}
\Tr\gamma_{\Phi}=\int_{\R^3}\chi^2_{R}\rho_{\gamma_{\Phi}}dx+\int_{\R^3}\eta^2_{R}\rho_{\gamma_{\Phi}}dx
\leq e^{4\tau R}\lambda+\Tr(\eta_{R}\gamma_{\Phi}\eta_{R}),
\end{align}
and
\begin{align}\label{2}
\Tr(\eta_{R/4}\gamma_{\Phi}\eta_{R/4})
&=\int_{\R^{3}}\chi_{R}^{2}(x)\eta_{R/4}^{2}(x)\rho_{\gamma_{\Phi}}(x)dx+\int_{\R^{3}}\eta_{R}^{2}(x)\rho_{\gamma_{\Phi}}(x)dx\nonumber\\
&\leq e^{4\tau R}\lambda+\Tr(\eta_{R}\gamma_{\Phi}\eta_{R}).
\end{align}
We thus deduce from  \eqref{decay13}--\eqref{2} that for any fixed sufficiently large $R>0$, 
\begin{align}\label{decay14}
&\ \     \Tr\Big(\big[L_{\gamma,\theta}-\mu_{\theta}-2(1+\kappa)(C_1\tau^{2}m^{-1}+\delta(R))\big]\eta_{R}\gamma_{\Phi}\eta_{R}\Big)\nonumber\\
&\leq 4\Big\{\Tr(L_{\gamma,\theta}g_{R}\gamma g_{R})+\big[\big(C_1\tau^{2}m^{-1}+\delta(R)-\mu_{\theta}\big) e^{4\tau R}+M_{R/4}\big]\lambda\Big\}\\
&=:N_{1}(R,\tau,\lambda)<\infty,\nonumber
\end{align}
where $C_1>0$ is independent of $R>0$.
	
We now estimate $\text{Tr}(\eta_R\gamma_\Phi\eta_R)$.  Note from  \eqref{kin-density} that
\begin{align*}\label{3.42a}
\Tr(L_{\gamma,\theta}\eta_{R}\gamma_{\Phi}\eta_{R})
&\geq \big\langle\eta_{R}\sqrt{\rho_{\gamma_{\Phi}}},\, L_{\gamma,\theta}\eta_{R}\sqrt{\rho_{\gamma_{\Phi}}}\big\rangle.
\end{align*}
This shows from \eqref{decay14} that for  any fixed  sufficiently large $R>0$, 
\begin{equation}\label{ess-property}
\begin{split}
&\big\langle\eta_{R}\sqrt{\rho_{\gamma_{\Phi}}},\, L_{\gamma,\theta}\eta_{R}\sqrt{\rho_{\gamma_{\Phi}}}\big\rangle\\
\leq& N_{1}(R,\tau,\lambda)+\big[\mu_{\theta}+2(\kappa+1)(C_1\tau^{2}m^{-1}+\delta(R))\big]\|\eta_{R}\sqrt{\rho_{\gamma_{\Phi}}}\|_{L^2}^2.
\end{split}\end{equation}
On the other hand,   let $\beta_{\theta}>0$ be given by Propositions \ref{prop:min-Gtheta}. Since $\theta\mapsto\beta_{\theta}$ is continuous and non-decreasing in $0\leq \theta\leq \widetilde{\theta}$, we deduce from \eqref{d.8} that $\sup_{0\leq\theta\leq \widetilde{\theta}}(\beta_\theta-m)=\beta_{\widetilde{\theta}}-m<0$. This implies from  Proposition \ref{prop:chemical-pot} that 
\begin{align}\label{3.44}
\mu_{\theta}&\leq\mu:=\sup_{\theta\in [0,\widetilde{\theta}]}\mu_{\theta}\leq\beta_{\widetilde{\theta}}-m<0,\ \   \forall\ 0< \theta\leq \min\{1,m/6,\widetilde{\theta}\}.
\end{align}
It then follows   from  Lemma \ref{lem:K-ess} that
\begin{equation*}\label{decay16}
\langle\eta_{R}\sqrt{\rho_{\gamma_{\Phi}}},L_{\gamma,\theta}\eta_{R}\sqrt{\rho_{\gamma_{\Phi}}}\rangle
\geq\frac{\mu_{\theta}}{2}\|\eta_{R}\sqrt{\rho_{\gamma_{\Phi}}}\|_{L^{2}}^{2}.
\end{equation*}
Together with \eqref{ess-property}, we  thus conclude that for  any fixed  sufficiently large $R>0$, 
\begin{align}\label{3.45}
\big[|\mu_\theta|-4(\kappa+1)\big(C_1\tau^{2}m^{-1}+\delta(R)\big)\big]\, \|\eta_{R}\sqrt{\rho_{\gamma_{\Phi}}}\|_{L^{2}}^{2}\leq 2N_{1}(R,\tau,\lambda).
\end{align}
By increasing $R>0$ if necessary, we can assume  from \eqref{3.44} that  $4(\kappa+1)\delta(R)\leq |\mu|/4$.
Denote
\begin{align}\label{tau}
\tau_{c_1}&:=\min\Big\{1, m/6, \theta, c_1\sqrt{(\kappa+1)^{-1}m|\mu|}\Big\}, \ \ \  c_1=\frac{1}{4\sqrt{C_1}},
\end{align}
where $C_1>0$ is as in \eqref{3.45}.
One can check  that  $4(\kappa+1)C_1\tau_{c_{1}}^{2}m^{-1}\leq |\mu|/4$.  Therefore,  substituting $\tau\equiv \tau_{c_{1}}$ into \eqref{3.45}, we derive that for  any fixed  sufficiently large $R>0$, 
\begin{align}\label{decay17}
&\|\eta_{R}\sqrt{\rho_{\gamma_{\Phi}}}\|_{L^{2}}^{2}
<\frac{4}{|\mu|}N_{1}(R,\tau_{c_{1}},\lambda)<\infty.
\end{align}

As a consequence of  \eqref{decay14} and \eqref{decay17}, we  immediately obtain that  for  any fixed  sufficiently large $R>0$, 
\begin{align*}
\Tr(\eta_{R_1}(T-\mu_{\theta})\eta_{R_1}\gamma_{\Phi})
&\leq \Big[\kappa\|V_{\theta}\|_{L^{\infty}}+2(\kappa+1)(C_1\tau_{c_{1}}^{2}m^{-1}+1/2)\Big]\Tr(\eta_{R_1}\gamma_{\Phi}\eta_{R_1})\nonumber\\	 &\ \     +N_1(R_1,\tau_{c_{1}},\lambda)<\infty,
\end{align*}
$i.e.,$ the   claim (\ref{st1}) holds true. This completes the proof of Step 1.

$Step\ 2$.  For $0< \theta\leq \min\{1,m/6,\widetilde{\theta}\}$, in this step we claim that  there exists a constant 
$\tau_{c_2}\in(0, \tau_{c_1}]$ such that for  any fixed  sufficiently large $R>0$, 
\begin{align}\label{st2}
(1-\Delta)^{1/4}\eta_{R_2}\gamma_{2\Phi}\eta_{R_2}(1-\Delta)^{1/4}\in\fS_{1}.
\end{align}

We employ the  boostrap argument to prove the claim (\ref{st2}). Indeed, it follows from Step 1 that for each $0<c_{2}\leq c_{1}$, if we substitute $\tau_{c_{2}}$ for $\tau_{c_1}$ in $\Phi$, then for any fixed  sufficiently large $R>0$, 
\begin{align}\label{3.49a}
(1-\Delta)^{1/4}\eta_{R}\gamma_{\Phi}\eta_{R}(1-\Delta)^{1/4}\in\fS_{1}.
\end{align}
Repeating the same analysis as in the proof of Proposition \ref{prop:esti-g-a-pos-a-wave}, where $\Phi$ and $B_{\Phi}(s)$ are replaced by $2\Phi$ and $B_{2\Phi}(s)$, respectively,  we thus conclude that for sufficiently large $R>0$,
\begin{align}\label{decay18}
\Tr\Big((L_{\gamma,\theta}-\mu_{\theta})\eta_{R}\gamma_{2\Phi}\eta_{R}-\big(C'_2\tau^{2}m^{-1}+\delta(R)\big)\gamma_{2\Phi}\Big)\leq \kappa \Ex_\theta(\eta_{R/4}\alpha_{2\Phi}\eta_{R/4}),
\end{align}
and
\begin{equation}\label{decay19}
\begin{split}
\varepsilon \Ex_\theta(\eta_{R}\alpha_{2\Phi}\eta_{R})
\leq &\big[ C'_2\tau^{2}m^{-1}+\delta(R)\big]\Tr(\eta_{R/4}\gamma_{2\Phi}\eta_{R/4})\\
&+\big( C'_2\tau^{2}m^{-1}+\delta(R)\big)e^{8\tau R}\lambda
+I'+II',
\end{split}\end{equation}
where
\begin{align}\label{3.53a}
I'&:=-\kappa\Re\big\langle\alpha_{R,2\Phi},\, \eta_{R}(x)\eta_{R}(y)e^{2\Phi(x)}\big((X_{\gamma,\theta})_{x}+(X_{\gamma,\theta})_{y}\big)\alpha\big\rangle,\nonumber\\
|II'|&:=\kappa|\Re\big\langle\alpha_{R,2\Phi},\, \eta_{R}(x)\eta_{R}(y)e^{2\Phi(x)}(\gamma_{x}+\gamma_{y})X_{\alpha,\theta}\big\rangle|\\
&\ \leq \delta(R)\Tr\big((L_{\gamma,\theta}-\mu_{\theta})\eta_{R}\gamma_{2\Phi}\gamma_{R}\big),\nonumber
\end{align}
and the constants $C'_2>0$ and $\epsilon>0$ are independent of $R>0$. 

We now estimate $I'$. Since $e^{\Phi}\gamma^{2}e^{\Phi}\leq \gamma_{\Phi}^{2}$, the same arguments of  \eqref{3.26f} and \eqref{kinetic-decay} give  that
\begin{equation}\label{3.49b}
\begin{split}
\|K^{-1/2}e^{\Phi}X_{\eta_{R}e^{\Phi}\gamma,\theta}e^{-\Phi}\|_{\fS_{2}}&\leq C\|K^{1/2}\eta_{R}e^{\Phi}\gamma\|_{\fS_{2}}\\
&\leq  C\Big(\int_{\R^{3}}\eta_{R}^{2}(x)\rho_{K^{1/2}\gamma_{\Phi}K^{1/2}}(x)dx\Big)^{1/2}+\delta(R),
\end{split}\end{equation}
where $K:=\sqrt{1-\Delta}$.
Note from \eqref{3.49a} that  there exist sufficiently large constants $R_1>0$  and $C(R_1)>0$ such that $
 C(R_1):=\text{Tr}\big(K\eta_{R_1}\gamma_{\Phi}\eta_{R_1}\big)<\infty$.
 We thus calculate that
\begin{align*}
\|\rho_{K^{1/2}\gamma_{\Phi} K^{1/2}}\|_{L^1}&=\text{Tr}\big(K^{1/2} \gamma_{\Phi}K^{1/2}\big)\\
&=\text{Tr}\big(K^{1/2}(1-\eta_{R_1}+\eta_{R_1}) \gamma_{\Phi}(1-\eta_{R_1}+\eta_{R_1})K^{1/2}\big)\\
&=\text{Tr}\big(K^{1/2}(1-\eta_{R_1}) \gamma_{\Phi}(1-\eta_{R_1})K^{1/2}\big)+\text{Tr}\big(K^{1/2}\eta_{R_1} \gamma_{\Phi}\eta_{R_1}K^{1/2}\big)\\
&\ \ \ \ +\text{Tr}\big(K^{1/2}(1-\eta_{R_1}) \gamma_{\Phi}\eta_{R_1}K^{1/2}\big)+\text{Tr}\big(K^{1/2}\eta_{R_1} \gamma_{\Phi}(1-\eta_{R_1})K^{1/2}\big)\\
&\leq \text{Tr}\big(K(1-\eta_{R_1}) \gamma_{\Phi}(1-\eta_{R_1})\big)+C(R_1)\\
&\ \ \ \ +2\big[\text{Tr}\big(K(1-\eta_{R_1}) \gamma_{\Phi}(1-\eta_{R_1})\big)\big]^{\frac{1}{2}}
\big[ \text{Tr}\big(K\eta_{R_1} \gamma_{\Phi}\eta_{R_1}\big)\big]^{\frac{1}{2}}\\
&:=\tilde{C}(R_1)<\infty,
\end{align*}
where we have used the fact that $(1-\eta_{R_1})\in C^\infty_0(\R^3)$ has bounded derivatives and compact support.
This then yields  from \eqref{3.49b} that
\begin{align*}
\|K^{-1/2}e^{\Phi}X_{\eta_{R}e^{\Phi}\gamma,\theta}e^{-\Phi}\|_{\fS_{2}}
&\to0\ \   \text{as}\ \ R\to\infty.
\end{align*}
Thus, similar to \eqref{decay5} and \eqref{esti-Xgty}, we derive  that
\begin{equation*}\label{3.52}
\begin{split}
&\ \     \big|\langle\alpha_{R,2\Phi},\, \eta_{R}(x)\eta_{R}(y)e^{2\Phi(x)}(X_{\gamma,\theta})_{x}\alpha\big|\nonumber\\
&=\big|\langle\alpha_{R,2\Phi},\, \eta_{R}(x)\eta_{R}(y)e^{\Phi(x)}(X_{\gamma e^{\Phi}\gamma, \theta})_{x}e^{-2\Phi}\alpha_{2\Phi}\big|\nonumber\\
&\leq \|K^{1/2}\alpha_{R,2\Phi}\|_{\fS_{2}}\|\alpha_{2\Phi}\eta_{R}\|_{\fS_{2}}\|K^{-1/2}e^{\Phi}X_{\eta_{R}\gamma_{\Phi},\theta}e^{-\Phi}\|\|e^{-\Phi}\|\\
&\leq \delta(R)\|K^{1/2}\alpha_{R,2\Phi}\|_{\fS_{2}}\|\alpha_{2\Phi}\eta_{R}\|_{\fS_{2}}\nonumber\\
&\leq \delta(R)\Big(\Tr\big((L_{\gamma,\theta}-\mu_{\theta})\eta_{R}\gamma_{2\Phi}\eta_{R}\big)+e^{8\tau R}\lambda\Big),\nonumber
\end{split}\end{equation*}
and
\begin{align*}
\big|\langle\alpha_{R,2\Phi},\eta_{R}(x)&\eta_{R}(y)e^{2\Phi(x)}(X_{\gamma,\theta})_{y}\alpha\rangle\big|\leq \delta(R)\Tr\big((L_{\gamma,\theta}-\mu_{\theta})\eta_{R}\gamma_{2\Phi}\eta_{R}\big)+e^{8\tau R}\lambda,
\end{align*}
which show that
\begin{align}\label{decay19.1}
|I'|&\leq \delta(R)\Tr\big((L_{\gamma,\theta}-\mu_{\theta})\eta_{R}\gamma_{2\Phi}\eta_{R}\big)+2e^{8\tau R}\lambda.
\end{align}

Substituting \eqref{3.53a} and \eqref{decay19.1} into \eqref{decay19}, we obtain that there exists a constant $C_2(\ \geq C_1>0)$, which is independent of $R>0$, such that for sufficiently large $R>0$,
\begin{equation*}\label{decay20}
\begin{split}
\Ex_\theta(\eta_{R/4}\alpha_{2\Phi}\eta_{R/4})
\leq&\delta(R)\Tr\big((L_{\gamma,\theta}-\mu_{\theta})\eta_{R/4}\gamma_{2\Phi}\eta_{R/4}\big)\\
&+C_{2}\tau^{2}m^{-1}\Tr(\eta_{R/4}\gamma_{2\Phi}\eta_{R/4})+C_{2}e^{8\tau R}\lambda.\\
\end{split}
\end{equation*}
Together with  \eqref{decay18}, the same arguments of \eqref{decay14} and \eqref{3.45} thus give that  for any fixed sufficiently large $R>0$,
\begin{align}\label{3.56}
 \Tr\Big(\big(L_{\gamma,\theta}-\mu_{\theta}-2C_2(1+\kappa)\tau^{2}m^{-1}-\delta(R)\big)\eta_{R}\gamma_{2\Phi}\eta_{R}\Big)\leq N_{2}(R,\tau,\lambda)<\infty,
\end{align}
and
\begin{equation}\label{decay21}
\begin{split}
\big[|\mu_{\theta}|-4C_{2}(1+\kappa)\tau^{2}m^{-1}-\delta(R)\big]\|\eta_{R}\sqrt{\rho_{\gamma_{2\Phi}}}\|_{L^{2}}^{2}\leq 2N_{2}(R,\tau,\lambda)<\infty,
\end{split}
\end{equation}
where $N_{2}(R,\tau,\lambda)$ is similar to $N_{1}(R,\tau,\lambda)$ in \eqref{decay14}. Choose sufficiently large $R>0$ and sufficiently small $0<c_{2}\leq c_{1}$ such that  $\delta(R)\leq |\mu|/4$ and  $4C_{2}(\kappa+1)\tau_{c_{2}}^{2}m^{-1}\leq |\mu|/4$, where the constant  $\mu<0$ is as in \eqref{3.44}. We then deduce from \eqref{decay21} that for any fixed sufficiently large $R>0$,
\begin{align}\label{decay22}
&|\mu|\|\eta_{R}\sqrt{\rho_{\gamma_{2\Phi}}}\|_{L^{2}}^{2}\leq \frac{4}{|\mu|}N_{2}(R,\tau_{c_{2}},\lambda)<\infty.
\end{align}
Together with \eqref{3.56}, we obtain \eqref{st2}. This proves Step 2.

$Step\ 3$. In this step, we prove that there exists $0<\theta_{c}:=\theta_{c}(m, \kappa, \lambda)\le \min\{1, m/6, \widetilde{\theta}\}$ such that if $\theta\in(0, \theta_c]$, then for any fixed sufficiently large $R>0$,
\begin{align}\label{3.54}
\eta_{R}\big(e^{3\theta|\cdot|}\gamma e^{3\theta|\cdot|},\ e^{3\theta|\cdot|}\alpha\big)\eta_{R}\in\cX.
\end{align}
We use the above bootstrap argument again to show that the estimates of \eqref{3.56} and \eqref{decay22} also hold true, when $2\Phi$ is replaced by $3\Phi$. We omit the detailed proof, since it is very similar to that of Step 2, except that the function $e^{8\tau R}$ is now replaced by $e^{12\tau R}$.

In summary, we obtain that if $(\gamma,\alpha)\neq(\gamma,0)\in\cK_{\lambda}$ is a minimizer of $I^{\rm HFB}_{m,\kappa,\theta}(\lambda)$ satisfying $0<\theta\leq\min\{1,m/6,\widetilde{\theta}\}$, then for any fixed sufficiently large $R>0$,
\begin{align}\label{exp-decay-min}
(1-\Delta)^{1/4}\eta_{R}e^{3\tau_{c_{3}}|\cdot|}\gamma e^{3\tau_{c_{3}}|\cdot|}\eta_{R}(1-\Delta)^{1/4}\in\fS_{1},
\end{align}
where
\begin{align*}
\tau_{c_3}:=\min\big\{1, m/6, \theta,c_3\sqrt{(\kappa+1)^{-1}m|\mu|}\big\}=\min\big\{\theta,c_3\sqrt{(\kappa+1)^{-1}m|\mu|}\big\},
\end{align*}
and the constant $\mu<0$ is given in \eqref{3.44}.
Select the critical Yukawa parameter
\begin{align}\label{3.75}
\theta_{c}=\theta_{c}(m, \kappa, \lambda):=\min\big\{1,m/6,\widetilde{\theta},c_{3}\sqrt{(\kappa+1)^{-1}m|\mu|}\big\}>0.
\end{align}
It thus follows that if  $\theta\in (0,\theta_{c}]$,  then $\theta= \tau_{c_3}$, and thus for any fixed sufficiently large $R>0$,
\begin{align}\label{decay23}
(1-\Delta)^{1/4}\eta_{R}e^{3\theta|\cdot|}\gamma e^{3\theta|\cdot|}\eta_{R}(1-\Delta)^{1/4}\in\fS_{1}.
\end{align}
By the operator inequality $\alpha\alpha^{*}\leq \gamma$, we hence obtain from \eqref{decay23} that  \eqref{3.54}  holds true.

$Step\ 4$. Note from Step 3 that  there exists a sufficiently large constant $\tilde{R}_0>1$ such that
\begin{align}\label{3.59a}
\text{Tr}\big(\sqrt{1-\Delta}\, \gamma_{\theta, \tilde{R}_0}\big)
<\infty,\ \ \ \text{where}\,\ \gamma_{\theta, \tilde{R}_0}:=\eta_{\tilde{R}_0}e^{3\theta|\cdot|}\gamma e^{3\theta|\cdot|}\eta_{\tilde{R}_0}.
\end{align}
This yields that there exist a sequence $\{n_j\}\subset\R$ and an orthonormal basis $\{e_j\}^\infty_{j=1}$ of $L^2(\R^3, \C^q)$ such that
\begin{align}\label{3.60a}
\gamma_{\theta, \tilde{R}_0}=\sum_{j=1}^{\infty}n_j|e_j\rangle \langle e_j|,\ \   \text{where}\,\ 0\leq n_j\leq1\ \text{and}\ \sum_{j=1}^{\infty}n_j=\text{Tr} (\gamma_{\theta, \tilde{R}_0})<\infty.
\end{align}
Since $\eta_R\eta_{\tilde{R}_0}=\eta_R$ for $R\geq2\tilde{R}_0$, we deduce from \eqref{3.22}, \eqref{3.59a} and \eqref{3.60a}  that for any  $R\geq2\tilde{R}_0$,
\begin{align*}
&\ \ \ \ \text{Tr}\big(\sqrt{1-\Delta}\, \eta_{R}e^{3\theta|\cdot|}\gamma e^{3\theta|\cdot|}\eta_{R}\big)\\
&=\text{Tr}\big(\sqrt{1-\Delta}\, \eta_{R}\gamma_{\theta, \tilde{R}_0}\eta_{R}\big)=\text{Tr}\big(\eta_{R}\sqrt{1-\Delta}\, \eta_{R}\gamma_{\theta, \tilde{R}_0}\big)\\
&=\sum_{j=1}^{\infty}n_j\big\langle e_j,\  \eta_{R}\sqrt{1-\Delta}\, \eta_{R}e_j\big\rangle =\sum_{j=1}^{\infty}n_j\big\|(1-\Delta)^{\frac{1}{4}}\eta_{R}e_j\big\|_{L^2}^2\\
&\leq 2\sum_{j=1}^{\infty}n_j\big\|\eta_{R}(1-\Delta)^{\frac{1}{4}}e_j\big\|_{L^2}^2
+2\sum_{j=1}^{\infty}n_j\Big\|\big((1-\Delta)^{\frac{1}{4}}\eta_{R}-\eta_{R}(1-\Delta)^{\frac{1}{4}}\big)e_j\Big\|_{L^2}^2\\
&\leq 2\sum_{j=1}^{\infty}n_j\big\|(1-\Delta)^{\frac{1}{4}}e_j\big\|_{L^2}^2
+2\big\|(1-\Delta)^{\frac{1}{4}}\eta_{R}-\eta_{R}(1-\Delta)^{\frac{1}{4}}\big\|^2\sum_{j=1}^{\infty}n_j\|e_j\|_{L^2}^2\\
&\leq 2\text{Tr}\big(\sqrt{1-\Delta}\, \gamma_{\theta, \tilde{R}_0}\big)+2C_1R^{-2}\sum_{j=1}^{\infty}n_j \le C_2<\infty,
\end{align*}
with the constants $C_2>C_1>0$ independent of $R\geq2\tilde{R}_0>0$. Consequently, 
$$
\sup\limits_{R\geq R_0:=2\tilde{R}_0}\big\|(1-\Delta)^{1/4} \eta_{R}e^{3\theta|\cdot|}\gamma e^{3\theta|\cdot|}\eta_{R}(1-\Delta)^{1/4}\big\|_{\fS_{1}}<\infty.
$$
Together with the fact that $\alpha\alpha^{*}\leq \gamma$, we thus conclude  that \eqref{3.36a} holds true.
This therefore completes the proof of Proposition \ref{pro3.6}.
\qed

\medskip
We now complete the proof of Theorem \ref{thm:main2}.

\medskip

\noindent \textbf{Proof of Theorem \ref{thm:main2}.}
By Proposition \ref{pro3.6}, we immediately obtain  that Theorem \ref{thm:main2} holds true for the case where $\alpha\neq0$.

For the case where $\alpha=0$, let $(\gamma, 0)$ be a minimizer of $I^{\rm HFB}_{m,\kappa,\theta}(\lambda)$, where $m>0, 0<\kappa<4/\pi$ and $0<\theta\leq \widetilde{\theta}$. By the proof of Proposition \ref{prop:chemical-pot}, we get that $M:=\text{Rank}(\gamma)<\infty$, and the range of $\gamma$ is spanned by the eigenfunctions  of the lowest $M$ eigenvalues  $\nu_{1}\leq \nu_{2}\leq\cdots\leq\nu_{M}<0$ for the operator  $H_{\gamma,\theta}$ on  $L^2(\R^3, \C^q)$.
We claim that
\begin{align}\label{3.59}
\nu_{j}<\beta_{\theta}-m,\ \  j=1, \cdots, M,
\end{align}
where $\beta_\theta$ is as in Proposition \ref{prop:min-Gtheta}.
In fact, the same argument of \eqref{HF-en-min} gives from  Lemma \ref{lem:prop-Imktheta-lamb} that there exists $\delta>0$ such that
\begin{align*}
I^{\rm HFB}_{m,\kappa,\theta}(\lambda-\delta)+\nu_{j}\delta
&\leq \mathcal{E}^{\rm HFB}_{m,\kappa,\theta}(\gamma,0)= I^{\rm HFB}_{m,\kappa,\theta}(\lambda)\\
&\leq I^{\rm HFB}_{m,\kappa,\theta}(\lambda-\delta)+I^{\rm HFB}_{m,\kappa,\theta}(\delta).
\end{align*}
Together with Proposition \ref{prop:min-Gtheta}, we then have
\begin{align*}
\nu_{j}\delta\leq I^{\rm HFB}_{m,\kappa,\theta}(\delta)<G_{\theta}(\delta)-m\delta=(\beta_{\theta}-m)\delta,
\end{align*}
which proves the claim (\ref{3.59}).

Employing a similar argument of Proposition \ref{pro3.6}, we further deduce from \eqref{3.59} that Theorem \ref{thm:main2} holds true for the case where $\alpha=0$.  This therefore finishes the proof of Theorem \ref{thm:main2}. \qed

\section{Existence of HFB minimizers}\label{sec:existence-nonexistence-HFB}

In this section, we prove Theorem \ref{thm:main1} on the existence of minimizers for the HFB problem $I^{\rm HFB}_{m,\kappa,\theta}(\lambda)$ defined in \eqref{eq:HFB-variational-problem},
where $m>0,\ 0<\kappa<4/\pi$, $0<\lambda<\lambda^{\rm HFB}$ and $0\leq\theta\leq\theta_c$. Here   $\lambda^{\rm HFB}:=\lambda^{\rm HFB}(\kappa)$ given by Proposition \ref{prop:Chandrasekhar} denotes the critical total mass of the HFB problem,  and $0<\theta_{c}=\theta_{c}(m,\kappa,\lambda)\leq\min\{1, m/6,\tilde{\theta}\}$ is as in Theorem \ref{thm:main2} and  Proposition \ref{pro3.6}.

Since the case $\theta=0$ of Theorem \ref{thm:main1}  was established in \cite{LenLew-10}, in the following we only prove Theorem \ref{thm:main1}  for the case $\theta\in(0, \theta_c]$ by using the concentration-compactness method.  Define $SO(3)$  the set of all rotations in $\R^3$ centered  at the origin.  To this end, we first address the following Newton-type theorem for the Yukawa potential.

\begin{lemma}\label{lem:shell-Yukawa}
Suppose $\psi\in L^1(\R^3)$ is a spherically symmetric positive  function, in the sense that $0\leq\psi(x)=\psi(Ux)$ for all $U\in SO(3)$ and $x\in\R^{3}$, and assume that $\theta>0$ and $\supp(\psi)\subset B_{R}$ holds for some $R>0$.  Then we have
\begin{align*}
\int_{\R^{3}}\frac{e^{-\theta|x-y|}}{|x-y|}\psi(y)dy=\frac{e^{-\theta|x|}}{\theta|x|}\int_{\R^{3}}\frac{\psi(y)\sinh(\theta|y|) }{|y|}dy,\ \  \forall\ |x|\geq R.
\end{align*}
\end{lemma}

\noindent \textbf{Proof.}
Let $x\in\R^{3}$ satisfy $|x|\geq R>0$.   Since $\psi$ is spherically symmetric, without loss of generality, we can assume that  $x$ is parallel to the $\hat{z}$-axis by using an appropriate rotation.  Let $(r,\varphi,\phi)$ be the spherical coordinate of $y\in\R^3$, where $r=|y|$, and the angles $ (\varphi,\phi)\in[0, \pi]\times [0,2\pi]$.  Under the assumption of Lemma \ref{lem:shell-Yukawa}, there exists a positive function $\Psi$ defined on $[0,\infty)$ with support within $[0,R]$, such that $\psi(y)=\Psi(r)$ holds for any $y\in\R^{3}$.  Then we define the variable
\begin{align*}
\xi^{2}&=|x-y|^{2}=|x|^{2}+r^{2}-2r|x|\cos\varphi,
\end{align*}
which gives that $\xi d\xi=r|x|\sin\varphi d\varphi$.  Therefore, one can calculate that  for $|x|\geq R$, 
\begin{align*} \int_{\R^{3}}\frac{e^{-\theta|x-y|}}{|x-y|}\psi(y)dy&=2\pi\int_{0}^{\infty}\int_{0}^{\pi}\frac{e^{-\theta\sqrt{|x|^{2}+r^{2}-2r|x|\cos\varphi}}}{\sqrt{|x|^{2}+r^{2}-2r|x|\cos\varphi}}\Psi(r)r^{2}\sin\varphi d\varphi dr\nonumber\\
	&=2\pi\int_{0}^{\infty}\Big(\int_{|r-|x||}^{r+|x|}\frac{e^{-\theta\xi}}{|x|}d\xi\Big)\Psi(r)r dr\nonumber\\
	&=\frac{2\pi}{\theta|x|}\int_{0}^{R}\Big(e^{-\theta(|x|-r)}-e^{-\theta(|x|+r)}\Big)\Psi(r)rdr\nonumber\\
	&=\frac{4\pi e^{-\theta|x|}}{\theta|x|}\int_{0}^{R}\Psi(r)r\sinh(\theta r)dr\nonumber\\
	&=\frac{e^{-\theta|x|}}{\theta|x|}\int_{\R^{3}}\frac{\psi(y)\sinh(\theta|y|)}{|y|}dy,
\end{align*}
which proves Lemma \ref{lem:shell-Yukawa}. \qed

\subsection{Proof of Theorem \ref{thm:main1}}

The purpose of this subsection is to address the proof of Theorem \ref{thm:main1}.  Let $\{(\gamma_{n},\alpha_{n})\}\subset\cK_{\lambda}$ be a minimizing sequence of $I^{\rm HFB}_{m,\kappa,\theta}(\lambda)$, where $m>0,\ 0<\kappa<4/\pi$, $0<\lambda<\lambda^{\rm HFB}$ and $0<\theta\leq\theta_c$, where $\theta_c>0$ is as in Theorem \ref{thm:main2}.  For each $n\geq 1$, we define the concentration function
\begin{align*}
Q_{n}(R)&:=\sup_{y\in\R^{3}}\int_{|x-y|\leq R}\rho_{\gamma_{n}}(x)dx.
\end{align*}
By the concentration-compactness theorem in \cite{Lions1}, it  follows that
\begin{align}\label{4.1:1}
\lambda_1&:=\lim_{R\rightarrow\infty}\limsup_{n\rightarrow\infty}Q_{n}(R)\in [0,\lambda].
\end{align}
We then have (cf. \cite{LenLew-10,Lions1}) the following operator version of the classical dichotomy result.

\begin{lemma}(\cite[Lemma 7.3]{LenLew-10})\label{lem:str-converge}
Up to a subsequence if necessary, there exist  $\gamma^{(1)}\in\fS_{1}$ satisfying $\Tr\gamma^{(1)}=\lambda_1$,  $\{y_{n}\}\subseteq\R^{3}$ and $\{R_{n}\}\subset\R_{+}$ satisfying $R_{n}\rightarrow\infty$ as $n\to\infty$ such that
\begin{align*}
K^{1/2}\tau_{y_{n}}\chi_{R_{n}}\gamma_{n}\chi_{R_{n}}\tau_{y_{n}}^{*}K^{1/2}\rightharpoonup K^{1/2}\gamma^{(1)}K^{1/2}\ \   \text{weakly-* in}\ \fS_{1}\  \text{as}\ n\to\infty,
\end{align*}
and
\begin{align*}
\tau_{y_{n}}\chi_{R_{n}}\gamma_{n}\chi_{R_{n}}\tau_{y_{n}}^{*}\rightarrow\gamma^{(1)}\ \   \text{strongly in}\ \fS_{1}\  \text{as}\ n\to\infty,
\end{align*}
where $K:=\sqrt{-\Delta+m^{2}}$, $(\tau_yf)(x):=f(x-y)$, $\chi_{R}(x):=\chi(x/R)$ for  $R>0$, and $\chi\in C_{0}^{\infty}(\R^{3};[0,1])$  is a smooth cutoff function with support on $B_1(0)$.
\end{lemma}

%
In the following, we shall rule out the vanishing case $\lambda_1=0$ and the dichotomy case $\lambda_1\in (0,\lambda)$, where the constant $\lam _1$ is as in (\ref{4.1:1}). This further implies  from Lemma \ref{lem:str-converge} and Corollary \ref{cor:conservation-mass} that  $I^{\rm HFB}_{m,\kappa,\theta}(\lambda)$ has a minimizer $(\gamma,\alpha)$, which thereby establishes Theorem \ref{thm:main1} in view of \cite[Section 7]{LenLew-10}. We first rule out the vanishing case by applying Proposition \ref{prop:min-Gtheta}.

\begin{prop}[No vanishing case]\label{prop:no-vanishing}
Suppose $m>0$, $0<\kappa<4/\pi$, $0<\lambda<\lambda^{\rm HFB}$ and $\theta\in (0,\theta_c]$, where $\theta_c>0$ is as in Theorem \ref{thm:main2}. Then the constant $\lam _1$ of (\ref{4.1:1}) satisfies $\lambda_1>0$.
\end{prop}

\noindent \textbf{Proof.}
Recall from Lemma  \ref{lem:coercive} that  the minimizing sequence $\{(\gamma_{n},\alpha_{n})\}\subset\cK_{\lambda}$  of $I^{\rm HFB}_{m,\kappa,\theta}(\lambda)$ is bounded uniformly  in $\cX$. This implies from  \eqref{kin-density} that $\{\sqrt{\rho_{\gamma_{n}}}\}$ is  bounded uniformly   in $H^{1/2}(\R^{3})$.  Moreover, note that
\begin{equation}\label{E-G-vanishing}
\begin{split}
\cE^{\rm HFB}_{m,\kappa,\theta}(\gamma_{n},\alpha_{n})
&=\cG_{\theta}(\gamma_{n},\alpha_{n})-m\lambda-\frac{\kappa}{2}D_{\theta}(\rho_{\gamma_{n}},\rho_{\gamma_{n}})+\frac{\kappa}{2}\Ex_\theta(\gamma_{n})\\
&\geq G_{\theta}(\lambda)-m\lambda-\frac{\kappa}{2}D_{\theta}(\rho_{\gamma_{n}},\rho_{\gamma_{n}})+\frac{\kappa}{2}\Ex_\theta(\gamma_{n}),
\end{split}
\end{equation}
where $\cG_{\theta}$ and $G_{\theta}$ are given by \eqref{min-Gtheta} and \eqref{G-func}, respectively.

On the contrary, suppose  $\{(\gamma_{n},\alpha_{n})\}$ is vanishing, $i.e.,$ $\lambda_1=0$.  Applying \cite[Lemma 7.2]{LenLew-10}, we then derive from \eqref{inter-bdd1} that
\begin{align*}
0&\leq \Ex_\theta(\gamma_{n})\leq D_{\theta}(\rho_{\gamma_{n}},\rho_{\gamma_{n}})\leq D(\rho_{\gamma_{n}},\rho_{\gamma_{n}})\rightarrow 0\ \   \text{as}\ \ n\to\infty,
\end{align*}
We hence obtain from \eqref{E-G-vanishing} that   $I^{\rm HFB}_{m,\kappa,\theta}(\lambda)\geq G_{\theta}(\lambda)-m\lambda$. Since $\theta\leq\widetilde{\theta}$, we deduce from  Proposition \ref{prop:min-Gtheta} that $I^{\rm HFB}_{m,\kappa,\theta}(\lambda)<G_{\theta}(\lambda)-m\lambda$,   a contradiction.  Therefore, this proves that $\lambda_{1}>0$.
\qed

In order to finish the proof of Theorem \ref{thm:main1}, it next remains to rule out the dichotomy case. Since the sequence $\{(\gamma_{n},\alpha_{n})\}$ is not   vanishing, we have $\lambda_{1}>0$.
Define
\begin{align}\label{3.8d} (\gamma_n^{(1)},\alpha_{n}^{(1)}):=\chi_{R_{n}}(\gamma_{n},\alpha_{n})\chi_{R_{n}},\ \   (\gamma_n^{(2)},\alpha_{n}^{(2)}):=\eta_{R_{n}}(\gamma_{n},\alpha_{n})\eta_{R_{n}},
\end{align}
where $\eta(x)=\sqrt{1-\chi^2(x)}$,   $\chi_{R}(x):=\chi(x/R)$  and $\{y_n\}\subset\R^3$ are as in Lemma \ref{lem:str-converge}.
Applying the same argument of \cite[Section 7, p.31--32]{LenLew-10}, one can obtain the following proposition, whose proof is omitted for simplicity.

\begin{prop}[No dichotomy case]\label{prop:dichotomy}
Suppose $m>0$, $0<\kappa<4/\pi$, $0<\lambda<\lambda^{\rm HFB}$ and $\theta\in (0, \theta_c]$, where $\theta_c>0$ is as in Theorem \ref{thm:main2}. Then we have
\begin{align*}
I^{\rm HFB}_{m,\kappa,\theta}(\lambda)&=I^{\rm HFB}_{m,\kappa,\theta}(\lambda_{1})+I^{\rm HFB}_{m,\kappa,\theta}(\lambda-\lambda_{1}),
\end{align*}
and $I^{\rm HFB}_{m,\kappa,\theta}(\lambda_{1})$ has a minimizer $(\gamma^{(1)},\alpha^{(1)})\in\cK_{\lambda_{1}}$.
\end{prop}

Note that the minimizer $(\gamma^{(1)},\alpha^{(1)})$ of $I^{\rm HFB}_{m,\kappa,\theta}(\lambda_{1})$ is the strong limit of $(\gamma_{n}^{(1)},\alpha_{n}^{(1)})$ and the weak limit of $(\gamma_{n},\alpha_{n})$ in $\cX$.  Moreover, the sequence $\big\{(\gamma_{n}^{(2)},\alpha_{n}^{(2)})\big\}$ is a minimizing sequence of $I^{\rm HFB}_{m,\kappa,\theta}(\lambda-\lambda_{1})$.  Thus, one of the following two cases must occur.
\begin{enumerate}
\item[(a)] Either $(\gamma_{n}^{(2)},\alpha_{n}^{(2)})$ is relatively compact  in $\cX$,  up to  translations: in this case, we further conclude  that up to a subsequence if necessary,  $(\gamma_{n}^{(2)},\alpha_{n}^{(2)})$ converges to a minimizer $(\gamma^{(2)},\alpha^{(2)})$ of $I^{\rm HFB}_{m,\kappa,\theta}(\lambda-\lambda_{1})$.

\item[(b)] Or $(\gamma_{n}^{(2)},\alpha_{n}^{(2)})$ is not relatively compact  in $\cX$, up to translations: in this case, it follows from Proposition \ref{prop:no-vanishing} that the vanishing case of $(\gamma_{n}^{(2)},\alpha_{n}^{(2)})$ cannot occur. We then repeat the above process to obtain that
\begin{align*}
I^{\rm HFB}_{m,\kappa,\theta}(\lambda)&=I^{\rm HFB}_{m,\kappa,\theta}(\lambda_{1})+I^{\rm HFB}_{m,\kappa,\theta}(\lambda_{2})+I^{\rm HFB}_{m,\kappa,\theta}(\lambda-\lambda_{1}-\lambda_{2})
\end{align*}
holds for some $0<\lambda_{2}<\lambda-\lambda_{1}$,  and $I^{\rm HFB}_{m,\kappa,\theta}(\lambda_{2})$ admits a minimizer $(\gamma^{(2)},\alpha^{(2)})\in\cK_{\lambda_{2}}$.
\end{enumerate}

\noindent  For each of above two cases, we conclude that
\begin{align}\label{section4:4.5}
I^{\rm HFB}_{m,\kappa,\theta}(\lambda)&=I^{\rm HFB}_{m,\kappa,\theta}(\lambda_{1})+I^{\rm HFB}_{m,\kappa,\theta}(\lambda_{2})+I^{\rm HFB}_{m,\kappa,\theta}(\lambda-\lambda_{1}-\lambda_{2})
\end{align}
holds for some $\lambda_{1},\lambda_{2}>0$ satisfying $\lambda_{1}+\lambda_{2}\leq \lambda$, and $I^{\rm HFB}_{m,\kappa,\theta}(\lambda_{i})$ possesses a minimizer $(\gamma^{(i)},\alpha^{(i)})$, $i=1,2$.
We finally derive the following binding inequality, which then yields Theorem \ref{thm:main1}.

\begin{prop}\label{prop:binding-esti}
Suppose $m>0$, $0<\kappa<4/\pi$ and $\theta\in (0,\theta_{c}]$, where $\theta_c>0$ is as in Theorem \ref{thm:main2}. Then we have
\begin{align}
I^{\rm HFB}_{m,\kappa,\theta}(\lambda_{1}+\lambda_{2})&<I^{\rm HFB}_{m,\kappa,\theta}(\lambda_{1})+I^{\rm HFB}_{m,\kappa,\theta}(\lambda_{2}),
\end{align}
where $\lambda_{1}>0$ and $\lambda_{2}>0$ are as in (\ref{section4:4.5}).
\end{prop}

\noindent \textbf{Proof.}
Set $\eta(x):=\sqrt{1-\chi^{2}(x)}$, where $\chi\in C_0^\infty(\R^3;[0,1])$ is a smooth cutoff function such that $\chi(x)=0$ for $|x|> 1$ and $\chi(x)=1$ for $|x|\leq 3/4$. Suppose $(\gamma^{(1)},\alpha^{(1)})$ and $(\gamma^{(2)},\alpha^{(2)})$ are a minimizer of $I^{\rm HFB}_{m,\kappa,\theta}(\lambda_{1})$ and $I^{\rm HFB}_{m,\kappa,\theta}(\lambda_{2})$, respectively.
We first claim that there exists a constant $C>0$ such that for sufficiently large $R>0$, 
\begin{align}\label{trial-error}
\cE^{\rm HFB}_{m,\kappa,\theta}(\chi_{R}\gamma^{(i)}\chi_{R},\chi_{R}\alpha^{(i)}\chi_{R})&\leq I^{\rm HFB}_{m,\kappa,\theta}(\lambda_{i})+Ce^{-4\theta R}, \ \  \ i=1,2,
\end{align}
where $\chi_{R}(x):=\chi(x/R)$ and $\eta_{R}(x):=\eta(x/R)$.

To prove the claim (\ref{trial-error}), we note from \cite[Lemma A.1]{LenLew-10} that
\begin{align}\label{4.11}
\Tr(T\gamma^{(1)})&\geq \Tr(T\chi_{R}\gamma^{(1)}\chi_{R})-\Tr(\cS_{R}\gamma^{(1)}),
\end{align}
where  $T:=K-m:=\sqrt{-\Delta+m^{2}}-m$, and $\cS_{R}$ is the following bounded positive operator
\begin{align*}
\cS_{R}&:=\frac{1}{\pi}\int_{0}^{\infty}\frac{1}{s+K^{2}}\big(|\nabla\chi_{R}|^{2}+|\nabla\eta_{R}|^{2}\big)\frac{1}{s+K^{2}}\sqrt{s}ds.
\end{align*}
Since $0<\theta\leq m/6$, one can calculate  from Lemma \ref{lem:resolv-s-bdd} that
\begin{align}\label{4.14a}
\big\|e^{-3\theta|\cdot|}\cS_{R}e^{-3\theta|\cdot|}\big\|
&\leq\frac{\|e^{-3\theta|\cdot|}\nabla\chi_{R}\|_{L^{\infty}}^{2}+\|e^{-3\theta|\cdot|}\nabla\eta_{R}\|_{L^{\infty}}^{2}}{\pi}\nonumber\\
&\ \        \cdot\int_{0}^{\infty}\|e^{3\theta|\cdot|}(s+K^{2})^{-1}e^{-3\theta|\cdot|}\|^{2}\sqrt{s}ds\\
&\leq \frac{4}{\pi}\big(\|e^{-3\theta|\cdot|}\nabla\chi_{R}\|_{L^{\infty}}^{2}+\|e^{-3\theta|\cdot|}\nabla\eta_{R}\|_{L^{\infty}}^{2}\big)\int_{0}^{\infty}\frac{\sqrt{s}}{(s+m^{2})^{2}}ds\nonumber\\
&\leq C_{1}\Big(\|e^{-3\theta|\cdot|}\nabla\chi_{R}\|_{L^{\infty}}^{2}+\|e^{-3\theta|\cdot|}\nabla\eta_{R}\|_{L^{\infty}}^{2}\Big),\nonumber
\end{align}
where $C_{1}>0$ is independent of $R>0$.  Note that $\nabla\chi_{R}$ and $\nabla\eta_{R}$ are supported on $\{3R/4\leq |x|\leq R\}$.  This implies that there exists a constant $C_{2}>0$ such that
\begin{align*}
\|e^{-3\theta|\cdot|}\nabla\chi_{R}\|_{L^{\infty}}^{2}+\|e^{-3\theta|\cdot|}\nabla\eta_{R}\|_{L^{\infty}}^{2}
&\leq e^{-9\theta R/2}\Big(\|\nabla\chi_{R}\|_{L^{\infty}}^{2}+\|\nabla\eta_{R}\|_{L^{\infty}}^{2}\Big)\nonumber\\
&\leq C_{2}R^{-2}e^{-9\theta R/2}.
\end{align*}
Together with  \eqref{4.14a}, we have
\begin{equation}\label{4.12h}
\begin{split}
\Tr(\cS_{R}\gamma^{(1)})&\leq \|e^{-3\theta|\cdot|}\cS_{R}e^{-3\theta|\cdot|}\|\|e^{3\theta|\cdot|}\gamma^{(1)}e^{3\theta|\cdot|}\|_{\fS_{1}}\\
&\leq C_1C_{2}\|e^{3\theta|\cdot|}\gamma^{(1)}e^{3\theta|\cdot|}\|_{\fS_{1}}R^{-2}e^{-9\theta R/2}.
\end{split}\end{equation}
As a consequence of \eqref{4.11} and \eqref{4.12h}, we obtain from Theorem \ref{thm:main2} that  for sufficiently large $R>0$,
\begin{align}\label{4.14f}
\Tr(T\chi_{R}\gamma^{(1)}\chi_{R})\leq \Tr(T\gamma^{(1)})+C_3R^{-2}e^{-4\theta R}.
\end{align}

Moreover,   utilizing  the Hardy-Kato inequality \eqref{hkt}, we deduce from \eqref{HFB-min-decay2} that  for sufficiently large $R>0$,
\begin{align}\label{4.9h}
&\Big|D_{\theta}(\rho_{\gamma^{(1)}},\rho_{\gamma^{(1)}})-D_{\theta}(\rho_{\chi_{R}\gamma^{(1)}\chi_{R}},\rho_{\chi_{R}\gamma^{(1)}\chi_{R}})\Big|\nonumber\\
\leq& \iinte\frac{e^{-\theta|x-y|}(1-\chi_{R}^{2}(x)\chi_{R}^{2}(y))}{|x-y|}\rho_{\gamma^{(1)}}(x)\rho_{\gamma^{(1)}}(y)dxdy\\
\leq& 2\iinte\frac{\eta_{R}^{2}(x)\rho_{\gamma^{(1)}}(x)\rho_{\gamma^{(1)}}(y)}{|x-y|}dxdy\leq C_{4}\int_{\R^{3}}\eta_{R}^{2}(x)\rho_{\gamma^{(1)}}(x)dx\nonumber\\
\leq & C_{4}\int_{|x|\geq3R/4}\rho_{\gamma^{(1)}}(x)dx\leq C_5e^{-4\theta R}.
\nonumber
\end{align}
The exchange energy can be  estimated by the same method.
As for the pairing term,  by  the relation $\alpha\alpha^{*}\leq \gamma$, one can calculate from \eqref{hkt} and  \eqref{HFB-min-decay}  that   for sufficiently large $R>0$,
\begin{align}\label{4.10}
&\iinte\frac{e^{-\theta|x-y|}\big(1-\chi_{R}^{2}(x)\chi_{R}^{2}(y)\big)}{|x-y|}|\alpha^{(1)}(x,y)|^{2}dxdy\nonumber\\
\leq& 2\iinte\frac{\eta_{R}^{2}(x)|\alpha^{(1)}(x,y)|^{2}}{|x-y|}dxdy\nonumber\\
\leq &2e^{-9\theta R/2}\iinte\frac{\eta_{R}^{2}(x)|\alpha_{\theta}^{(1)}(x,y)|^{2}}{|x-y|}dxdy\\
\leq &\pi e^{-9\theta R/2}\Tr(K\eta_{R}\alpha_{\theta}^{(1)}(\alpha_{\theta}^{(1)})^{*}\eta_{R})\leq C_{6}e^{-9\theta R/2}\Tr(K\eta_{R}\gamma_{\theta}^{(1)}\eta_{R})\nonumber\\
\leq& C_7e^{-4\theta R},
\nonumber
\end{align}
where $\alpha_{\theta}^{(1)}:=e^{3\theta|\cdot|}\alpha$ and $\gamma_{\theta}^{(1)}:=e^{3\theta|\cdot|}\gamma e^{3\theta|\cdot|}$.  Therefore,  we conclude from  \eqref{4.14f}--\eqref{4.10}  that the claim \eqref{trial-error}  holds true.

We now define the trial state
\begin{align*}
(\gamma_{R},\alpha_{R})&:=U\chi_{R}(\gamma^{(1)},\alpha^{(1)})\chi_{R}U^{*}+\tau_{3R\vec{v}}V\chi_{R}(\gamma^{(2)},\alpha^{(2)})\chi_{R}V^{*}\tau_{3R\vec{v}}^{*},
\end{align*}
where $U,V\in SO(3)$ are some rotations, and $\vec{v}\in\R^{3}$ is a  unit vector.  This trial state is clearly in $\cK$, since $\chi_{R}$ and $\tau_{3R\vec{v}}\chi_{R}\tau_{3R\vec{v}}^{*}$ have disjoint supports. Because $\cE^{\rm HFB}_{m,\kappa,\theta}$ admits the  invariance of  translations and rotations, and
\begin{align*}
\Tr(\gamma_{R})&=\Tr(\chi_{R}\gamma^{(1)}\chi_{R})+\Tr(\chi_{R}\gamma^{(2)}\chi_{R})\leq \lambda_{1}+\lambda_{2},
\end{align*}
we deduce from Lemma \ref{lem:prop-Imktheta-lamb} (ii) that
\begin{align*}
I^{\rm HFB}_{m,\kappa,\theta}(\lambda_{1}+\lambda_{2})
&\leq \int_{SO(3)}\int_{SO(3)}\cE^{\rm HFB}_{m,\kappa,\theta}(\gamma_{R})dUdV\nonumber\\
&\leq \cE^{\rm HFB}_{m,\kappa,\theta}(\chi_{R}\gamma^{(1)}\chi_{R})+\cE^{\rm HFB}_{m,\kappa,\theta}(\chi_{R}\gamma^{(2)}\chi_{R})\nonumber\\
&\ \    -\kappa\iinte\frac{e^{-\theta|x-y|}f(x)g(y-3R\vec{v})}{|x-y|}dxdy,
\end{align*}
where
\begin{align*}
f(x)&=\int_{SO(3)}\chi_{R}^{2}(x)\rho_{\gamma^{(1)}}(Ux)dU>0\ \ \ \text{and}\ \ \ g(y)=\int_{SO(3)}\chi_{R}^{2}(x)\rho_{\gamma^{(2)}}(Vy)dV >0
\end{align*}
are both spherically symmetric   functions with supports contained in $B_{R}(0)$.
This yields from  \eqref{trial-error} that for sufficiently large $R>0$,
\begin{equation}\label{subadditivity}
\begin{split}	
I^{\rm HFB}_{m,\kappa,\theta}(\lambda_{1}+\lambda_{2})
&\leq I^{\rm HFB}_{m,\kappa,\theta}(\lambda_{1})+I^{\rm HFB}_{m,\kappa,\theta}(\lambda_{2})+Ce^{-4\theta R}\\
&\ \   -\kappa\iinte\frac{e^{-\theta|x-y|}f(x)g(y-3R\vec{v})}{|x-y|}dxdy.
\end{split}\end{equation}
By Lemma \ref{lem:shell-Yukawa}, we have
\begin{align}\label{cross-term}
&\iinte\frac{e^{-\theta|x-y|}f(x)g(x-3R\vec{v})}{|x-y|}dxdy\nonumber\\
=&\int_{\R^{3}}\Big(\int_{\R^{3}}\frac{e^{-\theta|x-y+3R\vec{v}|}}{|x-y+3R\vec{v}|}g(y)dy\Big)f(x)dx\\
&=\lambda_{\rm eff}^{2}\int_{\R^{3}}\frac{e^{-\theta|x+3R\vec{v}|}}{\theta|x+3R\vec{v}|}f(x)dx\geq\tilde{C}\lambda_{1}\lambda_{\rm eff}^{2}R^{-1}e^{-3\theta R}>0,\nonumber
\end{align}
where the constant $\lambda_{\rm eff}^{2}$ satisfies
\begin{align*}
\lambda_{\rm eff}^{2}&:=\int_{\R^{3}}g(y)\sinh(\theta|y|)\frac{dy}{4|y|}.
\end{align*}
Substituting \eqref{cross-term} into \eqref{subadditivity}, we further obtain that for sufficiently large $R>0$,
\begin{align*}
I^{\rm HFB}_{m,\kappa,\theta}(\lambda_{1}+\lambda_{2})
&\leq I^{\rm HFB}_{m,\kappa,\theta}(\lambda_{1})+I^{\rm HFB}_{m,\kappa,\theta}(\lambda_{2})
+Ce^{-4\theta R}-\tilde{C}\kappa\lambda_{1}\lambda_{\rm eff}^{2}R^{-1}e^{-3\theta R}\nonumber\\
&<I^{\rm HFB}_{m,\kappa,\theta}(\lambda_{1})+I^{\rm HFB}_{m,\kappa,\theta}(\lambda_{2}),
\end{align*}
which therefore completes the proof of Proposition \ref{prop:binding-esti}. \qed

Since the binding inequality of Proposition \ref{prop:binding-esti} contradicts with Proposition \ref{prop:dichotomy},   the dichotomy case is ruled out. This therefore completes the proof of Theorem \ref{thm:main1}.

\section{Existence and non-existence of HF minimizers}\label{sec:existence-nonesistence-HF}
Let $I^{\rm HF}_{m,\kappa,\theta}(N)$ be defined in \eqref{problem}, where $m>0,\ \kappa>0,\ \theta\geq0$  and $2\leq N\in\mathbb{N}$. The main purpose of this section is to establish Theorem \ref{th1} on the existence and nonexistence of minimizers for the problem $I^{\rm HF}_{m,\kappa,\theta}(N)$.  For simplicity, we denote
\begin{align*}
D(\rho,\rho):=\iinte\frac{\rho(x)\rho(y)}{|x-y|}dxdy,
\ \
\Ex(\gamma):=\iinte\frac{|\gamma(x,y)|^2}{|x-y|}dxdy.
\end{align*}
We begin with the following energy estimates.

\begin{lemma}\label{lem2.1}  Suppose $m>0$, $2\leq N\in\mathbb{N}$, and $\kappa_N^{\rm HF}>0$ is as in \eqref{eq:GN-HF}. Then the problem $I^{\rm HF}_{m,\kappa,\theta}(N)$ defined in \eqref{problem} satisfies the following properties:
\begin{enumerate}
\item [$(i)$] There exists a constant $C(N)>0$,  depending only on $N$, such that if $0< \kappa \leq \kappa^{\rm HF}_N$, then
\begin{equation}\label{3.1a}
-mN\leq I^{\rm HF}_{m,\kappa,\theta}(N)<0\ \ \text{for\ all}\  \,   0\leq\theta\leq \kappa mC(N).
\end{equation}

\item [$(ii)$] If $\kappa>\kappa^{\rm HF}_N$, then $I^{\rm HF}_{m,\kappa,\theta}(N)=-\infty$ for any $\theta\geq0$.
\end{enumerate}
\end{lemma}

\noindent \textbf{Proof.}
 Using
$\rho_\gamma(x)\rho_\gamma(y)-|\gamma(x,y)|^2 \ge 0$ and 
the definition of $\kappa^{\rm HF}_N$, we obtain that for any given  $\gamma\in\mathcal{P}_N$, 
\begin{align}\label{3.2}
\mathcal{E}^{\rm HF}_{m,\kappa,\theta}(\ga)
&\geq \Tr\big(\sqrt{-\Delta}\,\ga\big)-\frac{\kappa}{2}\Big[D(\rho_{\gamma},\rho_{\gamma})-\mathrm{Ex}(\gamma)\Big] -\Tr (m\ga)\\
&\geq\Big(1-\frac{\kappa}{\kappa^{\rm HF}_N}\Big)\Tr\big(\sqrt{-\Delta}\,\ga\big)-mN\geq -mN,\ \ \forall\ 0<\kappa \le  \kappa^{\rm HF}_N,\ \ \theta\geq0,\nn
\end{align}
where $\mathcal{P}_N$ given by \eqref{set}. 
Moreover, for any fixed $2\leq N\in\mathbb{N}$, we choose an operator $\gamma=\sum_{j=1}^{N}|u_j\rangle\langle u_j|\in\mathcal{P}_N$  and denote 
$$
\gamma_t:=\sum_{j=1}^{N}t^3|u_j(t\cdot)\rangle\langle u_j(t\cdot)|,\ \   \text{where}\ t=\kappa m \frac{D(\rho_\gamma, \rho_\gamma)-\text{Ex}(\gamma)}{2\, \Tr(-\Delta \gamma)}>0.
$$ 
Using the inequality $e^{-s}\geq 1-s$ for $s\geq0$ and the  operator inequality $
	\sqrt{-\Delta+m^2}-m\leq \frac{-\Delta}{2m}$ for $m>0$, we  deduce  that for any $0< \kappa \leq \kappa^{\rm HF}_N$,
\begin{align*}\label{2.12a}
-mN&\leq I^{\rm HF}_{m,\kappa,\theta}(N)\leq\mathcal{E}^{\rm HF}_{m,\kappa,\theta}(\gamma_t)\\
&\leq\frac{1}{2m}\Tr\big(-\Delta\gamma_t\big)
-\frac{\kappa}{2}\iinte\frac{1-\theta|x-y|}{|x-y|}\Big(\rho_{\gamma_t}(x)\rho_{\gamma_t}(y)-|\gamma_t(x,y)|^2\Big)dxdy\nonumber\\
&=\frac{t^2}{2m}\Tr\big(-\Delta\gamma\big)-\frac{\kappa t}{2}\iinte\frac{1-\theta t^{-1}|x-y|}{|x-y|}\Big(\rho_{\gamma}(x)\rho_{\gamma}(y)-|\gamma(x,y)|^2\Big)dxdy\nonumber\\[1mm]
&=-\frac{m\kappa^2\Big[D(\rho_{\gamma},\rho_{\gamma})-\mathrm{Ex}(\gamma)\Big]^2}{8\Tr\big(-\Delta\gamma\big)}+\frac{\kappa \theta}{2}\big(N^2-N\big) <0 
\end{align*}
provided that $0\leq\theta\leq \kappa mC(N)$, where the  constant $C(N)>0$ depending only on $N$. This proves Lemma \ref{lem2.1} (i).


To obtain (ii), we use Proposition \ref{th2.1} and  take $\gamma^{(N)}\in\mathcal{P}_N$ as an optimizer of $\kappa^{\rm HF}_N$. Denoting $\gamma^{(N)}_t(x,y):=t^3\gamma^{(N)}(tx, ty)$, $t>0$ and using the operator inequality
\begin{equation*}
	\sqrt{-\Delta+m^2}-m\leq \sqrt{-\Delta},
\end{equation*}
we then deduce that for any $\kappa> \kappa^{\rm HF}_N$ and $\theta\geq0$,
\begin{align*}
&I^{\rm HF}_{m,\kappa,\theta}(N)\leq\mathcal{E}^{\rm HF}_{m,\kappa,\theta}\big(\gamma^{(N)}_t\big)\\
\leq& \mathrm{Tr}\big(\sqrt{-\Delta}\gamma^{(N)}_t\big)-\frac{\kappa}{2}\iinte\frac{e^{-\theta|x-y|}}{|x-y|}\Big(\rho_{\gamma^{(N)}_t}(x)\rho_{\gamma^{(N)}_t}(y)-|\gamma^{(N)}_t(x,y)|^2\Big)dxdy\\
=&-t\Big[\frac{\kappa-\kappa^{\rm HF}_N}{2}\iinte\frac{1}{|x-y|}\Big(\rho_{\gamma^{(N)}}(x)\rho_{\gamma^{(N)}}(y)-|\gamma^{(N)}(x,y)|^2\Big)dxdy\\
&\ \   \ \   \  -\frac{\kappa}{2}\iinte\frac{1-e^{-\theta t^{-1}|x-y|}}{|x-y|}\Big(\rho_{\gamma^{(N)}}(x)\rho_{\gamma^{(N)}}(y)-|\gamma^{(N)}(x,y)|^2\Big)dxdy\Big]\\
=&-t\Big[\frac{\kappa-\kappa^{\rm HF}_N}{2}\iinte\frac{1}{|x-y|}\Big(\rho_{\gamma^{(N)}}(x)\rho_{\gamma^{(N)}}(y)-|\gamma^{(N)}(x,y)|^2\Big)dxdy-o(1)\Big]\\[1mm]
=&-\infty\ \ \text{as}\ \ t\to\infty.
\end{align*}
This completes the proof of Lemma \ref{lem2.1}.\qed

\begin{lemma}\label{lem3.2} For $2\leq N\in\mathbb{N}$, let $\gamma$  be a minimizer of $I^{\rm HF}_{m,\kappa,\theta}(N)$, where $m>0$, $\kappa>0$, $0\leq\theta\leq \kappa m C(N)$, and $C(N)>0$ is as in Lemma \ref{lem2.1}. Then  $\gamma$  can be written as $\gamma=\sum_{j=1}^N|u_j\rangle \langle u_j|$, where the orthonormal system  $(u_1, \cdots, u_N)$ solves the following fermionic  system
\begin{align*}\label{3.10b}
H_{\gamma}u_j:=&\big(\sqrt{-\Delta+m^2}-m\big)u_j- \kappa \inte\frac{e^{-\theta|x-y|}\rho_\gamma(y)}{|x-y|}dy\, u_j\\
&+ \kappa \inte\frac{e^{-\theta|x-y|}\gamma(x,y)}{|x-y|}	u_j(y)dy
=\mu_j u_j\ \ \text{in}\, \ \R^3, \ \ j=1,\cdots, N,
\end{align*}
and $\mu_1 \leq \mu_2\leq\cdots\leq\mu_N< 0$ are  $N$ negative eigenvalues  of the operator $H_{\gamma}$ on $L^2(\R^3, \mathbb{C})$.
\end{lemma}
\vspace{-0.2cm}
Since the proof is similar to that of \cite[Theorem 4 (iii)]{LenLew-10}, we omit the details for simplicity. Applying above lemmas, we now address the existence of Theorem \ref{th1} as follows.

\vspace{.15cm}

\noindent\textbf{Proof of Theorem \ref{th1} (i) and (iii).} Following Lemmas \ref{lem2.1}  and \ref{lem3.2}, we only need to prove that $I^{\rm HF}_{m,\kappa,\theta}(N)$ has a minimizer for any $\kappa\in(0,\,  \kappa^{\rm HF}_N)$ and $\theta\in\big[0, \kappa mC(N)\big]$, where $m>0,\ 2\leq N\in\mathbb{N}$,  $\kappa^{\rm HF}_N>0$ and $C(N)>0$ are given by \eqref{eq:GN-HF} and \eqref{3.1a}, respectively. Essentially, let $\{\gamma_n\}$ be a minimizing sequence of $I^{\rm HF}_{m,\kappa,\theta}(N)$ for the above case.  It then follows from \eqref{3.1a} and \eqref{3.2} that $-\infty<I^{\rm HF}_{m,\kappa,\theta}(N)<0$, and  the sequence $\big\{\Tr(\sqrt{-\Delta}\, \gamma_n)\big\}$ is bounded uniformly in $n$. Therefore,  similar to the proof of \cite[Theorem 4]{LenLew-10}, one can further prove the existence of minimizers for $I^{\rm HF}_{m,\kappa,\theta}(N)$ for the above case. The proof of Theorem \ref{th1} (i) and (iii) is therefore complete. \qed

\subsection{Non-existence of minimizers}

This subsection is devoted to the proof of Theorem \ref{th1} (ii), which focuses on the nonexistence of minimizers for $I^{\rm HF}_{m,\kappa^{\rm HF}_N,\theta}(N)$ defined in (\ref{problem}), where $2\leq N\in\mathbb{N}$ is arbitrary. According to Theorem \ref{th1} (i),  if $\kappa \in (0,\kappa^{\rm HF}_N)$, then the existence of minimizers for $I^{\rm HF}_{m,\kappa,\theta}(N)$ holds for all sufficiently small $\theta \ge 0$, where $\kappa^{\rm HF}_N\in(0, \infty)$ is the optimal constant of \eqref{eq:GN-HF}.  Moreover,  recall from \cite[Lemma 3.3]{CGNO-25} that $I^{\rm HF}_{m,\kappa^{\rm HF}_N,0}(N)$ has no minimizer. Inspired by these facts, we thus expect that for sufficiently small $\theta>0$, the problem $I^{\rm HF}_{m,\kappa^{\rm HF}_N,\theta}(N)$ does not have any minimizer.

As stated in Theorem \ref{th1} (ii), we shall prove that there exists a constant $\theta^{*}=\theta^{*}(m, N)>0$ such that  for any  $0\leq\theta\leq \theta^{*}$, the problem $I^{\rm HF}_{m,\kappa^{\rm HF}_N,\theta}(N)$ does not admit any minimizer. To this end, by contradiction, suppose there exists a sequence $\{\theta_n\}$ satisfying $\theta_n\searrow0$ as $n\to\infty$ such that $I^{\rm HF}_{m,\kappa^{\rm HF}_N,\theta_n}(N)$ admits a minimizer $\gamma_{\theta_n}$ for any sufficiently large $n>0$. Following Lemma \ref{lem3.2},  any minimizer $\gamma_{\theta_n}$ of $I^{\rm HF}_{m,\kappa^{\rm HF}_N,\theta_n}(N)$ can be  written as $\gamma_{\theta_n}=\sum_{j=1}^N\big|u_j^{\theta_n}\big\rangle \big\langle u_j^{\theta_n}\big|$, where the orthonormal system $(u^{\theta_n}_1, \cdots, u^{\theta_n}_N)$ satisfies
\begin{equation}\label{3.11b}
\begin{split}
	&\big(\sqrt{-\Delta+m^2}-m\big)u_j^{\theta_n}-\kappa^{\rm HF}_N\inte\frac{e^{-\theta_n|x-y|}\rho_{\gamma_{\theta_n}}(y)}{|x-y|}dy\, u_j^{\theta_n}\\
	&+\kappa^{\rm HF}_N\inte\frac{e^{-\theta_n|x-y|}\gamma_{\theta_n}(x,y)}{|x-y|}	u_j^{\theta_n}(y)dy=\mu_j^{\theta_n} u_j^{\theta_n}\ \ \text{in}\, \ \R^3
\end{split}
\end{equation}
for some  $\mu_1^{\theta_n} \leq\cdots\leq\mu_N^{\theta_n}< 0$.
The following lemmas present the analytical properties of blow-up minimizers $(u^{\theta_n}_1, \cdots, u^{\theta_n}_N)$ as  $n\to\infty$.

\begin{lemma}\label{lem2.2}
For any given $m>0 $ and $ 2\leq N\in\mathbb{N}$,  suppose $\gamma_{\theta_n}=\sum_{j=1}^N\big|u_j^{\theta_n}\big\rangle \big\langle u_j^{\theta_n}\big|$ is a minimizer of $I^{\rm HF}_{m,\kappa^{\rm HF}_N,\theta_n}(N)$, where $(u_1^{\theta_n}, \cdots,  u_N^{\theta_n})$ satisfies \eqref{3.11b} and $\theta_n\searrow0$ as $n\to\infty$. If $\sup\limits_{n}\Tr\big(\sqrt{-\Delta}\, \gamma_{\theta_n}\big)=\infty$, then up to a subsequence if necessary, there exist a sequence $\{z_n\}\subset\R^3$ and an orthonormal  system $(w_1,\cdots, w_N)$ such that for $j=1,\cdots,N$, 
\begin{align}\label{2.62a}
w_j^{\theta_n}(x):=\epsilon_{\theta_n}^{\frac{3}{2}}u_j^{\theta_n}\big(\epsilon_{\theta_n}(x+z_n)\big)\to w_j\ \ \text{strongly\ in}\ \ H^{\frac{1}{2}}(\R^3, \C)\ \text{as}\ n\to\infty,
\end{align}
where $0<\epsilon_{\theta_n}:=\big(\Tr(\sqrt{-\Delta}\gamma_{\theta_n})\big)^{-1}\to0$ as $n\to\infty$, and $\gamma^*:=\sum_{j=1}^N|w_j\rangle\langle w_j|$ is an optimizer of $\kappa^{\rm HF}_N$.
\end{lemma}

\noindent\textbf{Proof.}  {\em Step 1.} In this step, we claim that 
\begin{align}\label{3.7a}
\lim\limits_{n\to\infty} I^{\rm HF}_{m,\kappa^{\rm HF}_N,\theta_n}(N)=-mN.
\end{align}

To prove the claim (\ref{3.7a}), suppose $\gamma^{(N)}=\sum_{j=1}^N|Q_j\rangle\langle Q_j|\in\mathcal{P}_N$ satisfying \eqref{decay} is an optimizer of $\kappa^{\rm HF}_N$, and denote $\gamma^{(N)}_{t_n}(x,y):=t_n^3\gamma^{(N)}(t_nx, t_ny)$, where $t_n>0$ satisfies $t_n\to\infty$ as $n\to\infty$. Using the inequality $e^{-s}\geq 1-s$ for $s\geq0$, we have 
\begin{align}\label{3.26}
&I^{\rm HF}_{m,\kappa^{\rm HF}_N,\theta_n}(N)\leq\mathcal{E}^{\rm HF}_{m,\kappa^{\rm HF}_N,\theta_n}\big(\gamma^{(N)}_{t_n}\big)\nonumber\\[1mm]
=&-mN+\Tr\Big[\Big(\sqrt{-\Delta+m^2}-\sqrt{-\Delta}\Big) \gamma^{(N)}_{t_n}\Big]\nonumber\\[1mm]
&+\frac{\kappa^{\rm HF}_N}{2}\iinte\frac{1-e^{-\theta_n|x-y|}}{|x-y|}\Big(\rho_{\gamma_{t_n}^{(N)}}(x)\rho_{\gamma_{t_n}^{(N)}}(y)-|\gamma_{t_n}^{(N)}(x,y)|^2\Big)dxdy\nonumber\\[1mm]
=&-mN+t_n\Tr\Big[\Big(\sqrt{-\Delta+t_n^{-2}m^2}-\sqrt{-\Delta}\, \Big) \gamma^{(N)}\Big]\\[1mm]
&+\frac{t_n\kappa^{\rm HF}_N}{2}\iinte\frac{1-e^{-\theta_nt_n^{-1}|x-y|}}{|x-y|}\Big(\rho_{\gamma^{(N)}}(x)\rho_{\gamma^{(N)}}(y)-|\gamma^{(N)}(x,y)|^2\Big)dxdy\nonumber\\[1.5mm]
\leq&-mN+\frac{t_n^{-1}m^2}{2}\Tr\frac{\gamma^{(N)}}{\sqrt{-\Delta}}+\frac{\theta_n\kappa^{\rm HF}_N}{2}\big(N^2-N\big),\ \ \forall\ n>0,\nonumber
\end{align}
where the last inequality follows from the operator inequality
$$
\sqrt{-\Delta+t_n^{-2}m^2}-\sqrt{-\Delta}=\frac{t_n^{-2}m^2}{\sqrt{-\Delta+t_n^{-2}m^2}+\sqrt{-\Delta}}\leq\frac{t_n^{-2}m^2}{2\sqrt{-\Delta}},\ \ \forall\ m,\, t_n>0.
$$
Since $\rho_{\gamma^{(N)}}$ satisfies the estimate \eqref{decay},  one gets from \eqref{hkt} that
\begin{align}\label{3.13c}
\Tr\frac{\gamma^{(N)}}{\sqrt{-\Delta}}\leq\frac{\pi}{2}\int_{\R^3}|x|\rho_{\gamma^{(N)}}dx<\infty.
\end{align}
Recall from \cite[Lemma 3.3]{CGNO-25} that $
I^{\rm HF}_{m,\kappa^{\rm HF}_N,0}(N)=-mN$ has\ no\ minimizer for all $2\leq N\in\mathbb{N}$. 
We  thus conclude from \eqref{3.26} and \eqref{3.13c} that
\begin{align}\label{3.6}
-mN=I^{\rm HF}_{m,\kappa^{\rm HF}_N,0}(N)\leq \lim\limits_{n\to\infty}I^{\rm HF}_{m,\kappa^{\rm HF}_N,\theta_n}(N)\leq-mN,
\end{align}
which yields that the claim \eqref{3.7a} holds true. 

{\em Step 2.}  Since $\sup\limits_{n}\Tr(\sqrt{-\Delta}\gamma_{\theta_n})=\infty$, we may suppose that
\begin{align}\label{3.12a}
\lim\limits_{n\to\infty}\epsilon_{\theta_n}:=\lim\limits_{n\to\infty}\big(\Tr(\sqrt{-\Delta}\gamma_{\theta_n})\big)^{-1}=0.
\end{align}
In this step, we claim that there exist a sequence $\{z_n\}\subset\R^3$ and a system $(w_1,\cdots, w_N)\in \big(H^{\frac{1}{2}}(\R^3, \C)\big)^N\backslash\{0\}$ such that up to a subsequence  if necessary,
\begin{eqnarray}\label{2.52}
w_j^{\theta_n}(x):=\epsilon_{\theta_n}^{\frac{3}{2}}u_j^{\theta_n}\big(\epsilon_{\theta_n}(x+z_n)\big)\rightharpoonup  w_j\ \ \mathrm{weakly\ in}\, \ H^{\frac{1}{2}}(\R^3, \C)\ \ \text{as}\ \ n\to\infty,
\end{eqnarray}
where
\begin{align}\label{3.26b}
\gamma^*:=\sum_{j=1}^N|w_j\rangle\langle w_j|\ \ \text{is\  an\ optimizer\ of}\  \kappa^{\rm HF}_N, \ \ \text{and}\ \   \|\gamma^*\|=1.
\end{align}

To prove the above claim, we define
\begin{equation}\label{3.10}
\tilde{\gamma}_{\theta_n}:=\sum_{j=1}^N\big|\tilde{w}_j^{\theta_n}\big\rangle \big\langle \tilde{w}_j^{\theta_n}\big|,\ \ \tilde{w}_j^{\theta_n}(x):=\epsilon_{\theta_n}^{\frac{3}{2}}u_j^{\theta_n}(\epsilon_{\theta_n}x),\ \ \forall\ n>0.
\end{equation}
We then have
\begin{equation}\label{3.15c}
\Tr\big(\sqrt{-\Delta}\, \tilde{\gamma}_{\theta_n}\big)=\sum_{j=1}^N\langle\tilde{w}_j^{\theta_n}, \sqrt{-\Delta}\, \tilde{w}_j^{\theta_n} \rangle=\epsilon_{\theta_n} \Tr\big(\sqrt{-\Delta}\, \gamma_{\theta_n}\big)=1,\ \ \forall\ n>0,
\end{equation}
and
\begin{equation}\label{2.48}
\begin{split}
&\iinte\frac{e^{-\theta_n\epsilon_{\theta_n}|x-y|}}{|x-y|}\Big(\rho_{\tilde{\gamma}_{\theta_n}}(x)\rho_{\tilde{\gamma}_{\theta_n}}(y)-|\tilde{\gamma}_{\theta_n}(x,y)|^2\Big)dxdy\\
=&\epsilon_{\theta_n}\iinte\frac{e^{-\theta_n|x-y|}}{|x-y|}\Big(\rho_{\gamma_{\theta_n}}(x)\rho_{\gamma_{\theta_n}}(y)-|\gamma_{\theta_n}(x,y)|^2\Big)dxdy,\ \ \forall\ n>0.
\end{split}
\end{equation}
Note from \eqref{eq:GN-HF} and \eqref{3.7a} that
\begin{align}\label{2.46a}
0\leq&\Tr\big(\sqrt{-\Delta}\, \gamma_{\theta_n}\big)-\frac{\kappa^{\rm HF}_N}{2}\iinte\frac{1}{|x-y|}\Big(\rho_{\gamma_{\theta_n}}(x)\rho_{\gamma_{\theta_n}}(y)-|\gamma_{\theta_n}(x,y)|^2\Big)dxdy\nonumber\\
\leq &\Tr\big(\sqrt{-\Delta}\, \gamma_{\theta_n}\big)-\frac{\kappa^{\rm HF}_N}{2}\iinte\frac{e^{-\theta_n|x-y|}}{|x-y|}\Big(\rho_{\gamma_{\theta_n}}(x)\rho_{\gamma_{\theta_n}}(y)-|\gamma_{\theta_n}(x,y)|^2\Big)dxdy\nonumber\\[1mm]
\leq&I^{\rm HF}_{m,\kappa^{\rm HF}_N,\theta_n}(N)+mN=o(1)\ \ \text{as}\ n\to\infty.
\end{align}
This implies from \eqref{3.15c}--\eqref{2.46a} that
\begin{equation}\label{2.49}
\lim\limits_{n\to\infty}\iinte\frac{e^{-\theta_n\epsilon_{\theta_n}|x-y|}}{|x-y|}\Big(\rho_{\tilde{\gamma}_{\theta_n}}(x)\rho_{\tilde{\gamma}_{\theta_n}}(y)-|\tilde{\gamma}_{\theta_n}(x,y)|^2\Big)dxdy=\frac{2}{\kappa^{\rm HF}_N}.
\end{equation}

Since it follows from \eqref{3.15c} that the sequence  $\big\{\rho^{1/2}_{\tilde{\gamma}_{\theta_n}}\big\}_n=\Big\{\big(\sum_{j=1}^N|\tilde{w}_j^{\theta_n}|^2\big)^{1/2}\Big\}_n$ is bounded uniformly in $H^{\frac{1}{2}}(\R^3, \C)$, where  $j=1, \cdots, N$,
we now prove that there exists a constant $R>0$ such that up to a subsequence if necessary,
\begin{equation}\label{vani}
\lim_{n\rightarrow \infty}\sup\limits_{z\in\R^3}\int_{B_R(z)} \rho_{\tilde{\gamma}_{\theta_n}}dx>0.
\end{equation}
In fact, on the contrary, suppose (\ref{vani}) is false. By the uniform boundedness of $\big\{\rho^{1/2}_{\tilde{\gamma}_{\theta_n}}\big\}$ in $H^{\frac{1}{2}}(\R^3, \C)$, we then conclude from  \eqref{inter-bdd1} and  \cite[Lemma 7.2]{LenLew-10} that
$$
0\leq\mathrm{Ex}_\theta(\tilde{\gamma}_{\theta_n})\leq D_\theta(\rho_{\tilde{\gamma}_{\theta_n}}, \rho_{\tilde{\gamma}_{\theta_n}})\to0\ \ \text{as}\ \ n\to\infty,
$$
which however contradicts with \eqref{2.49}. Hence, \eqref{vani} holds true.
As a consequence of \eqref{vani},  there exists  a sequence $\{z_n\}\subset\R^3$ such that 
\begin{equation}\label{2.11}
\lim_{n\rightarrow \infty}\int_{B_R(z_n)}\rho_{\tilde{\gamma}_{\theta_n}}dx=\lim\limits_{n\to\infty}\int_{B_R(z_n)} \sum_{j=1}^N|\tilde{w}_j^{\theta_n}|^2dx>0.
\end{equation}
Following the uniform  boundedness of $\{\tilde{w}_j^{\theta_n}\}_n$  in $H^{\frac{1}{2}}(\R^3, \C)$ for $j=1,\cdots,N$,  we therefore derive from \eqref{2.11} that up to a subsequence if necessary, there exists a system $(w_1,\cdots, w_N)\in \big(H^{\frac{1}{2}}(\R^3, \C)\big)^N\backslash\{0\}$ such that for $j=1, \cdots, N$,
\begin{eqnarray}\label{2.52a}
w_j^{\theta_n}(x):=\tilde{w}_j^{\theta_n}(x+z_n)\rightharpoonup  w_j\ \ \mathrm{weakly\ in}\, \ H^{\frac{1}{2}}(\R^3,\C)\ \ \text{as}\ \ n\to\infty.
\end{eqnarray}
This proves the weak convergence \eqref{2.52}.

To prove (\ref{3.26b}), we next define
\begin{equation*}\label{3.14a}
\gamma^*_{\theta_n}:=\sum_{j=1}^N\big|w_j^{\theta_n}\big\rangle \big\langle w_j^{\theta_n}\big|=\sum_{j=1}^N\epsilon_{\theta_n}^3\big|u_j^{\theta_n}\big(\epsilon_{\theta_n}(\cdot+z_n)\big)\big\rangle \big\langle u_j^{\theta_n}\big(\epsilon_{\theta_n}(\cdot+z_n)\big)\big|,
\end{equation*}
where $z_n\in\R^3$ is as in \eqref{2.52}. We then deduce from \eqref{3.11b} that
\begin{align}\label{2.50}
&\big(\sqrt{-\Delta+m^2\epsilon_{\theta_n}^2}-m\epsilon_{\theta_n}\big)w_j^{\theta_n}-\kappa^{\rm HF}_N\inte\frac{e^{-\theta_n\epsilon_{\theta_n}|x-y|}\rho_{\gamma^*_{\theta_n}}(y)}{|x-y|}dy\, w_j^{\theta_n}\\
&+\kappa^{\rm HF}_N\inte\frac{e^{-\theta_n\epsilon_{\theta_n}|x-y|}\, \gamma^*_{\theta_n}(x,y)}{|x-y|}w_j^{\theta_n}(y)dy
=\epsilon_{\theta_n} \mu_j^{\theta_n}w_j^{\theta_n}
\ \ \text{in}\, \ \R^3, \ \ j=1,\cdots, N,\nonumber
\end{align}
where $
\mu_1^{\theta_n}\leq\cdots\leq\mu_N^{\theta_n}<0$ for any $n>0$.
This shows from \eqref{eq:GN-HF} and \eqref{3.15c} that
\begin{align*}
\lim\limits_{n\to\infty}\sum_{j=1}^N\epsilon_{\theta_n}\mu_j^{\theta_n}
&\geq\lim\limits_{n\to\infty}\Big[-\Tr\big(\sqrt{-\Delta}\, \gamma^*_{\theta_n}\big)-m\epsilon_{\theta_n}N\Big]\\
&=\lim\limits_{n\to\infty}\Big[-\Tr\big(\sqrt{-\Delta}\, \tilde{\gamma}_{\theta_n}\big)-m\epsilon_{\theta_n}N\Big]=-1>-\infty,
\end{align*}
where $\tilde{\gamma}_{\theta_n}$ is as in \eqref{3.10}.
We then derive that the sequence $\big\{\epsilon_{\theta_n}\mu_j^{\theta_n}\big\}_n$ is bounded uniformly for all $j=1, \cdots, N$.
Thus,  up to a subsequence  if necessary,  we have
\begin{equation}\label{3.21}
-\infty<\lim\limits_{n\to\infty}\epsilon_{\theta_n} \mu_j^{\theta_n}=\mu^*_j\leq0,\ \ j=1,\cdots,N.
\end{equation}

Note  that  $\gamma^*:=\sum_{j=1}^N|w_j\rangle\langle w_j|\neq0$, where the nonzero system $(w_1, \cdots,w_N)$ is as in \eqref{2.52}.  By \eqref{3.12a}, \eqref{2.52},  \eqref{2.50} and \eqref{3.21},  the same argument of proving \cite[Section 4]{CGNO-25} then gives that for $j=1, \cdots, N$,
\begin{equation}\label{2.56}
\begin{split}
\sqrt{-\Delta}\, w_j-\kappa^{\rm HF}_N\inte\frac{\rho_{\gamma^*}(y)}{|x-y|}dy\, w_j+\kappa^{\rm HF}_N\inte\frac{\gamma^*(x,y)}{|x-y|}w_j(y)dy
=\mu^*_jw_j 
\end{split}
\end{equation}
in $\R^3$. Using \eqref{2.56}, we can further prove  (see also \cite[Eq. (4.9)]{CGNO-25}) the following Pohozaev-type identity
\begin{equation}\label{2.57}
\begin{split}
\Tr\big(\sqrt{-\Delta}\, \gamma^*\big)-\frac{5\kappa^{\rm HF}_N}{4}\Big[D(\rho_{\gamma^*}, \rho_{\gamma^*})-\text{Ex}(\gamma^*)\Big]=\frac{3}{2}\sum_{j=1}^N\mu^*_j\inte |w_j|^2dx.
\end{split}
\end{equation}
As a consequence of \eqref{2.56}, \eqref{2.57} and the definition of $\kappa^{\rm HF}_N$, one  gets that
\begin{equation}\label{3.34}
\|\gamma^*\|\, \Tr\big(\sqrt{-\Delta}\, \gamma^*\big)
\geq\frac{\kappa^{\rm HF}_N}{2}\Big(D(\rho_{\gamma^*},\rho_{\gamma^*})-\text{Ex}(\gamma^*)\Big)=\Tr\big(\sqrt{-\Delta}\, \gamma^*\big),
\end{equation}
and the equality occurs as $\|\gamma^*\|\leq\liminf\limits_{n\to\infty}\|\gamma^*_{\theta_n}\|=1$.
This hence gives that \eqref{3.26b} holds true, and the proof of Step 2 is therefore done.

{\em Step 3.} The aim of this step is to establish the strong convergence \eqref{2.62a}. By Proposition \ref{th2.1} (ii), we deduce from \eqref{3.26b} that  $\inte\rho_{\gamma^*}dx=N$, 
which then implies from \eqref{2.52}  that
\begin{align}\label{3.26a}
\|w_j\|_2^2=1=\lim\limits_{n\to\infty}\|w_j^{\theta_n}\|_2^2,\ \ j=1,\cdots,N.
\end{align}
 By  \eqref{2.52} again and the Br\'ezis-Lieb lemma, we thus obtian that for all $r\in[2, 3),$
\begin{align}\label{3.36}
w_j^{\theta_n}\to w_j\ \ \text{in}\ \ L^r(\R^3, \C), \ \       j=1,\cdots,N.
\end{align}
Together with  the Hardy-Littlewood-Sobolev inequality, we  conclude  that
\begin{equation}\label{3.38}
\begin{split}
&\lim\limits_{n\to\infty}\iinte\frac{e^{-\theta_n\epsilon_{\theta_n}|x-y|}}{|x-y|}\Big(\rho_{\gamma^*_{\theta_n}}(x)\rho_{\gamma^*_{\theta_n}}(y)-|\gamma^*_{\theta_n}(x,y)|^2\Big)dxdy\\
&=D(\rho_{\gamma^*},\rho_{\gamma^*})-\text{Ex}(\gamma^*).
\end{split}
\end{equation}
Moreover, it follows from \eqref{3.15c}--\eqref{2.46a} that
\begin{equation}\label{3.39a}
\begin{split}
&\Tr\big(\sqrt{-\Delta}\, \gamma^*_{\theta_n}\big)-\frac{\kappa^{\rm HF}_N}{2}\iinte\frac{e^{-\theta_n\epsilon_{\theta_n}|x-y|}}{|x-y|}\Big(\rho_{\gamma^*_{\theta_n}}(x)\rho_{\gamma^*_{\theta_n}}(y)-|\gamma^*_{\theta_n}(x,y)|^2\Big)dxdy\\
&=o(\epsilon_{\theta_n})\ \ \text{as}\ \ n\to\infty.
\end{split}
\end{equation}
We therefore deduce from \eqref{3.26b}, \eqref{3.38} and \eqref{3.39a} that
$$
\Tr\big(\sqrt{-\Delta}\, \gamma^*\big)\leq\liminf\limits_{n\to\infty}\Tr\big(\sqrt{-\Delta}\, \gamma^*_{\theta_n}\big)=\frac{\kappa^{\rm HF}_N}{2}\Big[D(\rho_{\gamma^*},\rho_{\gamma^*})-\text{Ex}(\gamma^*)\Big]=\Tr\big(\sqrt{-\Delta}\, \gamma^*\big).
$$
Together with \eqref{3.36}, this proves \eqref{2.62a}. The proof of Lemma \ref{lem2.2} is complete. \qed

We next study the uniformly decaying property of $\{w_j^{\theta_n}\}_n$ as $n\to\infty$, where $j=1, \cdots, N$.

\begin{lemma}\label{lem3.4}
Under the assumptions of Lemma \ref{lem2.2},  let $\{w_j^{\theta_n}\}_n$ be given by \eqref{2.62a} for $j=1, \cdots, N$, where $\theta_n\searrow0$ as $n\to\infty$. Then there exists a constant $C>0$, which is independent of $n>0$, such that for sufficiently large $n>0$,
\begin{equation}\label{3.31}
\sum_{j=1}^N|w_j^{\theta_n}(x)|\leq C\big(1+|x|\big)^{-4}\ \ \,\text{in}\ \ \R^3.
\end{equation}
\end{lemma}

\noindent\textbf{Proof.}  We first claim that for $j=1, \cdots, N$,
\begin{equation}\label{3.23}
\big\{\big\|w_j^{\theta_n}\big\|_{L^\infty}\big\}\ \ \text{is\ bounded\ uniformly\ for\ sufficiently\ large}\ n>0,
\end{equation}
and
\begin{equation}\label{3.24}
\lim\limits_{|x|\to\infty}|w_j^{\theta_n}(x)|=0 \ \ \mathrm{uniformly\ for \ sufficiently\ large}\ n>0,
\end{equation}
We shall prove the above claim by three steps.

{\em Step 1.} We prove that
\begin{equation}\label{3.9a}
\big\{w_j^{\theta_n}\big\}_n\ \ \text{is\ bounded\ uniformly\ in}\ H^1(\R^3, \C),\ \ j=1, \cdots,N. 
\end{equation}
Actually, since $\{w_j^{\theta_n}\}_n$ is given by \eqref{2.62a}, we first note from \eqref{2.50} that
\begin{align}\label{3.34a}
& \Tr\big(-\Delta+m^2\epsilon^2_{\theta_n}\big)\gamma^*_{\theta_n}\nonumber
\\
=&\sum_{j=1}^N\Big\langle\sqrt{-\Delta+m^2\epsilon^2_{\theta_n}}\,  w_j^{\theta_n},\  \sqrt{-\Delta+m^2\epsilon^2_{\theta_n}}\,  w_j^{\theta_n}\Big\rangle\nonumber\\
\leq&m^2\epsilon_{\theta_n}^2N+\epsilon_{\theta_n}^2\sum_{j=1}^N\big|\mu_j^{\theta_n}\big|^2+|\kappa^{\rm HF}_N|^2\inte\rho_{\gamma^*_{\theta_n}}(x)\, \Big|\inte\frac{\rho_{\gamma^*_{\theta_n}}(y)}{|x-y|}dy\Big|^2dx\nonumber\\
&+|\kappa^{\rm HF}_N|^2\sum_{j=1}^N\inte\Big|\inte\frac{\big|\gamma^*_{\theta_n}(x,y)w_j^{\theta_n}(y)\big|}{|x-y|}dy\Big|^2dx+2m\epsilon_{\theta_n}^2 \sum_{j=1}^N\mu_j^{\theta_n}\nonumber\\
&+2m\kappa^{\rm HF}_N\epsilon_{\theta_n}\big[D(\rho_{\gamma^*_{\theta_n}},\rho_{\gamma^*_{\theta_n}})-\text{Ex}(\gamma^*_{\theta_n})\big]\\
&+2\kappa^{\rm HF}_N\epsilon_{\theta_n}\sum_{j=1}^N\big|\mu_j^{\theta_n}\big|\iinte\frac{\rho_{\gamma^*_{\theta_n}}(y)|w_j^{\theta_n}(x)|^2}{|x-y|}dxdy\nonumber\\
&+2\kappa^{\rm HF}_N\epsilon_{\theta_n}\sum_{j=1}^N\big|\mu_j^{\theta_n}\big|\iinte\frac{\big|\gamma^*_{\theta_n}(x,y)\overline{w_j^{\theta_n}(x)}\, w_j^{\theta_n}(y)\big|}{|x-y|}dxdy\nonumber\\
&+2|\kappa^{\rm HF}_N|^2\inte\Big[\inte\frac{\rho_{\gamma^*_{\theta_n}}(y)}{|x-y|}dy\inte\frac{\big|\gamma^*_{\theta_n}(x,y)\big|^2}{|x-y|}dy\Big]dx,\nonumber
\end{align}
where $\gamma^*_{\theta_n}=\sum_{j=1}^N\big|w_j^{\theta_n}\big\rangle \big\langle w_j^{\theta_n}\big|$, $\epsilon_{\theta_n}>0$ and $\mu_j^{\theta_n}<0$ are as in \eqref{3.21}. Using the Hardy-Kato inequality \eqref{hkt},
one can deduce from \eqref{2.62a}  that  there exists a constant $C>0$, independent of $n>0$, such that for all $n>0$,
\begin{equation}\label{3.11}
\begin{split}
0\leq\sup\limits_{x\in\R^3}\inte\frac{\rho_{\gamma^*_{\theta_n}}(y)}{|x-y|}dy
=&\sum_{j=1}^N\sup\limits_{x\in\R^3}\inte\frac{\big|w_j^{\theta_n}(y)\big|^2}{|x-y|}dy\\
\leq&\frac{\pi}{2}\sum_{j=1}^N\big\langle w_j^{\theta_n},  \sqrt{-\Delta}\, w_j^{\theta_n}\big\rangle=\frac{\pi}{2}\Tr\big(\sqrt{-\Delta}\, \gamma^*_{\theta_n}\big)<C,
\end{split}
\end{equation}
and
\begin{align}\label{3.12}
0\leq&\inte\frac{\big|\gamma^*_{\theta_n}(x,y)w_j^{\theta_n}(y)\big|}{|x-y|}dy \leq\sum_{k=1}^N\big|w_k^{\theta_n}(x)\big|\inte\frac{\big|\overline{w_k^{\theta_n}(y)}\, w_j^{\theta_n}(y)\big|}{|x-y|}dy\nonumber\\
\leq&\sum_{k=1}^N\big|w_k^{\theta_n}(x)\big|\Big(\inte\frac{\big|w_k^{\theta_n}(y)\big|^2}{|x-y|}dy
\Big)^{\frac{1}{2}}\Big(\inte\frac{\big|w_j^{\theta_n}(y)\big|^2}{|x-y|}dy\Big)^{\frac{1}{2}}\\
\leq&\frac{\pi}{2}\sum_{k=1}^N\big|w_k^{\theta_n}(x)\big|\big(w_k^{\theta_n}, \sqrt{-\Delta}\, w_k^{\theta_n}\big)^{\frac{1}{2}}\big(w_j^{\theta_n}, \sqrt{-\Delta}\, w_j^{\theta_n}\big)^{\frac{1}{2}}\nonumber\\
\leq&\frac{\pi}{2}\Tr\big(\sqrt{-\Delta}\, \gamma^*_{\theta_n}\big)\ \sum_{k=1}^N\big|w_k^{\theta_n}(x)\big|<C\sum_{k=1}^N\big|w_k^{\theta_n}(x)\big|\ \ \text{in}\ \R^3,\ \ j=1, \cdots, N.\nonumber
\end{align}
Applying \eqref{3.21}, we then obtain from \eqref{3.34a}--\eqref{3.12} that the sequence $\big\{\Tr\big(-\Delta\gamma^*_{\theta_n}\big)\big\} $ is bounded uniformly in $n$. Thus, \eqref{3.9a} holds true.

{\em Step 2.} We prove that
\begin{align}\label{3.13}
f_j^{\theta_n}\to   f_j\ \ \text{strongly\ in}\ L^r(\R^3)\ \text{as}\ n\to\infty,\ \ \forall\ 2\leq r<6,\ \ j=1,\cdots,N,
\end{align}
where
\begin{equation}\label{3.9}
\begin{split}
\frac{f_j^{\theta_n}(x)}{\kappa^{\rm HF}_N}:=&\inte\frac{\rho_{\gamma^*_{\theta_n}}(y)}{|x-y|}dy\, |w_j^{\theta_n}(x)|+\inte\frac{|\gamma^*_{\theta_n}(x,y)|}{|x-y|}|w_j^{\theta_n}(y)|dy,\\
\frac{f_j(x)}{\kappa^{\rm HF}_N}:=&\inte\frac{\rho_{\gamma^*}(y)}{|x-y|}dy\, |w_j(x)|+\inte\frac{|\gamma^*(x,y)|}{|x-y|}|w_j(y)|dy,
\end{split}
\end{equation}
$\gamma^*:=\sum_{j=1}^N|w_j\rangle\langle w_j|$, and $w_j$ is as in \eqref{2.62a}.

By the $H^{\frac{1}{2}}$-uniform convergence of $\big\{w_j^{\theta_n}\big\}_n$ as $n\to\infty$ for $j=1,\cdots,N$, we obtain from \eqref{hkt} that  for $j=1,\cdots,N$,
$$
w_j^{\theta_n}\to w_j\ \ \text{strongly\ in}\ L^r(\R^3)\ \ \text{as}\ \ n\to\infty,\ \ 2\leq r\leq3,
$$
and
\begin{align*}
\sup\limits_{x\in\R^3}\inte\frac{\big|w_j^{\theta_n}(y)-w_j(y)\big|^2}{|x-y|}dy
\leq&\frac{\pi}{2}\big\langle \big(w_j^{\theta_n}-w_j\big), \sqrt{-\Delta}\, \big(w_j^{\theta_n}-w_j\big)\big\rangle
\\
=&o(1)\ \ \text{as}\ \ n\to\infty.
\end{align*}
By the Hardy-Littlewood-Sobolev inequality, we then calculate from \eqref{3.11} that
\begin{align}\label{3.13b}
&\inte\Big(\inte\frac{\rho_{{\gamma^*_{\theta_n}}}(y)}{|x-y|}dy\, |w_j^{\theta_n}(x)|-\inte\frac{\rho_{\gamma^*}(y)}{|x-y|}dy\, |w_j(x)|\Big)^2dx\nonumber\\[1mm]
\leq&2\inte\Big|\inte\frac{\rho_{{\gamma^*_{\theta_n}}}(y)}{|x-y|}dy\Big|^2\, \big|w_j^{\theta_n}(x)-w_j(x)\big|^2dx\nonumber\\[1mm]
&+2\inte\Big|\inte\frac{\rho_{{\gamma^*_{\theta_n}}}(y)-\rho_{\gamma^*}(y)}{|x-y|}dy\Big|^2\, \big|w_j(x)\big|^2dx \\
\leq&2\inte\Big[\sum_{k=1}^N\inte\frac{\big( |w_k^{\theta_n}(y)|+|w_k(y)|\big)\ \big| w_k^{\theta_n}(y)-w_k(y) \big|}{|x-y|}dy\Big]^2\, \big|w_j(x)\big|^2dx+o(1)\nn\\
\leq&2^N\sum_{k=1}^N\inte\Big[\inte\frac{\big(|w_k^{\theta_n}(y)|+|w_k(y)|\big)^2}{|x-y|}dy\inte\frac{\big|w_k^{\theta_n}(y)-w_k(y)\big|^2}{|x-y|}dy\Big]\, \big|w_j(x)\big|^2dx+o(1)\nonumber\\
\leq&\Big[D(\rho_{\gamma^*_{\theta_n}}, \rho_{\gamma^*})+D(\rho_{\gamma^*}, \rho_{\gamma^*})\Big]o(1)+o(1)\nonumber\\
=&o(1)\ \ \text{as}\ \ n\to\infty,\ \ j=1,\cdots,N.\nonumber
\end{align}
Similarly, we have
\begin{align}\label{3.14}
&\inte\Big|\inte\frac{|\gamma^*_{\theta_n}(x,y)|\, |w_j^{\theta_n}(y)|-|\gamma^*(x,y)|\, |w_j(y)|}{|x-y|}dy\Big|^2dx\nonumber\\[1mm]
\leq&2\inte\Big(\inte\frac{\big|\gamma^*_{\theta_n}(x,y)\big|\, \big|w_j^{\theta_n}(y)-w_j(y)\big|}{|x-y|}dy\Big)^2dx\nonumber\\[1mm]
&+2\inte\Big(\inte\frac{\big|\gamma^*_{\theta_n}(x,y)-\gamma^*(x,y)\big|\, |w_j(y)|}{|x-y|}dy\Big)^2dx\\[1mm]
\leq&2\inte\Big[\inte\frac{\big|\gamma^*_{\theta_n}(x,y)\big|^2}{|x-y|}dy\inte\frac{\big|w_j^{\theta_n}(y)-w_j(y)\big|^2}{|x-y|}dy\Big]dx\nonumber\\[2mm]
&+2\inte\Big[\sum_{k=1}^N\inte\frac{\big|w_j(y)w_k^{\theta_n}(x)\big|\, \big|w_k^{\theta_n}(y)-w_k(y)\big|}{|x-y|}dy\nonumber\\
&\ \   \ \   \ \   +\sum_{k=1}^N\inte\frac{\big|w_j(y)w_k(y)\big|\, \big|w_k^{\theta_n}(x)-w_k(x)\big|}{|x-y|}dy\Big]^2dx \to 0\nonumber
\end{align}
as $n\to \infty$ for $j=1,\cdots,N$. We hence conclude from \eqref{3.13b} and \eqref{3.14} that
\begin{equation}\label{3.9M}
f_j^{\theta_n}\to   f_j\ \   \text{strongly\ in}\ L^2(\R^3)\ \ \text{as}\ \ n\to\infty,\ \ j=1,\cdots,N.
\end{equation}

Moreover,  it yields from \eqref{3.11} and  \eqref{3.12} that
\begin{align}\label{3.42}
f_j^{\theta_n}(x)\leq 2C\kappa^{\rm HF}_N\sum_{k=1}^N \big|w_k^{\theta_n}(x)\big|\ \ \text{in}\ \ \R^3,\ \ j=1,\cdots,N
\end{align}
where the constant $C>0$ is independent of $n>0$.  Therefore, using  the interpolation inequality and the uniform boundedness of $\{w_j^{\theta_n}\}_n$ in $H^1(\R^3, \C)$,
we deduce from (\ref{3.9M}) and \eqref{3.42} that the convergence \eqref{3.13} holds true.

{\em Step 3.} In this step, we prove the claims \eqref{3.23} and \eqref{3.24}. Let $G_j^{\theta_n}$ be the Green's function of the operator  $$\sqrt{-\Delta+m^2\epsilon_{\theta_n}^2}-\epsilon_{\theta_n}m-\epsilon_{\theta_n}\mu_j^{\theta_n}\ \ \text{in}\ \ \R^3,\ \  j=1, \cdots, N,
$$
where $\mu_j^{\theta_n}\in\R$ is as in \eqref{2.50}. Since
$\gamma^*=\sum_{j=1}^N|w_j\rangle\langle w_j|$ is  an optimizer of $\kappa^{\rm HF}_N$,  by Proposition \ref{th2.1} (ii), we can deduce from \eqref{3.21} and \eqref{2.56} that
\begin{align}\label{3.49}
\lim\limits_{n\to\infty}\big(\epsilon_{\theta_n}m+\epsilon_{\theta_n}\mu_j^{\theta_n}\big)=\mu^*_j<0,\ \ j=1,\cdots,	N.
\end{align}
Following the argument of \cite[Lemma 2.4]{CheGuoLiu-24}, we have for any sufficiently large $n>0$,
\begin{align}\label{g1}
0<G_j^{\theta_n}(x)<C_g|x|^{-2} \ \ \text{in}\ \ \R^3,
\end{align}
where $C_g>0$ is independent of $j,m, n$ and $ x$. Note from \eqref{2.50} that for any $n>0$,
\begin{align*}
|w_j^{\theta_n}(x)|
\leq&\big(G_j^{\theta_n}\ast f_j^{\theta_n}\big)(x)\ \ \text{in}\ \R^3, \ \ \text{where}\ f_j^{\theta_n}\ \text{is\ as\ in}\ \eqref{3.9}.
\end{align*}
We thus calculate from  \eqref{3.13} and \eqref{g1}  that  for $ j=1, \cdots, N$,
\begin{align*}
&\|w_j^{\theta_n}(x)\|_\infty \leq \sup\limits_{x\in\R^3}\Big[C_g\int_{|x-y|\leq1}|x-y|^{-2}f_j^{\theta_n}(y) dy
+C_g\int_{|x-y|>1}|x-y|^{-2}f_j^{\theta_n}(y) dy\\
\leq&C_g\big\|f_j^{\theta_n}\big\|_5\Big(\int_{|y|\leq1}|y|^{-\frac{5}{2}}dy\Big)^{\frac{4}{5}}
+C_g\big\|f_j^{\theta_n}\big\|_{5/2}\Big(\int_{|y|>1}|y|^{-\frac{10}{3}}dy\Big)^{\frac{3}{5}} \le C<\infty
\end{align*}
uniformly\ for  sufficiently\ large\ $n>0$, and
\begin{align*}
\lim\limits_{|x|\to\infty}|w_j^{\theta_n}(x)|
\leq&\lim\limits_{R\to\infty}\lim\limits_{|x|\to\infty}\int_{\R^3}G_j^{\theta_n}(x-y)f_j^{\theta_n}(y) dy\\
\leq&\lim\limits_{R\to\infty}\lim\limits_{|x|\to\infty}\Big(\int_{|y|\leq R} G_j^{\theta_n}(x-y)f_j^{\theta_n}(y) dy+\int_{|y|\geq R} G_j^{\theta_n}(x-y)f_j^{\theta_n}(y)dy\Big)\\[1.5mm]
=&\lim\limits_{R\to\infty}\lim\limits_{|x|\to\infty}\int_{|y|\geq R} G_j^{\theta_n}(x-y)f_j^{\theta_n}(y)dy\\[1.5mm]
\leq
&\lim\limits_{R\to\infty}\lim\limits_{|x|\to\infty}\int_{\{y:\, |y|\geq R, \, |x-y|\leq 1\}}
G_j^{\theta_n}(x-y)f_j^{\theta_n}(y)dy\\
&+\lim\limits_{R\to\infty}\lim\limits_{|x|\to\infty}\int_{\{y:\, |y|\geq R, \, |x-y|>1\}}
G_j^{\theta_n}(x-y)f_j^{\theta_n}(y)dy\\
\leq&C_g(8\pi)^{\frac{4}{5}}\lim\limits_{R\to\infty}\big\|f_j^{\theta_n}\big\|_{L^5(B_R^c)}+C_g(12\pi)^{\frac{3}{5}}\lim\limits_{R\to\infty}\big\|f_j^{\theta_n}\big\|_{L^{\frac{5}{2}}(B_R^c)} =0
\end{align*}
{uniformly\ for \ sufficiently\ large}\ $n>0$. Hence,  \eqref{3.23} and \eqref{3.24} hold true.

Applying \eqref{2.50}, \eqref{3.23},  \eqref{3.24} and \eqref{3.49}, the similar proof of \cite[(3.25)]{CGNO-25} then yields that  \eqref{3.31} holds true.  The proof of Lemma \ref{lem3.4} is therefore complete. \qed

\vspace{.20cm}

\noindent\textbf{Proof of Theorem \ref{th1} (ii).}  For any fixed $m>0$ and $2\leq N\in\mathbb{N}$, according to \cite[Lemma 3.3]{CGNO-25}, it suffices to prove that there exists a small constant $\theta^*=\theta^*(m, N)>0$ such that $I^{\rm HF}_{m,\kappa^{\rm HF}_N,\theta}(N)$ does not admit any minimizer for any $0<\theta\leq \theta^*$.

To prove the above result, on the contrary, suppose there exists a sequence $\{\theta_n\}\subset(0, 1)$ such that $I^{\rm HF}_{m,\kappa^{\rm HF}_N,\theta_n}(N)$ admits minimizers for any $n>0$, where $\theta_n\searrow0$ as $n\to\infty$.
Let $\gamma_{\theta_n}=\sum_{j=1}^N\big|u_j^{\theta_n}\big\rangle \big\langle u_j^{\theta_n}\big|$ be a minimizer of $I^{\rm HF}_{m,\kappa^{\rm HF}_N,\theta_n}(N)$ satisfying \eqref{3.11b}. Since   $\sup\limits_{n}\Tr\big(\sqrt{-\Delta}\, \gamma_{\theta_n}\big)>0$ holds true, we next proceed the rest proof by  separately ruling out the following two different cases: $\sup\limits_{n}\Tr\big(\sqrt{-\Delta}\, \gamma_{\theta_n}\big)=\infty$ and $\sup\limits_{n}\Tr\big(\sqrt{-\Delta}\, \gamma_{\theta_n}\big)<\infty$.

$Case\ 1$:   $\sup\limits_{n}\Tr\big(\sqrt{-\Delta}\, \gamma_{\theta_n}\big)=\infty$. In this case, set $\gamma^*_{\theta_n}:=\sum_{j=1}^N|w^{\theta_n}_j\rangle\langle w_j^{\theta_n}|$, where $w_j^{\theta_n}(x)=\epsilon_{\theta_n}^{\frac{3}{2}}u_j^{\theta_n}\big(\epsilon_{\theta_n}(x+z_n)\big)$ and $\epsilon_{\theta_n}>0$ are as in Lemma \ref{lem2.2}. Since $e^{-s}\leq 1-s+\frac{s^2}{2}$ holds for any $s\geq0$, we  calculate that 
\begin{equation*}\label{3.29}
\begin{split}
&\Tr\Big[\big(\sqrt{-\Delta+m^2}-\sqrt{-\Delta}\, \big)\gamma_{\theta_n}\Big]\nonumber\\[1mm]
=&\epsilon_{\theta_n}^{-1}\Tr\Big[\big(\sqrt{-\Delta+\epsilon_{\theta_n}^{2}m^2}-\sqrt{-\Delta}\, \big)\gamma^*_{\theta_n}\Big]\\
=&\Tr\frac{\epsilon_{\theta_n}m^2\, \gamma^*_{\theta_n}}{\sqrt{-\Delta+\epsilon_{\theta_n}^{2}m^2}+\sqrt{-\Delta}}
\geq \Tr\frac{\epsilon_{\theta_n}m^2\, \gamma^*_{\theta_n}}{2\sqrt{-\Delta+\epsilon^2_{\theta_n}m^2}},\ \ \forall\ n>0,\nonumber
\end{split}
\end{equation*}
and
\begin{equation*}\label{3.30}
\begin{split}
&\iinte\frac{1-e^{-\theta_n|x-y|}}{|x-y|} \Big[\rho_{\gamma_{\theta_n}}(x)\rho_{\gamma_{\theta_n}}(y)-|\gamma_{\theta_n}(x,y)|^2\Big]dxdy\nonumber\\
=&\iinte\frac{1-e^{-\theta_n\epsilon_{\theta_n}|x-y|}}{\epsilon_{\theta_n}|x-y|} \Big[\rho_{\gamma^*_{\theta_n}}(x)\rho_{\gamma^*_{\theta_n}}(y)-|\gamma^*_{\theta_n}(x,y)|^2\Big]dxdy\nonumber\\[1mm]
\geq&\iinte\frac{\epsilon_{\theta_n}\theta_n|x-y|-\frac{1}{2}\epsilon^2_{\theta_n}\theta^2_n|x-y|^2}{\epsilon_{\theta_n}|x-y|} \Big[\rho_{\gamma^*_{\theta_n}}(x)\rho_{\gamma^*_{\theta_n}}(y)-|\gamma^*_{\theta_n}(x,y)|^2\Big]dxdy\\[1mm]
=&\theta_n\big(N^2-N\big)-\frac{\epsilon_{\theta_n}\theta_n^2}{2}\iinte|x-y| \Big[\rho_{\gamma^*_{\theta_n}}(x)\rho_{\gamma^*_{\theta_n}}(y)-|\gamma^*_{\theta_n}(x,y)|^2\Big]dxdy,\ \ \forall\ n>0.\nonumber
\end{split}
\end{equation*}
By the definition of $\kappa^{\rm HF}_N$, we then derive that
\begin{align}\label{3.27}
I^{\rm HF}_{m,\kappa^{\rm HF}_N,\theta_n}(N)
=&\mathcal{E}^{\rm HF}_{m,\kappa^{\rm HF}_N,\theta_n}(\gamma_{\theta_n})\nonumber\\
=&-mN+\Tr\Big[\big(\sqrt{-\Delta+m^2}-\sqrt{-\Delta}\, \big)\gamma_{\theta_n}\Big]\nonumber\\[1mm]
&+\Tr\big(\sqrt{-\Delta}\, \gamma_{\theta_n}\big)-\frac{\kappa^{\rm HF}_N}{2}\Big(D(\rho_{\gamma_{\theta_n}},\rho_{\gamma_{\theta_n}})-\text{Ex}(\gamma_{\theta_n})\Big)\\[1mm]
&+\frac{\kappa^{\rm HF}_N}{2}\iinte\frac{1-e^{-\theta_n|x-y|}}{|x-y|} \Big[\rho_{\gamma_{\theta_n}}(x)\rho_{\gamma_{\theta_n}}(y)-|\gamma_{\theta_n}(x,y)|^2\Big]dxdy\nonumber\\
\geq&-mN+\frac{\epsilon_{\theta_n}m^2}{2}\Tr\frac{\gamma^*_{\theta_n}}{\sqrt{-\Delta+\epsilon_{\theta_n}^2m^2}}+\frac{\theta_n \kappa^{\rm HF}_N}{2}\big(N^2-N\big)\nonumber\\
&-\frac{\kappa^{\rm HF}_N\epsilon_{n}\theta_n^2}{4}\iinte|x-y| \rho_{\gamma^*_{\theta_n}}(x)\rho_{\gamma^*_{\theta_n}}(y)dxdy,\ \ \forall\ n>0.\nonumber
\end{align}

On the other hand, let $\gamma^*=\sum_{j=1}^N|w_j\rangle\langle w_j|$ be given by Lemma \ref{lem2.2}.  Since $\gamma^*\in\mathcal{P}_N$ is an optimizer of $\kappa^{\rm HF}_N$, we can choose $\gamma^{(N)}=\gamma^*$ and $t_n=\epsilon_{\theta_n}^{-2}>0$ into \eqref{3.26}. This implies from \eqref{3.26} and \eqref{3.27}  that
\begin{equation}\label{3.51a}
\begin{split}
\frac{\epsilon_{\theta_n}m^2}{2}\Tr\frac{\gamma^*}{\sqrt{-\Delta}}
\geq&\frac{m^2}{2}\Tr\frac{\gamma^*_{\theta_n}}{\sqrt{-\Delta+\epsilon_{\theta_n}^2m^2}}\\
&-\frac{\kappa^{\rm HF}_N\theta_n^2}{4}\iinte|x-y|\rho_{\gamma^*_{\theta_n}}(x)\rho_{\gamma^*_{\theta_n}}(y)dxdy,\ \ \forall\ n>0.
\end{split}
\end{equation}
By Sobolev's embedding  theorem and  Plancherel's theorem, we deduce from \eqref{2.62a} and \eqref{3.31} that for $    j=1,\cdots,N,$
\begin{eqnarray*}
|(w_j^{\theta_n})^\wedge|^2\to  |\hat{w}_j|^2\ \ \mathrm{strongly\ in}\, \ L^{1}(\R^3)\ \ \text{as}\ \ n\to\infty,
\end{eqnarray*}
and
\begin{equation*}
\big\{(w_j^{\theta_n})^\wedge\big\}_n \ \ \text{is\ bounded\ uniformly\ in}\ L^\infty(\R^3, \C).
\end{equation*}
Applying the interpolation inequality, we then derive from above  that for $ j=1,\cdots,N,$
\begin{eqnarray}\label{3.51}
|(w_j^{\theta_n})^\wedge|^2\to  |\hat{w}_j|^2\ \ \mathrm{strongly\ in}\, \ L^{r}(\R^3)\ \ \text{as}\ \ n\to\infty,\ \ \forall\ r\in[1, \infty).
\end{eqnarray}
Applying H\"older's inequality and the dominated convergence theorem, we hence calculate from \eqref{3.51} that
\begin{align}\label{5.53}
\Big|\Tr\frac{\gamma_{\theta_n}^*}{\sqrt{-\Delta}}-\Tr\frac{\gamma^*}{\sqrt{-\Delta}}\Big|
\leq&\sum_{j=1}^N\Big|\inte|\xi|^{-1}\Big(|(w_j^{\theta_n})^\wedge(\xi)|^2-|\hat{w}_j(\xi)|^2\Big)d\xi\Big|\nonumber\\
\leq&\sum_{j=1}^N\Big|\int_{|\xi|\geq1}\Big(|(w_j^{\theta_n})^\wedge(\xi)|^2-|\hat{w}_j(\xi)|^2\Big)d\xi\Big|\nonumber\\
&+\sum_{j=1}^N\Big|\int_{|\xi|\leq1}|\xi|^{-1}\Big(|(w_j^{\theta_n})^\wedge(\xi)|^2-|\hat{w}_j(\xi)|^2\Big)d\xi\Big|\\
\leq&o(1)+\big\||\xi|^{-1}\big\|_{L^2(B_1)}\sum_{j=1}^N\big\||(w_j^{\theta_n})^\wedge|^2- |\hat{w}_j|^2\big\|_2\nonumber\\
=&o(1)\ \ \text{as}\ \ n\to\infty,\nonumber
\end{align}
and
\begin{align}\label{5.54}
&\iinte|x-y|\Big(\rho_{\gamma_{\theta_n}^*}(x)\rho_{\gamma_{\theta_n}^*}(y)-|\gamma_{\theta_n}^*(x,y)|^2\Big)dxdy\\
=&\iinte|x-y|\Big(\rho_{\gamma^*}(x)\rho_{\gamma^*}(y)-|\gamma^*(x,y)|^2\Big)dxdy+o(1)\ \ \text{as}\ \ n\to\infty.\nonumber
\end{align}
It thus yields from \eqref{3.51a} that
\begin{equation*}\label{3.53}
\begin{split}
0=&\lim\limits_{n\to\infty}\frac{\epsilon_{\theta_n}m^2}{2}\Tr\frac{\gamma^*}{\sqrt{-\Delta}}\\
\geq&\lim\limits_{n\to\infty}\Big[\frac{m^2}{2}\Tr\frac{\gamma^*_{\theta_n}}{\sqrt{-\Delta+\epsilon_{\theta_n}^2m^2}}-\frac{\kappa^{\rm HF}_N\theta_n^2}{4}\iinte|x-y|\rho_{\gamma^*_{\theta_n}}(x)\rho_{\gamma^*_{\theta_n}}(y)dxdy\Big]\\
=&\frac{m^2}{2}\Tr\frac{\gamma^*}{\sqrt{-\Delta}}>0,
\end{split}
\end{equation*}
a contradiction. Therefore, Case 1 cannot occur.

$Case\ 2$:   $
\sup\limits_{n}\Tr\big(\sqrt{-\Delta}\gamma_{\theta_n}\big)<\infty.
$
In this case, recall from \eqref{3.11b} that
\begin{equation}\label{3.8a}
\begin{split}
&\big(\sqrt{-\Delta+m^2}-m\big)u_j^{\theta_n}-\kappa^{\rm HF}_N\inte\frac{e^{-\theta_n|x-y|}\rho_{\gamma_{\theta_n}}(y)}{|x-y|}dy\, u_j^{\theta_n}\\&+\kappa^{\rm HF}_N\inte\frac{e^{-\theta_n|x-y|}\, \gamma_{\theta_n}(x,y)}{|x-y|}u_j^{\theta_n}(y)dy
= \mu_j^{\theta_n}u_j^{\theta_n}\ \ \text{in}\,\ \R^3,\ \ j=1,\cdots, N,
\end{split}
\end{equation}
where $\mu_1^{\theta_n}\leq\cdots\leq\mu_N^{\theta_n}<0$.
It then follows from \eqref{eq:GN-HF} that  there exists a constant $\mu_j\leq0$ such that up to a subsequence  if necessary,
\begin{equation}\label{3.10a}
-\infty<\lim\limits_{n\to\infty}\mu_j^{\theta_n}=\mu_j\leq0,\ \ j=1,\cdots, N.
\end{equation}
Moreover, since it yields from \eqref{3.6} that $\lim\limits_{n\to\infty}I^{\rm HF}_{m,\kappa^{\rm HF}_N,\theta_n}(N)<0$, the same argument of \eqref{2.52} yields that  there exist a sequence $\{\tilde{z}_n\}\subset\R^3$ and a nonzero system $(u_1,\cdots, u_N)$ $\in \big(H^{\frac{1}{2}}(\R^3, \C)\big)^N$ such that up to a subsequence  if necessary,
\begin{eqnarray}\label{3.11a}
\tilde{u}_j^{\theta_n}(x):=u_j^{\theta_n}(x+\tilde{z}_n)\rightharpoonup  u_j \ \mathrm{weakly\ in}\, \ H^{\frac{1}{2}}(\R^3,\C)\ \text{as}\ n\to\infty, \   j=1,\cdots,N.
\end{eqnarray}
Consequently, similar to  \eqref{2.56} and \eqref{2.57}, we derive from \eqref{3.8a}--\eqref{3.11a}  that
\begin{equation*}\label{3.8b}
\begin{split}
&\big(\sqrt{-\Delta+m^2}-m\big)u_j-\kappa^{\rm HF}_N\inte\frac{\rho_{\gamma}(y)}{|x-y|}dy\, u_j\\
&+\kappa^{\rm HF}_N\inte\frac{ \gamma(x,y)}{|x-y|}u_j(y)dy
=\mu_ju_j\ \ \text{in}\,\ \R^3,\ \ j=1,\cdots, N,
\end{split}
\end{equation*}
and 
\begin{equation*}\label{3.13a}
\begin{split}
&\Tr\big(\sqrt{-\Delta+m^2}\, \gamma\big)-\frac{3m}{2}\sum_{j=1}^N\inte |u_j|^2dx-\frac{5\kappa^{\rm HF}_N}{4}\Big[D(\rho_\gamma, \rho_\gamma)-\text{Ex}(\gamma)\Big]\\
&+\frac{m^2}{2}\Tr\frac{\gamma}{\sqrt{-\Delta+m^2}}=\frac{3}{2}\sum_{j=1}^N\mu_j\inte |u_j|^2dx,
\end{split}
\end{equation*}
where $\gamma:=\sum_{j=1}^N|u_j\rangle\langle u_j|\neq0.$ 
This gives that
\begin{equation}\label{3.15a}
\begin{split}
\Tr\big(\sqrt{-\Delta+m^2}\, \gamma\big)+\Tr\frac{m^2\, \gamma}{\sqrt{-\Delta+m^2}}=\frac{\kappa^{\rm HF}_N}{2}\Big[D(\rho_\gamma, \rho_\gamma)-\text{Ex}(\gamma)\Big].
\end{split}
\end{equation}

Define
\begin{equation}\label{bn}
\begin{split}
\frac{c^*_N}{2}:=\inf\limits_{\gamma'\in\mathcal{R}_N}F(\gamma'):=\inf\limits_{\gamma'\in\mathcal{R}_N}\frac{\|\gamma'\|\Big[\mathrm{Tr}(\sqrt{-\Delta+m^2}\,  \gamma')+\Tr\frac{m^2\, \gamma'}{\sqrt{-\Delta+m^2}}\Big]}{D(\rho_{\gamma'},\rho_{\gamma'})-\mathrm{Ex}(\gamma')},
\end{split}
\end{equation}
where  $\mathcal{R}_N$ is as in \eqref{rn1}. The definitions of $\kappa^{\rm HF}_N$ and $c^*_N$ yield that $\kappa^{\rm HF}_N\leq c_N^*$ holds for any $2\leq N\in\mathbb{N}$. On the other hand, take $\gamma^{(N)}\in\mathcal{R}_N$ as an optimizer of $\kappa^{\rm HF}_N$, and denote $\gamma_t^{(N)}(x, y)=t^3\gamma^{(N)}(tx, ty)$ for $t>0$. We then  have
\begin{equation*}
\begin{split}
\frac{c^*_N}{2}\leq\lim\limits_{t\to\infty}F\big(\gamma_t^{(N)}\big)
=&\lim\limits_{t\to\infty}\frac{\|\gamma^{(N)}\|\Big[\mathrm{Tr}\big(\sqrt{-\Delta+m^2t^{-2}}\,  \gamma^{(N)}\big)+t^{-2}\Tr\frac{m^2\, \gamma^{(N)}}{\sqrt{-\Delta+m^2t^{-2}}}\Big]}{D(\rho_{\gamma^{(N)}},\rho_{\gamma^{(N)}})-\mathrm{Ex}(\gamma^{(N)})}\\
=&\frac{\|\gamma^{(N)}\|\mathrm{Tr}(\sqrt{-\Delta}\,  \gamma^{(N)})}{D(\rho_{\gamma^{(N)}},\rho_{\gamma^{(N)}})-\mathrm{Ex}(\gamma^{(N)})}=\frac{\kappa_N^{\rm HF}}{2},
\end{split}
\end{equation*}
which thus yields that
\begin{equation}\label{3.16a}
\begin{split}
c^*_N=\kappa^{\rm HF}_N.
\end{split}
\end{equation}

As a consequence of \eqref{3.15a}--\eqref{3.16a},  we obtain that
\begin{equation*}\label{3.18}
\begin{split}
&\|\gamma\|\Big[\mathrm{Tr}\big(\sqrt{-\Delta+m^2}\,  \gamma\big)+\Tr\frac{m^2\, \gamma}{\sqrt{-\Delta+m^2}}\Big]\\
\geq&\frac{c^*_N}{2}\Big[D(\rho_{\gamma},\rho_{\gamma})-\mathrm{Ex}(\gamma)\Big] =\Tr\big(\sqrt{-\Delta+m^2}\, \gamma\big)+\Tr\frac{m^2\, \gamma}{\sqrt{-\Delta+m^2}}\\
\geq& \|\gamma\|\Big[\Tr\big(\sqrt{-\Delta+m^2}\, \gamma\big)+\Tr\frac{m^2\, \gamma}{\sqrt{-\Delta+m^2}}\Big],
\end{split}
\end{equation*}
where the last inequality follows from the fact that $$\|\gamma\|=\max\limits_{j}\|u_j\|_2^2\leq\max\limits_{j}\lim\limits_{n\to\infty}\|\tilde{u}_j^{\theta_n}\|_2^2=1.$$
This shows that $\|\gamma\|=1$ and $\gamma$ is an optimizer of (\ref{bn}), and hence
$$
\frac{c^*_N}{2}=\frac{\|\gamma\|\Big[\mathrm{Tr}\big(\sqrt{-\Delta+m^2}\,  \gamma\big)+\Tr\frac{m^2\, \gamma}{\sqrt{-\Delta+m^2}}\Big]}{D(\rho_{\gamma},\rho_{\gamma})-\mathrm{Ex}(\gamma)}>\frac{\|\gamma\|\mathrm{Tr}\big(\sqrt{-\Delta}\,  \gamma\big)}{D(\rho_{\gamma},\rho_{\gamma})-\mathrm{Ex}(\gamma)}\geq \frac{\kappa_N^{\rm HF}}{2}.
$$
which however contradicts with \eqref{3.16a}. We thus conclude that Case 2 cannot occur, either. The proof of Theorem \ref{th1} (ii) is  complete.\qed

\appendix

\section{Proof of Propositions \ref{prop:Chandrasekhar} and  \ref{prop:min-Gtheta}}\label{app:pf-Chandra-Yukawa}
The purpose of this appendix is to prove Propositions \ref{prop:Chandrasekhar} and  \ref{prop:min-Gtheta}.
\vspace{.15cm}

\noindent\textbf{Proof of Proposition \ref{prop:Chandrasekhar}.}  Let $\lambda^{\text{HFB}}:=\lambda^{\text{HFB}}(\kappa)>0$ be given by \cite[Proposition 2.1]{LenLew-10}. We remark that Proposition \ref{prop:Chandrasekhar}
was already proved in \cite[Proposition 2.1]{LenLew-10} for the case $\theta=0$. Since $\cE^{\rm HFB}_{0,\kappa,\theta}(\gamma,\alpha)\geq \cE^{\rm HFB}_{0,\kappa,0}(\gamma,\alpha)$, it suffices to prove Proposition \ref{prop:Chandrasekhar} (ii) for $\theta>0$. 
For $\lambda>\lambda^{\rm HFB}$, since $I^{\rm HFB}_{0,\kappa,0}(\lambda)=-\infty$, there exists  $(\gamma,\alpha)\in\cK_\lambda$ such that $\cE^{\rm HFB}_{0,\kappa,0}(\gamma,\alpha)<0$ holds for $\lambda>\lambda^{\rm HFB}$. 
If we denote $ \gamma_{\beta}(x,y):=\beta^{3}\gamma(\beta x,\beta y)$ and $\alpha_{\beta}(x,y):=\beta^{3}\alpha(\beta x,\beta y)$ for  $\beta>0$, then using
\begin{align}\label{a.8}
\sup_{\beta>0}&\Big(\beta\, \big|\cE^{\rm HFB}_{0,\kappa,\beta^{-1}\theta}(\gamma,\alpha)-\cE^{\rm HFB}_{0,\kappa,0}(\gamma,\alpha)\big|\Big)\nonumber\\
&\leq\sup_{\beta>0}\iinte\left|\frac{1-e^{-\beta^{-1}\theta|x-y|}}{\beta^{-1}|x-y|}\right|\big(\rho_{\gamma}(x)\rho_{\gamma}(y)+|\alpha(x,y)|^{2}\big)dxdy\\
&\leq\theta\iinte\big(\rho_{\gamma}(x)\rho_{\gamma}(y)+|\alpha(x,y)|^{2}\big)dxdy<\infty\nonumber
\end{align}
we conclude that 
\begin{align*}
\cE^{\rm HFB}_{0,\kappa,\theta}(\gamma_{\beta},\alpha_{\beta})
&=\beta\cE^{\rm HFB}_{0,\kappa, \beta^{-1}\theta}(\gamma,\alpha)\nonumber\\
&\leq\beta\cE^{\rm HFB}_{0,\kappa,0}(\gamma,\alpha)+\sup_{\beta>0}\beta\big|\cE^{\rm HFB}_{0,\kappa,\beta^{-1}\theta}(\gamma,\alpha)-\cE^{\rm HFB}_{0,\kappa,0}(\gamma,\alpha)\big|\\
&
\rightarrow -\infty\ \ \text{ as }\ \ \beta\rightarrow\infty.\nonumber
\end{align*}
Thus $I^{\rm HFB}_{0,\kappa,\theta}(\lambda)=-\infty$ holds for $\lambda>\lambda^{\rm HFB}$.  $\hfill\qed$

The rest part of this Appendix is to prove Proposition \ref{prop:min-Gtheta}, for which we need the following two lemmas.

\begin{lemma}\label{lem:Gtheta-eigen}
Let $\fH$ be either $L_{\rm odd}^{2}(\R^{3})$ or $L^{2}(\R^{3})$.  For $m>0$ and $0<\kappa<4/\pi$, consider the pseudo-relativistic Schr\"odinger operator with Yukawa potential $K_{\theta}:=\sqrt{-\Delta+m^{2}}-\frac{\kappa e^{-\theta|x|}}{2|x|}$ defined in the sense of quadratic forms on $L^2(\R^3)$ with form domain $H^{\frac{1}{2}}(\R^3)$. Then we have
\begin{enumerate}
\item[$(i)$] For every $\theta\geq 0$, we have
\begin{align}\label{d.1}
\sigma_{\fH,\rm ess}(K_{\theta})=[m,\infty),
\end{align}
where $\sigma_{\fH,\rm ess}(K_{\theta})$ denotes the essential spectrum of $K_{\theta}$ on  $\fH$.

\item[$(ii)$] There exists a constant $\theta^{*}:=\theta^{*}(m,\kappa)>0$ such that 
\begin{align*}
m(1-\kappa\pi/4)<\beta_{\theta}:=\inf\sigma_{\fH}(K_{\theta})<m,\ \   \forall\  \theta\in [0,\theta^{*}],
\end{align*}
and there is a real-valued eigenfunction $f_{\theta}$ for $K_{\theta}$ on $\fH$.  Moreover, $\beta_{\theta}$ is continuous and non-increasing in $\theta\in[0,\theta^{*}]$.
\end{enumerate}
\end{lemma}

\noindent \textbf{Proof.}
We shall only prove the case where $\fH=L^{2}(\R^{3})$, since the case $\fH=L_{\rm odd}^{2}(\R^{3})$ can be established  in a similar way.  Throughout this proof, for simplicity, we drop the symbol $\fH$ from all subscripts below.

(i).  We first claim that  $\sigma(K_{\theta})\subseteq [0,\infty)$. Indeed,  since $0<\kappa<4/\pi$, by the following Hardy-Kato inequality (cf. \cite[Chapter V.5.4]{hkt})
\begin{align}\label{hkt}
\frac{e^{-\theta|x|}}{|x|} =W_\theta(x) \le \frac{1}{|x|}\leq\frac{\pi}{2}\sqrt{-\Delta}\ \  \, \text{on}\ \ L^2(\R^3),
\end{align}
we have
\begin{align*}
\langle\psi,K_{\theta}\psi\rangle=\big\langle\psi,\sqrt{-\Delta+m^{2}}\psi\big\rangle-\frac{\kappa}{2}\langle\psi,\, W_{\theta}\psi\rangle\geq \left(1-\frac{\kappa\pi}{4}\right)\langle\psi,\sqrt{-\Delta}\psi\rangle\geq 0
\end{align*}
for all $\psi\in\sD(K_{\theta})$, 
where $K_{\theta}=K-\frac{\kappa}{2}W_{\theta}$ with $K:=\sqrt{-\Delta+m^{2}}$. 
If $C:=(K_{\theta}+1)^{-1}-(K+1)^{-1}$ is compact, then we further have $\sigma_{\rm ess}(K_{\theta})=\sigma_{\rm ess}(K)=[m,\infty)$, and thus \eqref{d.1} holds true, which therefore completes the proof of Lemma \ref{lem:Gtheta-eigen} (i).

The rest is to prove that $C:=(K_{\theta}+1)^{-1}-(K+1)^{-1}$ is compact.  Actually, the second resolvent identity yields that
\begin{align*}
C=\frac{\kappa}{2}(K+1)^{-1}Vw_{\theta}V(K_{\theta}+1)^{-1},
\end{align*}
where $V(x):=|x|^{-1/2}$ and $w_{\theta}(x):=e^{-\theta|x|}$. To prove the compactness of $C$, since $w_{\theta}$ defines a bounded operator on $L^{2}(\R^{3})$, it suffices to prove that $V(K_{\theta}+1)^{-1}$ is bounded on $L^{2}(\R^{3})$ and $(K+1)^{-1}V$ is compact on $L^{2}(\R^{3})$.

Following \cite{Herbst-77}, we first note that $(K_{\theta}+1)^{-1}$ maps $L^{2}(\R^{3})$ into $ \sD(K_{\theta})\subseteq \sD(K_{\theta}^{1/2})=\sD(K^{1/2})\subseteq \sD(V)$. By the closed graph theorem, this implies that $V(K_{\theta}+1)^{-1}$ extends to a bounded operator on $L^{2}(\R^{3})$. On the other hand, since $(K+1)^{-1}V$ can be defined as the adjoint of the operator $V(K+1)^{-1}$, as before it is also bounded. Hence, in order to further show that  $(K+1)^{-1}V$ is compact on $L^{2}(\R^{3})$, it next suffices to show that $V$ is relatively compact with respect to $K$.

Let $\{u_{n}\}\subset L^{2}(\R^{3})$ satisfy $u_{n}\rightharpoonup 0$ weakly in $L^{2}(\R^{3})$ as $n\to\infty$. Then $\psi_{n}:=(K+1)^{-1}u_{n}\rightharpoonup 0$ weakly in $\sD(K)=H^{1}(\R^{3})$ as $n\to\infty$.  By Sobolev's compact embedding theorem,  we have
\begin{align}\label{d.4}
\psi_{n}\rightarrow 0\  \ \text{ strongly in }L^{s}_{loc}(\R^{3})\ \  \text{as}\ \ n\to\infty, \ \  2\leq s<6.
\end{align}
Note that  $V=V_{R,1}+V_{R,2}$ holds for any $R>0$, where $V_{R,1}:=V1_{|x|\leq R}\in L^{p}(\R^{3})$ holds for any  $1\leq p\leq 4$, and $V_{R,2}:=V1_{|x|\geq R}\in L^{\infty}(\R^{3})$ satisfies $\|V_{R,2}\|_{L^{\infty}}\leq R^{-1/2}$.
By  H\"older's  inequality, we then get from \eqref{d.4} that for any $R>0$,
\begin{align}\label{d.5}
\lim\limits_{n\to\infty}\|V_{R,1}\psi_{n}\|_{L^{2}}
&\leq \lim\limits_{n\to\infty}\|V_{R,1}\|_{L^{4}}\, \big\|1_{|x|\leq R}\psi_{n}\big\|_{L^{4}}=0.
\end{align}
Moreover, since  $\{\psi_{n}\}$ is uniformly bounded in $L^{2}(\R^{3})$, we have $\sup\limits_{n}\|\psi_{n}\|_{L^{2}}\leq C<\infty$, and thus
\begin{align}\label{d.6}
\lim\limits_{n\to\infty}\|V_{R,2}\psi_{n}\|_{L^{2}}
&\leq \lim\limits_{n\to\infty}\|V_{R,2}\|_{L^{\infty}}\|\psi_{n}\|_{L^{2}}\leq CR^{-1/2}\rightarrow 0\ \   \text{as}\ \ R\to\infty.
\end{align}
As a consequence of \eqref{d.5} and \eqref{d.6}, we obtain that up to a subsequence if necessary, $V\psi_{n}\rightarrow 0$ strongly in $L^{2}(\R^{3})$ as $n\to\infty$. This implies that $V$ is relatively compact with respect to $K$, and we are done.

(ii). Since $\sup_{s\geq 0}\frac{1-e^{-s}}{s}=1$, we have
$$
K_{\theta}=K_{0}+\frac{\kappa(1-e^{-\theta|x|})}{2|x|}\leq K_{0}+\frac{\kappa\theta}{2},
$$
which shows that
\begin{align}\label{inf-spec-esti}
\inf\sigma(K_{0})\leq\beta_{\theta}:=\inf\sigma(K_{\theta})&\leq \inf\sigma(K_{0})+\frac{\kappa\theta}{2}.
\end{align}
By the min-max principle and the operator inequality $K_{0}\leq -\frac{1}{2m}\Delta+m-\frac{\kappa}{2|x|}$, we obtain that $C:= C(m,\kappa):=\inf\sigma(K_{0})<m$.  This then implies from \eqref{hkt} and  \eqref{inf-spec-esti} that
\begin{align}\label{d.8}
m(1-\kappa\pi/4)<\beta_{\theta}&\leq C+\frac{\kappa\theta}{2}<m,\ \   \forall\ 0\leq \theta\leq \theta^{*}:=(m-C)\kappa^{-1}.
\end{align}
Since $\sigma_{\rm ess}(K_{\theta})=[m,\infty)$ by \eqref{d.1}, we conclude from \eqref{d.8} that $\beta_{\theta}$ must be an eigenvalue of $K_{\theta}$.  Moreover, one can calculate that for $\theta_{1},\theta_{2}\in [0,\theta^{*}]$,
\begin{align*}
\|K_{\theta_{1}}-K_{\theta_{2}}\|\leq \frac{\kappa}{2}|\theta_{1}-\theta_{2}|,\ \   K_{\theta_{1}}-K_{\theta_{2}}\geq 0\ \, \text{ if }\theta_{1}\geq \theta_{2}.
\end{align*}
Applying the standard spectral theory, we then obtain that the function $\theta\mapsto\beta_{\theta}$ is continuous and non-increasing in $[0,\theta^{*}]$. This therefore completes the proof of  Lemma \ref{lem:Gtheta-eigen}.  \qed

\begin{lemma}\label{lem:Ktheta-eigen-decay}
Let $\fH$ be either $L_{\rm odd}^{2}(\R^{3})$ or $L^{2}(\R^{3})$.  Suppose $u$ is an eigenfunction with eigenvalue $E<0$ of $T_{\theta}:=K_{\theta}-m$ in $\fH$, where $\theta\in [0,\theta^{*}]$ and $\theta^{*}$ is the critical Yukawa parameter given in Lemma \ref{lem:Gtheta-eigen}.  Then we have
\begin{enumerate}
\item[$(i)$] $\chi u\in H^{s}(\R^{3})$ holds for any $s\geq 1$ and $\chi\in C_{0}^{\infty}(\Omega)$, where $\Omega\subset\R^{3}\setminus\{0\}$ denotes any bounded domain with smooth boundary.

\item[$(ii)$] $u\in \sD(e^{\beta|\cdot|})$ holds for $\beta<\nu_{E}$, where $\nu_{E}:=\sqrt{|m^{2}-(E+m)^{2}|}$.

\item[$(iii)$] For each $R>0$ and $\beta<\nu_{E}$, there exists a  constant $C_{R,\beta}>0$ such that
\begin{align*}\label{eigen-exp-decay}
|u(x)|&\leq C_{R,\beta}e^{-\beta|x|}\quad\text{ for }\ |x|\geq R.
\end{align*}
\end{enumerate}
\end{lemma}

\noindent \textbf{Proof.}
Since the proof of this lemma is  analogous to \cite{DalSoeSto-08} with minor modifications,   we omit the details  for simplicity.  We also refer to \cite[p.196]{RB4} for a beautiful proposition due to O'Connor, which is the key of the argument in \cite{DalSoeSto-08}.\qed

We are now ready to complete the proof of Proposition \ref{prop:min-Gtheta}.
\vspace{.20cm}

\noindent \textbf{Proof of Proposition \ref{prop:min-Gtheta}.}
Since the case $\theta=0$ was proved in \cite{LenLew-10}, we focus on the case $\theta>0$. Consider $\theta\in (0,{\theta}^*]$, where $\theta^*=\theta^*(m,\kappa)>0$ is given by Lemma \ref{lem:Gtheta-eigen}. By Lemmas \ref{lem:Gtheta-eigen}, \ref{lem:Ktheta-eigen-decay}, and  the argument of \cite[Appendix C]{LenLew-10}, the relation \eqref{Gtheta-eigen} holds true for any  $\theta\in (0,{\theta}^*]$.
We next prove that  \eqref{min-Gtheta-relation} holds true.

By Lemma \ref{lem:Gtheta-eigen}, we can take a real-valued eigenfunction $f_{\theta}\in\fH$ of $K_{\theta}$ associated with the  eigenvalue $\beta_{\theta}$ for $\theta\in [0,\tilde{\theta}]$, where $\tilde{\theta}\in(0, \theta^{*}]$ is to be determined later.  For each $L>0$, we then construct
\begin{align*}
\alpha_{L}(x,y):=\lambda^{1/2}\chi_{L}(x)f_{\theta}(x-y)\chi_{L}(y)\Sigma,
\end{align*}
where  $\Sigma:=1$ if $q=1$, and
\begin{align*}
\Sigma:=\frac{1}{\sqrt{2}}
\begin{pmatrix}
0	&	1	&	0	&	\cdots	&	0\\
-1	&	0	&	0	&	\cdots	&	0\\
0	&	0	&	0	&	\cdots	&	0\\
\vdots	&	\vdots	&	\vdots	&	\ddots	&	\vdots\\
0	&	0	&	0	&	\cdots	&	0
\end{pmatrix}_{q\times q}
\ \   \text{if}\ \ q\geq2.
\end{align*}
Here 
$\chi_{L}(x):=L^{-3/4}\chi(x/L)$, $\chi\in C_{0}^{\infty}(\R^{3})$ is a   non-negative function satisfying  $\int_{\R^{3}}\chi^{4}dx=1$ and $\|\chi\|_{L^{\infty}}=1$.
Note from Lemma \ref{lem:Ktheta-eigen-decay} (iii) that  $f_{\theta}$ decays exponentially. This implies that  
\begin{equation}\label{a.12a}
\|\alpha_{L}\|\leq C\lambda^{1/2}\|f_{\theta}\|_{L^{1}}\|\chi_{L}\|_{L^{\infty}}^{2}=O(\lambda^{1/2}L^{-3/2})\ \   \text{as}\ \ L\to\infty,
\end{equation}
and hence  $\|\alpha_{L}\|\rightarrow 0$ as $L\rightarrow\infty$.  Thus, we can define $\gamma_{L}$ as the unique non-negative trace-class operator solving the following equation: for sufficiently large $L>0$,
\begin{align}\label{aL-gL-rel}
\gamma_{L}(1-\gamma_{L})&=\alpha_{L}\alpha_{L}^{*}.
\end{align}
Note that $\gamma_{L}\leq 2\alpha_{L}\alpha_{L}^{*}$ holds for  sufficiently large $L>0$. We then have $\|\gamma_{L}\|=O(\lambda L^{-3})$ as $L\to\infty$,  which further yields from \eqref{a.12a} and \eqref{aL-gL-rel} that 
\begin{align}\label{b.8}
\Tr(\gamma_{L}^{2})=O(\lambda L^{-3})\ \   \text{as}\ \  L\to\infty.
\end{align}
Since $\int_{\R^{3}}\chi^{4}dx=1$, we obtain that $\Tr(\alpha_{L}\alpha_{L}^{*})=\lambda+O(\lambda L^{-2})$ as $L\to\infty$. This thus implies from \eqref{aL-gL-rel} and \eqref{b.8} that
\begin{align}\label{b.9}
\|\gamma_{L}\|\leq \Tr(\gamma_{L})=\lambda+O(\lambda L^{-2})\ \ \text{as} \ \ L\to\infty.
\end{align}
The same argument of \cite[Appendix C]{LenLew-10} therefore gives that  for sufficiently large $L>0$,
\begin{align}\label{b.10}
(\gamma_{L},\alpha_{L})\in \cK,\ \   \cG_{\theta}(\gamma_L,\alpha_L)\leq \beta_\theta \lambda+O(\lambda L^{-2}),
\end{align}
where $\beta_\theta>0$ is given in Lemma \ref{lem:Gtheta-eigen}.

Since
\begin{align*}
	\Tr(K\gamma_{L}^{2})&\leq 4\Tr(K(\alpha_{L}\alpha_{L}^{*})^{2})\leq 4\|\alpha_{L}\|^{2}\Tr(K\alpha_{L}\alpha_{L}^{*})=O(\lambda L^{-3})\ \   \text{as}\ \ L\to\infty,
\end{align*}
where $K=\sqrt{-\Delta+m^{2}}$, it implies that
\begin{equation}\label{d.10}
	\Ex_{\theta}(\gamma_{L})\leq \Ex_{0}(\gamma_{L})=\int_{\R^{3}\times\R^{3}}\frac{|\gamma_{L}(x,y)|^{2}}{|x-y|}dxdy\leq \frac{\pi}{2}\Tr(K\gamma_{L}^{2})=O(\lambda L^{-3})
\end{equation}
as $L\to\infty$. Moreover,
setting
\begin{align*}
	\varphi_{L}(x):=&L^3\int_{\R^{3}}|f_{\theta}(x-zL)|^{2}\chi^{2}(z)dz,
\end{align*}
we then deduce  from \eqref{aL-gL-rel} and $\|f_\theta\|_{L^2}=1$ that  $\rho_{\gamma_{L}}\geq \rho_{\alpha_{L}\alpha_{L}^{*}}= L^{-3/2}\chi_{L}^{2}\varphi_{L}$ and
\begin{align*}
\varphi_{L}(Lx)\rightarrow\chi^{2}(x)\ \   \text{strongly in}\ \,  L^2(\R^3)\ \, \text{as} \ L\rightarrow\infty.
\end{align*}
Since  $\|\chi\|_{L^{4}}=1$,  we now calculate that
\begin{align}\label{d.11}
D_{\theta}(\rho_{\gamma_{L}},\rho_{\gamma_{L}})
&\geq D_{\theta}(\rho_{\alpha_{L}\alpha_{L}^{*}},\rho_{\alpha_{L}\alpha_{L}^{*}})
=\frac{\lambda^2}{L}D_{L\theta}\big(\chi^{2}\varphi_{L}(L\cdot),\chi^{2}\varphi_{L}(L\cdot)\big)\nonumber\\
&\geq \frac{\lambda^2}{L}D_{L\theta}(\chi^{4},\chi^{4})-\frac{2\lambda^2}{L}\|W_{L\theta}\|_{L^{2}}\|\varphi_{L}(L\cdot)-\chi^{2}\|_{L^{2}}\nonumber\\
&\geq\frac{\lambda^2}{L}D_{L\theta}(\chi^{4},\chi^{4})-o(\lambda^2L^{-3/2})\\
&=\frac{\lambda^2}{L}D(\chi^{4},\chi^{4})-\frac{\lambda^2}{L}\iinte\frac{1-e^{-L\theta|x-y|}}{|x-y|}\chi^{4}(x)\chi^{4}(y)dxdy-o(\lambda^2L^{-3/2})\nonumber\\
&\geq \frac{\lambda^2}{L}D(\chi^{4},\chi^{4})-\theta\lambda^2-o(\lambda^2L^{-3/2})\ \   \text{as}\ \ L\to\infty,\nonumber
\end{align}
where $W_{\theta}(x)=e^{-\theta|x|}/|x|$.
We hence deduce from \eqref{Gtheta-eigen} and \eqref{b.9}--\eqref{d.11} that  for sufficiently large $L>0$,
\begin{equation*}
\begin{split}
I^{\rm HFB}_{m,\kappa,\theta}(\lambda)
\leq&G_{\theta}(\lambda)-m\lambda-\frac{\kappa\lambda^{2}}{2L}D(\chi^{4},\chi^{4})+\frac{\kappa\theta\lambda^{2}}{2}+o(\lambda^2L^{-3/2})+O(\lambda L^{-2}).
\end{split}
\end{equation*}
This further yields that there exists a sufficiently large $L>0$, which depends on $\lambda$, such that
\begin{equation}\label{min-G-I}
\begin{split}
I^{\rm HFB}_{m,\kappa,\theta}(\lambda)
<G_{\theta}(\lambda)-m\lambda-\frac{\kappa\lambda^{2}}{4L}D(\chi^{4},\chi^{4})+\frac{\kappa\theta\lambda^{2}}{2}.
\end{split}
\end{equation}
Choose sufficiently large $L>0$ so that \eqref{min-G-I} holds true, and take $\theta\in [0,\theta^{*}]$  so that
\begin{equation}\label{d.13}
\frac{1}{4L}D(\chi^{4},\chi^{4})\geq \theta,
\end{equation}
where $\theta^{*}>0$ is as in Lemma \ref{lem:Gtheta-eigen}.
We then  deduce from \eqref{min-G-I} and \eqref{d.13} that
\begin{align}\label{b.16}
I^{\rm HFB}_{m,\kappa,\theta}(\lambda)<G_{\theta}(\lambda)-m\lambda.
\end{align}
Choose $\tilde{\theta}\in(0, \theta^{*}]$ to be the largest value so that \eqref{b.16} holds true.  Thus, we  obtain that  \eqref{min-Gtheta-relation} holds for $\theta\in (0,\tilde{\theta}]$. This therefore proves Proposition \ref{prop:min-Gtheta}.\qed

\end{document}